\pdfoutput=1
\documentclass[11pt,a4paper]{article}

\usepackage[margin=1in]{geometry}
\usepackage[T1]{fontenc}
\usepackage{lmodern}
\usepackage{amsmath,amssymb,amsthm,mathtools}
\usepackage{mathrsfs}
\usepackage{bm}
\usepackage{booktabs}
\usepackage{array}
\usepackage{graphicx}
\usepackage{multicol}
\usepackage{url}
\usepackage{fancyvrb}
\usepackage{newunicodechar}
\usepackage{xcolor}
\usepackage[colorlinks=true,linkcolor=blue!70!black,citecolor=blue!70!black,urlcolor=blue!70!black]{hyperref}
\usepackage{microtype}

\usepackage[affil-it]{authblk}

\hypersetup{
  pdftitle={Guerra Interpolation and Optimal Transport: Kantorovich Bounds, Brenier Maps and the Replicon},
  pdfauthor={Elena Agliari, Adriano Barra, Andrea Ladiana, Andrea Lepre, Diego Pallara}
}

\DeclareUrlCommand\lean{\urlstyle{tt}}
\makeatletter\g@addto@macro\UrlBreaks{\do\_\do\.}\makeatother
\newcolumntype{L}[1]{>{\raggedright\arraybackslash}p{#1}}
\newcommand{\doi}[1]{doi:\href{https://doi.org/#1}{\nolinkurl{#1}}}

\newunicodechar{¹}{\ensuremath{^{1}}}
\newunicodechar{×}{\ensuremath{\times}}
\newunicodechar{Σ}{\ensuremath{\Sigma}}
\newunicodechar{Φ}{\ensuremath{\Phi}}
\newunicodechar{Ω}{\ensuremath{\Omega}}
\newunicodechar{α}{\ensuremath{\alpha}}
\newunicodechar{β}{\ensuremath{\beta}}
\newunicodechar{ε}{\ensuremath{\varepsilon}}
\newunicodechar{η}{\ensuremath{\eta}}
\newunicodechar{ι}{\ensuremath{\iota}}
\newunicodechar{ν}{\ensuremath{\nu}}
\newunicodechar{ρ}{\ensuremath{\rho}}
\newunicodechar{σ}{\ensuremath{\sigma}}
\newunicodechar{τ}{\ensuremath{\tau}}
\newunicodechar{φ}{\ensuremath{\varphi}}
\newunicodechar{⁻}{\ensuremath{^{-}}}
\newunicodechar{₀}{\ensuremath{_{0}}}
\newunicodechar{₁}{\ensuremath{_{1}}}
\newunicodechar{₂}{\ensuremath{_{2}}}
\newunicodechar{ℕ}{\ensuremath{\mathbb{N}}}
\newunicodechar{ℝ}{\ensuremath{\mathbb{R}}}
\newunicodechar{→}{\ensuremath{\to}}
\newunicodechar{∀}{\ensuremath{\forall}}
\newunicodechar{∂}{\ensuremath{\partial}}
\newunicodechar{∃}{\ensuremath{\exists}}
\newunicodechar{∈}{\ensuremath{\in}}
\newunicodechar{∑}{\ensuremath{\sum}}
\newunicodechar{√}{\ensuremath{\surd}}
\newunicodechar{∧}{\ensuremath{\wedge}}
\newunicodechar{∫}{\ensuremath{\int}}
\newunicodechar{≠}{\ensuremath{\ne}}
\newunicodechar{≤}{\ensuremath{\le}}
\newunicodechar{⊕}{\ensuremath{\oplus}}
\newunicodechar{⨆}{\ensuremath{\bigsqcup}}

\newtheorem{theorem}{Theorem}[section]
\newtheorem{proposition}[theorem]{Proposition}
\newtheorem{lemma}[theorem]{Lemma}

\theoremstyle{definition}

\theoremstyle{remark}
\newtheorem{remark}[theorem]{Remark}

\newcommand{\R}{\mathbb{R}}

\newcommand{\E}{\mathbb{E}}
\newcommand{\Var}{\operatorname{Var}}
\newcommand{\Ent}{\operatorname{Ent}}
\newcommand{\KL}{\mathrm{KL}}
\newcommand{\Ps}{\mathscr{P}}
\newcommand{\W}{\mathcal{W}_{2}}
\newcommand{\Wone}{\mathcal{W}_{1}}
\newcommand{\sech}{\operatorname{sech}}
\newcommand{\lch}{\operatorname{lch}}
\newcommand{\HL}{\mathrm{HL}}
\newcommand{\hlRS}{\mathrm{HL}_{\mathrm{RS}}}
\newcommand{\hlK}{\mathrm{HL}_{K}}
\newcommand{\RS}{\mathrm{RS}}
\newcommand{\Gu}{\mathrm{Gu}}
\newcommand{\LAT}{\Lambda_{\mathrm{AT}}}
\newcommand{\LL}{\Lambda_{\mathrm{L}}}
\newcommand{\dd}{\mathrm{d}}
\newcommand{\eps}{\varepsilon}

\title{\textbf{\Large Guerra Interpolation and Optimal Transport:\\[0.3ex]
Kantorovich Bounds, Brenier Maps and the Replicon}}

\author[a,b,c]{Elena~Agliari}
\author[b,c,d,e]{Adriano~Barra}
\author[b,d]{Andrea~Ladiana}
\author[a,b]{Andrea~Lepre}
\author[f,g,h]{Diego~Pallara}

\affil[a]{\small Dipartimento di Matematica, Sapienza Universit\`a di Roma, Rome, Italy}
\affil[b]{\small Istituto Nazionale d'Alta Matematica, GNFM, Rome, Italy}
\affil[c]{\small CNR-Nanotec, Unit\`a di Lecce, Lecce, Italy}
\affil[d]{\small Dipartimento di Scienze di Base e Applicate all'Ingegneria, Sapienza Universit\`a di Roma, Rome, Italy}
\affil[e]{\small Istituto Nazionale di Fisica Nucleare, Sezione di Roma 1, Rome, Italy}
\affil[f]{\small Dipartimento di Matematica e Fisica ``Ennio De Giorgi'', Universit\`a del Salento, Lecce, Italy}
\affil[g]{\small Istituto Nazionale d'Alta Matematica, GNAMPA, Lecce, Italy}
\affil[h]{\small Istituto Nazionale di Fisica Nucleare, Sezione di Lecce, Lecce, Italy}

\date{November 13, 2026}

\begin{document}

\maketitle

\begin{abstract}
Since Guerra's work, mean-field spin glasses have been studied through Hamilton--Jacobi equations. Here, using the Curie--Weiss and the Sherrington--Kirkpatrick models as prototypical examples, we show that his celebrated interpolation method is itself an optimal transport computation.

After a linear change of variables the finite-$N$ pressure becomes an action $u$ with $\partial_t u = -A/2$ and $\partial_x u = -B$, where $B^2 \le A$, so $u$ is a classical subsolution of $\partial_t u+\frac12(\partial_x u)^2 \le 0$. Integrating along a straight segment gives Guerra's sum rule, and dropping its remainder makes $u(0,\cdot)$ and $u(t,\cdot)$ an admissible pair for the dual Kantorovich problem with quadratic cost.
For the Curie--Weiss model and $t > 0$, $T_t(y) = y-t\tanh y$ is an optimal transport map from every source with finite first moment if and only if $t \le 1$, and the Kantorovich potentials are the limiting pressures.
For the replica-symmetric Sherrington--Kirkpatrick model, the cavity map is injective on the physical half-line and the curvature of its Brenier potential is the longitudinal eigenvalue, which exceeds the replicon eigenvalue $\Lambda_{\mathrm{AT}}$: transport of Dirac masses does not detect the de Almeida--Thouless line. The replicon appears along the $\mathcal{W}_2$ geodesic that splits a Dirac mass into two atoms, as the second variation of a second-order surrogate of the two-atom Parisi functional.
From a Dirac initial datum, Guerra's $K$-step bound takes Hopf--Lax form, and the Parisi formula becomes a Hopf--Lax formula over atomic measures. With a $\mathcal{W}_1$-Lipschitz bound for its initial datum, proved by comparing Cole--Hopf layers, we let $K\to\infty$ and recover at Dirac data the Hopf--Lax form of the Parisi formula due to Mourrat and to Mourrat and Panchenko, with a rate for equal-mass approximations.
All numbered statements are formalized in Lean~4 with Mathlib.
\end{abstract}

\vspace{0.8em}


\vspace{1em}
\hrule

\newpage

\noindent\includegraphics[width=\textwidth]{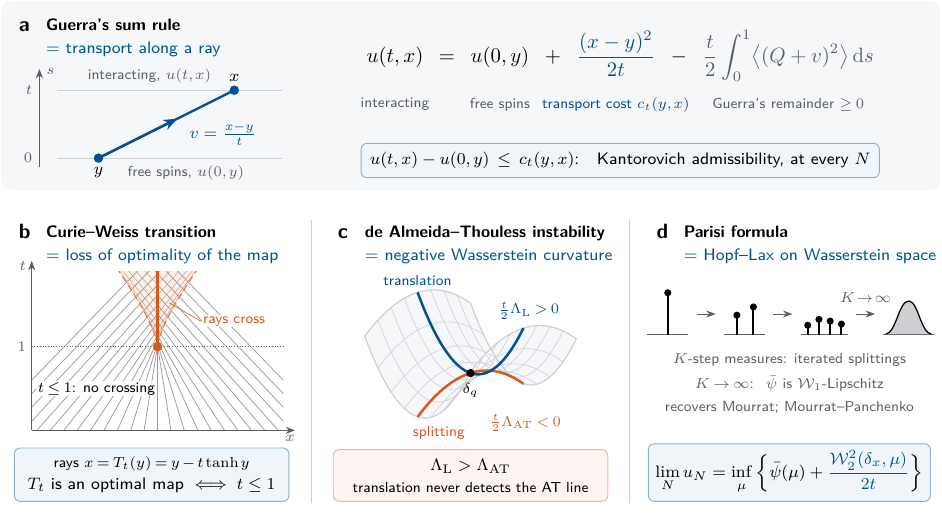}

\vspace{0.5em}
{\small\noindent\textbf{Graphical abstract.}
\textbf{(a)}~Guerra's interpolation moves the effective field along a straight ray from $(0,y)$ to $(t,x)$. In the sum rule for the action $u$, Guerra's quadratic term is the transport cost $c_t(y,x) = (x-y)^2/(2t)$ and the remainder has a sign, so $u(0,\cdot)$ and $u(t,\cdot)$ form an admissible pair for the dual Kantorovich problem at every $N$ (Lemma~\ref{lem:sumrule-path}).
\textbf{(b)}~Curie--Weiss model: the rays $x = T_t(y)$ in the $(x,t)$ plane. For $t \le 1$ they do not cross and $T_t$ is an optimal transport map; for $t > 1$ they cross inside the caustic (dashed), and on the Maxwell line $\{x = 0,\ t > 1\}$ the Hopf--Lax objective has two maximizers (Theorem~\ref{thm:cw-brenier}).
\textbf{(c)}~Sherrington--Kirkpatrick model: at a Dirac mass $\delta_q$ the Hessian of the second-order surrogate of the Parisi functional is diagonal, with entry $\frac{t}{2}\LL$ along translations, positive on the physical half-line, and entry $\frac{t}{2}\LAT$ along the $\W$-geodesic that splits $\delta_q$ into two atoms, negative beyond the de~Almeida--Thouless line (Theorems~\ref{thm:latala-guerra} and~\ref{thm:twoatom}; the surface is schematic).
\textbf{(d)}~Iterated splittings give the $K$-step order parameters; as $K\to\infty$ the Parisi formula takes the Hopf--Lax form of Mourrat and of Mourrat and Panchenko \cite{Mourrat2022,MourratPanchenko2020}, started at $\delta_x$ (Theorem~\ref{thm:klimit}).\par}

\vspace{0.6em}
\hrule
\vspace{0.2em}

{\small
\begin{multicols}{2}[\section*{Contents}]
\makeatletter\@starttoc{toc}\makeatother
\end{multicols}}


\section{Introduction}\label{sec:intro}

\subsection{Guerra's interpolation and optimal transport}\label{sec:guerra-ot}
Determining the limiting free energy of mean-field spin systems, where all spins interact pairwise, has long been a core objective of equilibrium statistical mechanics. 
In disordered media like spin glasses, reaching a complete proof took decades of effort and inspired much of the modern theory \cite{MPV1987,TalagrandI,PanchenkoBook}.

A fundamental turning point came with Guerra's interpolation technique \cite{Guerra2001,GuerraToninelli,Guerra2003}. 
Rather than confronting the interacting system directly, the method couples it continuously to an uncoupled reference model governed by effective fields, tracking the free-energy drift along the path. The resulting sum rule decomposes the free energy into a solvable term, a quadratic correction, and a remainder of definite sign.
In this way Guerra and Toninelli proved the existence of the thermodynamic limit of the Sherrington--Kirkpatrick model \cite{GuerraToninelli}, and Guerra proved that Parisi's solution \cite{Parisi1979,Parisi1980} is a bound for its free energy \cite{Guerra2003}, which Talagrand then showed to be an equality \cite{Talagrand2006}.

Already in his paper on sum rules \cite{Guerra2001}, Guerra observed that once the strength of the interaction and that of the effective field are promoted to independent variables, a fictitious time and a fictitious space, the free energy obeys a partial differential equation of Hamilton--Jacobi type; a problem of statistical mechanics becomes a problem of classical mechanics.
This viewpoint was developed in the following decade, mainly by Barra and Genovese \cite{Barra2008,GenoveseBarra2009} and their collaborators \cite{BarraDiBiasioGuerra2010,BDT2013,Moro1,Moro2}. In their work the free energy is a mechanical action, the order parameter is a velocity field obeying a Burgers equation, phase transitions appear as shocks, and each step of replica symmetry breaking adds a spatial dimension to the mechanical problem.
About ten years later the programme was taken up in probability theory, first by Mourrat \cite{Mourrat2021,Mourrat2021Nonconvex,Mourrat2022,Mourrat2025} and by Mourrat and Panchenko \cite{MourratPanchenko2020}, and then by their collaborators \cite{ChenXia2025}; see \cite{DominguezMourrat2024} for an account. These works gave the approach a general mathematical framework, extended it to a wide class of models, and showed that the Parisi formula itself can be read as the solution of a Hamilton--Jacobi equation on a space of probability measures.

Mean-field models are by now well known to admit a partial differential equation description, but Guerra's interpolation technique has not yet been identified directly as an optimal transport computation. 
Yet the parallel is natural: Guerra's method shifts a reference system toward the interacting state along a prescribed trajectory with a quadratic penalty, while optimal transport \cite{Villani2009} governs mass transportation under minimal cost, where Brenier's theorem characterizes the optimal quadratic maps \cite{Brenier1991}.

In this paper we fill this gap and show that Guerra's interpolation strategy and quadratic optimal transport are linked by an exact dictionary:
\begin{itemize}
\item \emph{Interpolation paths and Kantorovich duality.} Advancing Guerra's interpolation parameter moves the effective field at constant velocity along a straight segment.
The quadratic correction in Guerra's bound is the quadratic transport cost of this trajectory, and Guerra's inequality is the admissibility constraint of Kantorovich's dual problem at every finite system size $N$.
\item \emph{Mean-field equations and Brenier maps.} The self-consistency equation of mean-field theory is the stationarity condition of the Hopf--Lax variational problem. Where the associated Brenier potential is convex, the map it defines is the optimal transport map of Brenier's theorem.
\item \emph{Caustics and the Curie--Weiss transition.} For the Curie--Weiss ferromagnet, the critical inverse temperature is the threshold at which the characteristic trajectories of the transport begin to cross, and beyond which the transport map ceases to be monotone and optimal.
\item \emph{The de~Almeida--Thouless instability and Wasserstein geodesics.} For the Sherrington--Kirkpatrick spin glass in a nonzero external field, transporting a single overlap value as a Dirac mass never detects the de~Almeida--Thouless (AT) instability \cite{AT1978}, because the longitudinal curvature of the Brenier potential remains strictly positive across the AT line.
The instability appears instead along the orthogonal $\W$-geodesic that splits a Dirac mass into two atoms at fixed barycenter: the splitting increases the transport cost and lowers the Cole--Hopf initial datum, and beyond the AT line the second effect prevails at second order.
\item \emph{Hierarchical splittings and the continuum limit.} For $K$-step replica symmetry breaking, Guerra's bounds take the form of Hopf--Lax inequalities over finitely supported overlap distributions.
Passing to the limit $K\to\infty$ via a $\Wone$-Lipschitz estimate for the initial datum recovers, for a Dirac initial measure, the Wasserstein Hopf--Lax formulation of the Parisi formula due to Mourrat and to Mourrat and Panchenko \cite{Mourrat2022,MourratPanchenko2020}.
\item \emph{Information-geometric counterpart.} At the microscopic level of spin configurations, Guerra's path is an exponential family ($e$-geodesic in information geometry) whose remainder decomposes into relative entropy and a mean-square error. For instance, in the Hubbard--Stratonovich representation of the Curie--Weiss model, Guerra's quadratic subtraction is the relative entropy of a Gaussian shift, which for Gaussian measures of equal variance coincides with a quadratic transport cost.
\end{itemize}
The rest of the introduction states these results precisely. Appendix~\ref{app:nutshell} summarizes the statistical-mechanics background and notation for readers coming from partial differential equations or optimal transport.

Finally, a remark on the choice of models. We treat one simple and one complex system, both at the mean-field level, for which the Curie--Weiss and the Sherrington--Kirkpatrick models are the natural prototypes. They serve here as paradigmatic examples: the bridge between Guerra's interpolation and optimal transport rests on the calculus of Section~\ref{sec:abstract}, which makes no reference to a specific model.

\subsection{Guerra's sum rules as Hamilton--Jacobi subsolutions}
For the Curie--Weiss (CW) model Guerra's interpolation gives the mean-field bound \cite{Barra2008}. For the Sherrington--Kirkpatrick (SK) model \cite{SK1975} it gives the replica-symmetric bound and, with Ruelle's cascades, the bounds with broken replica symmetry that lead to the Parisi formula \cite{Guerra2003,Talagrand2006}. We read these bounds through classical optimal transport \cite{Brenier1991,McCann1997,AGS2008,Villani2009,Santambrogio2015}.

In the CW model let $u = S_N := -\alpha_N$, where $\alpha_N = N^{-1}\log Z_N$ is the finite-$N$ statistical pressure, that is, $-\beta$ times the free energy per spin at inverse temperature $\beta$. In the SK model let $u = u_N := 2F_N-x-t/2$, where $F_N = N^{-1}\E\log Z_N$ is the quenched pressure in the presence of an additional Gaussian field of variance $x$ (Section~\ref{sec:sk}); following the spin-glass literature \cite{TalagrandI,PanchenkoBook}, we call $F_N$ the free energy.
In both models,
\begin{equation}\label{eq:intro-pde}
   \partial_tu = -\tfrac12A,\qquad \partial_xu = -B,\qquad B^2 \le A,
\end{equation}
where $B = \langle Q\rangle$ and $A = \langle Q^2\rangle$ are the first two moments of the order parameter under the Gibbs measure: the magnetization $Q = m(\sigma)$ in CW, and the overlap $Q = R(\sigma,\tau)$ of two replicas in SK, where the Gibbs average is also averaged over the disorder. So $u$ is a classical subsolution of the inviscid Hamilton--Jacobi equation:
\[
   \partial_tu+\tfrac12(\partial_xu)^2 = -\tfrac12\big(A-B^2\big) \le 0.
\]
Along the segment from $(0,y)$ to $(t,x)$, travelled at velocity $v = (x-y)/t$, the chain rule and a completion of the square give
\begin{equation}\label{eq:intro-sumrule}
   u(t,x) = u(0,y)+\frac{(x-y)^2}{2t}-\frac{t}{2}\int_0^1\big\langle(Q+v)^2\big\rangle_{(st,\,y+s(x-y))}\,\dd s .
\end{equation}
This is Guerra's sum rule.
For $t > 0$ its remainder is nonpositive, and so \begin{equation}\label{eq:intro-kantorovich}
   u(t,x)-u(0,y) \le \frac{(x-y)^2}{2t}\qquad\text{for all }x,y\in\R .
\end{equation}
In the language of optimal transport, \eqref{eq:intro-kantorovich} says that $u(0,\cdot)$ and $u(t,\cdot)$ form an admissible pair for the dual Kantorovich problem with cost $c_t(y,x) = (x-y)^2/(2t)$.
Taking the infimum over $y$ gives the Hopf--Lax bound $u(t,x) \le \inf_y\{u(0,y)+(x-y)^2/(2t)\}$ \cite{Hopf1965,Lax1957}, an instance of the fact that a subsolution lies below the Hopf--Lax solution with the same initial datum \cite[Section~3.3]{Evans2010}. The initial datum does not depend on $N$: $u(0,y) = -\lch(y)$ with $\lch(y) := \log(2\cosh y)$ in CW, and $u(0,w) = 2\E\lch(h+\sqrt{w}Z)-w$ in SK, where $Z$ is a standard Gaussian variable and $h$ the external field.

In this formulation:
\begin{itemize}
\item Guerra's quadratic correction is the transport cost. In CW the correction $-\frac{t}{2}M^2$ equals $-c_t(x+tM,x)$. In SK the correction $\frac{t}{4}(1-q)^2$ becomes $c_t(w,x)$ after the change of variables, with $w = x+tq$.
\item Guerra's CW bound is a lower bound on $\alpha_N$ and his SK bound is an upper bound on $F_N$. Both are the upper bound \eqref{eq:intro-kantorovich} on $u$. The signs differ because $\partial_t\alpha_N = \frac12\langle m^2\rangle$ in CW, whereas Gaussian integration by parts gives $\partial_tF_N = \frac14\big(1-\langle R^2\rangle\big)$ in SK: the second moment of the order parameter enters the two derivatives with opposite signs.
\item The mean-field self-consistency equations, $M = \tanh(x+tM)$ in CW and $q = \E\tanh^2(h+\sqrt{x+tq}\,Z)$ in SK, are the stationarity condition $x = T_t(y) := \Psi_t'(y)$ of the Hopf--Lax objective, where $\Psi_t(y) = \frac12y^2+t\,u(0,y)$. Where $\Psi_t$ is convex, $T_t$ is the gradient of a convex function, that is, a Brenier map \cite{Brenier1991}.
\end{itemize}

\subsection{Where optimal transport enters}\label{sec:ot}
Beyond the pointwise weak-duality inequality \eqref{eq:intro-kantorovich}, the optimal-transport formulation organizes the main results of the paper as follows.

\smallskip\noindent\emph{(i) Optimal maps in the Curie--Weiss model} (Section~\ref{sec:cw}). Here $T_t(y) = y-t\tanh y$ and $\Psi_t(y) = \frac12y^2-t\lch(y)$.
For $0 < t \le 1$ the potentials $-\lch$ and $-\HL_{\mathrm{val}}(t,\cdot)$, where $\HL_{\mathrm{val}}(t,x) = \lim_N\alpha_N(t,x)$, satisfy the admissibility inequality with equality on the graph of $T_t$. This dual certificate shows that $T_t$ is an optimal transport map from every finite source measure with finite first moment, and it identifies the Kantorovich potentials with the statistical pressures of the model at times $0$ and $t$, the latter in the limit $N\to\infty$.
The contact set of the certificate, the graph of $T_t$, consists of the pairs $(y,x)$ for which $M = \tanh y$ solves the mean-field equation $M = \tanh(x+tM)$ with $y = x+tM$. For $t > 1$ the map $T_t$ is not monotone, and a two-point source shows that it is not optimal. In the $(x,t)$ plane the characteristics $x = T_t(y)$ issued from $\pm y^{*}$, where $y^{*} > 0$ solves $y^{*} = t\tanh y^{*}$, meet on the half-line $\{x = 0,\ t > 1\}$.
This is a Maxwell line, along which the Hopf--Lax objective has the two maximizers $\pm y^{*}$, the two signs of the spontaneous magnetization (Figure~\ref{fig:cw-char}).

\begin{figure}[ht]
\centering
\includegraphics[width=\textwidth]{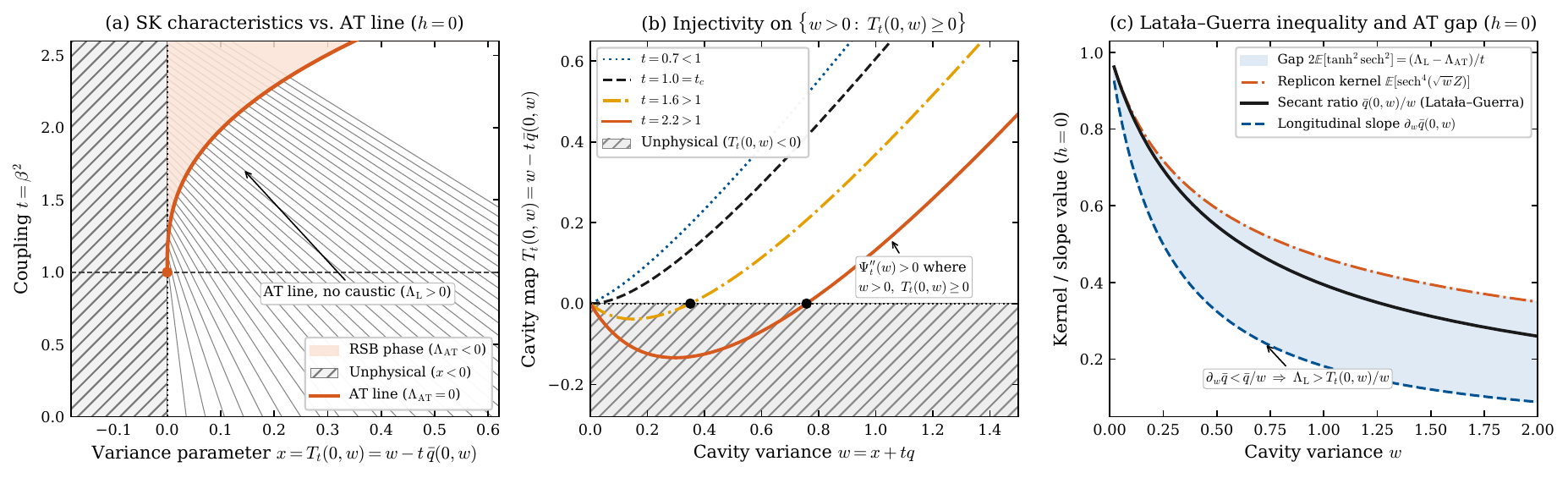}
\caption{The replica-symmetric SK model at $h = 0$. \textbf{(a)}~Characteristics $x = T_t(0,w) = w-t\bar{q}(0,w)$ in the $(x,t)$ plane, for $w\in[0,1.25]$.
The thick curve is the AT line $\LAT = 0$, above which (shaded) $\LAT < 0$; the hatched half-plane $x < 0$ is outside the physical domain. In $\{x > 0\}$ the characteristics do not cross, even above the AT line, where $\LL > 0$ (Theorem~\ref{thm:latala-guerra}); the characteristic issued from $w = 0$ is the line $x = 0$. \textbf{(b)}~$T_t(0,\cdot)$ for $t = 0.7$, $1$, $1.6$ and $2.2$; the dots mark the positive solutions $w^{*}$ of $T_t(0,w) = 0$ for $t > 1$.
The map is strictly increasing where $w > 0$ and $T_t(0,w) \ge 0$. \textbf{(c)}~The slope $\partial_w\bar{q}(0,w)$, the ratio $\bar{q}(0,w)/w$ and $\E\sech^4(\sqrt{w}Z)$. The Lata{\l}a--Guerra inequality is $\partial_w\bar{q} < \bar{q}/w$; the shaded band between $\partial_w\bar{q}$ and $\E\sech^4$ has width $(\LL-\LAT)/t = 2\E[\tanh^2\sech^2]$.}
\label{fig:sk-char}
\end{figure}

\smallskip\noindent\emph{(ii) No caustics in the replica-symmetric SK model} (Section~\ref{sec:sk}). The transport variable is now the cavity variance $w = x+tq$, and $T_t(h,w) = w-t\bar{q}(h,w)$ with $\bar{q}(h,w) = \E\tanh^2(h+\sqrt{w}Z)$.
From the Lata{\l}a--Guerra inequality $w\,\partial_w\bar{q} < \bar{q}$ \cite{Guerra2001}, \cite[Proposition~A.14.1]{TalagrandI} we deduce that $T_t(h,\cdot)$ is injective on the physical half-line $\{w > 0:T_t(h,w) \ge 0\}$, that the Brenier potential is strictly convex there, and that every positive solution $w$ of the replica-symmetric equation $T_t(h,w) = x$ is the strict minimizer of the Hopf--Lax objective over positive variances.
The curvature $\LL = \partial_w^2\Psi_t$ is, up to the factor $2/t$, the second derivative of the replica-symmetric functional along the replica-symmetric (longitudinal) direction, and with $Y = h+\sqrt{w}Z$,
\[
   \LL-\LAT = 2t\,\E\big[\tanh^2(Y)\sech^2(Y)\big] > 0,
\]
where $\LAT$ is the replicon eigenvalue of the de~Almeida--Thouless Hessian \cite{AT1978}.
The characteristics therefore cross the de~Almeida--Thouless (AT) line $\LAT = 0$ without forming caustics, and the transport of Dirac masses $\delta_q$ does not detect the AT instability (Figure~\ref{fig:sk-char}).

\smallskip\noindent\emph{(iii) The replicon as a Wasserstein second variation} (Section~\ref{sec:twoatom}). The instability lives in a direction orthogonal to the curve of Dirac masses.
Splitting $\delta_{\bar{m}}$ into $m\delta_a+(1-m)\delta_b$ at fixed barycenter $\bar{m} = ma+(1-m)b$ is a geodesic for the quadratic Wasserstein distance $\W$, and by Pythagoras' theorem it is orthogonal to the translations of $\delta_{\bar{m}}$. A Riccati identity for the derivatives of $\lch$ shows that a second-order surrogate $\mathcal{J}_2$ of the two-atom Parisi functional has zero first variation along the splitting and second variation $\frac{t}{2}\LAT\,m(1-m)$.
In orthonormal coordinates given by translation and splitting, its Hessian is diagonal with entries $\frac{t}{2}\LL$ and $\frac{t}{2}\LAT$. Under a cubic bound on the remainder this gives Toninelli's instability \cite{Toninelli2002} and shows that the Parisi functional is not displacement convex, in McCann's sense \cite{McCann1997}, along the splitting when $\LAT < 0$.
For the SK model the Parisi functional is convex along linear interpolations of measures \cite{AuffingerChen2015}; if $q^{*}$ solves the replica-symmetric equation and $\nu$ has a smaller value than $\delta_{q^{*}}$, the value along $(1-\lambda)\delta_{q^{*}}+\lambda\nu$ therefore decreases at first order in $\lambda$. Along the $\W$-geodesics above, $\delta_{q^{*}}$ is a critical point and the instability is of second order.

\smallskip\noindent\emph{(iv) Dirac data and the limit $K\to\infty$} (Section~\ref{sec:kstep}). For $K$-step replica symmetry breaking the transport variables are the atoms of the Parisi measure, and the subsolution calculus holds coordinatewise with weights. From a Dirac mass the only coupling is the product coupling.
The cost in Guerra's $K$-step bound is therefore $\W^2(\delta_x,\nu_t)/(2t)$, where $\nu_t$ is the image of the Parisi measure under $q\mapsto x+tq$, and the Parisi formula over atomic measures becomes a Hopf--Lax formula on Wasserstein space started at $\delta_x$. To pass to all measures we prove that the initial datum is $1$-Lipschitz for the distance $\Wone$, comparing Cole--Hopf layers along the monotone coupling built by the north-west corner rule.

\subsection{Related literature and outline of results}\label{sec:sources}
We briefly review how the results proved in Sections~\ref{sec:abstract}--\ref{sec:kstep} relate to the existing literature.

\smallskip\noindent\emph{Sum rules and Kantorovich duality.} The interpolation identities are due to Guerra \cite{Guerra2001,Guerra2003}; for the CW model they are worked out in \cite{Barra2008}.
Lemma~\ref{lem:sumrule-path}, Proposition~\ref{prop:dual-certificate} and Theorem~\ref{thm:cw-hopflax} write them as Kantorovich inequalities. The limit $\alpha_N\to\HL_{\mathrm{val}}$ in Theorem~\ref{thm:cw-hopflax}(b) is the classical variational formula for the CW pressure \cite{EllisNewman1978,Ellis1985}, here with the error bound $\log(N+1)/N$.

\smallskip\noindent\emph{Curie--Weiss.} Theorem~\ref{thm:cw-brenier} proves optimality of $T_t$ for $0 < t \le 1$ by a dual certificate, and non-optimality for $t > 1$. Proposition~\ref{prop:cw-entropy} applies the relative-entropy identities for exponential families (Proposition~\ref{prop:exp-geodesic}; see \cite{AmariNagaoka2000}) to Guerra's interpolation.
Theorem~\ref{thm:cw-fleming} combines the Hubbard--Stratonovich transform \cite{EllisNewman1978,Ellis1985} with the Gibbs variational inequality restricted to Gaussian shifts, a static form of Fleming's logarithmic transformation \cite{Fleming1978} and of the Bou\'e--Dupuis formula \cite{BoueDupuis1998}.

\smallskip\noindent\emph{Replica-symmetric SK.} The replicon eigenvalue $\LAT$ is that of de~Almeida and Thouless \cite{AT1978}.
For the Lata{\l}a--Guerra lemma see \cite{Guerra2001} and \cite[Proposition~A.14.1]{TalagrandI}; Theorem~\ref{thm:latala-guerra}(a) includes a proof. Theorems~\ref{thm:lamL-vs-lamAT} and~\ref{thm:latala-guerra} relate the two eigenvalues to the Brenier potential and to the cavity map.

\smallskip\noindent\emph{Two atoms.} Toninelli proved the instability of the replica-symmetric solution beyond the AT line by a two-atom perturbation \cite{Toninelli2002}; see also \cite{Alessia2024,Chen2021,JagannathTobasco2017,Talagrand2006,TalagrandI}.
Lemma~\ref{lem:riccati-collapse} and Theorem~\ref{thm:twoatom} compute the second variation of the surrogate $\mathcal{J}_2$ along the splitting geodesic, and derive Toninelli's instability and the failure of displacement convexity from the cubic remainder bound. For linear interpolations, Panchenko \cite{Panchenko2005} proved convexity of the Parisi functional between stochastically ordered measures and asked whether it holds in general;
Auffinger and Chen proved strict convexity \cite{AuffingerChen2015}, and Jagannath and Tobasco gave a dynamic programming proof \cite{JagannathTobasco2016}. Further properties of Parisi measures are proved in \cite{Talagrand2006Measures,AuffingerChen2015PTRF,Chen2017}, and uniqueness for enriched convex vector models in \cite{ChenIssa2025}.

\smallskip\noindent\emph{$K$ steps and $K\to\infty$.} The background of Section~\ref{sec:kstep} is Parisi's ansatz \cite{Parisi1979,Parisi1980}, Guerra's broken replica symmetry bound \cite{Guerra2003}, the extended variational principle of Aizenman, Sims and Starr \cite{AizenmanSimsStarr2003}, built on Ruelle's cascades \cite{Ruelle1987}, and the proofs of the Parisi formula by Talagrand \cite{Talagrand2006} and Panchenko \cite{Panchenko2014}; see \cite{PanchenkoBook}.
Mourrat showed that the Parisi formula is a Hamilton--Jacobi equation on the Wasserstein space of probability measures on $\R_{+}$ \cite{Mourrat2022}, following \cite{Mourrat2021}, and Mourrat and Panchenko proved the corresponding Hopf--Lax formula for product reference measures and initial measures of bounded support \cite{MourratPanchenko2020}. The program is developed in \cite{Mourrat2021Nonconvex,ChenXia2025,Mourrat2025} and presented in \cite{DominguezMourrat2024};
Hamilton--Jacobi equations and Hopf--Lax formulas on Wasserstein spaces are studied for their own sake in \cite{GangboNguyenTudorascu2008,AmbrosioFeng2014,Ghoussoub2017}. Theorems~\ref{thm:kstep-hopflax} and~\ref{thm:dirac-hl} write Guerra's $K$-step bound in Hopf--Lax form at Dirac data. Proposition~\ref{prop:psi-lip} is Guerra's continuity estimate \cite{Guerra2003} in the $\Wone$ form of \cite[Proposition~2.1]{Mourrat2022}; the proof given here compares Cole--Hopf layers.
Theorem~\ref{thm:klimit} recovers, at Dirac data, the Hopf--Lax formula of \cite[Theorem~1.1]{Mourrat2022} and \cite{MourratPanchenko2020}, with an explicit rate for the equal-mass approximations; Remark~\ref{rem:mourrat-dict} gives the dictionary.

\smallskip\noindent\emph{Recent work.} Lopatto \cite{Lopatto2026}, through the characterization of Parisi measures of Jagannath and Tobasco \cite{JagannathTobasco2017}, and Kusuoka and Nakajima \cite{KusuokaNakajima2026}, with an $O(N^{-1})$ bound on the overlap fluctuations that does not use the Parisi formula, proved that the SK model with a positive external field is replica symmetric in the whole region predicted by de~Almeida and Thouless.
Chatterjee \cite{Chatterjee2026} gives a historical account of the rigorous theory.

\subsection{Formalization in Lean~4}\label{sec:formal}
All numbered statements (theorems, propositions and lemmas) are proved in Lean~4 \cite{Lean4} with Mathlib \cite{Mathlib}, in the library \texttt{GuerraHJ}: Lean and Mathlib \texttt{v4.28.0}, 18 modules, an axiom check and a root file, 7{,}891 lines in total, 519 theorems and lemmas, no \texttt{sorry}.
The axiom check runs \texttt{\#print axioms} on 125 principal theorems, which depend only on \texttt{propext}, \texttt{Classical.choice} and \texttt{Quot.sound}. Appendix~\ref{app:lean} lists, for each numbered statement, the Lean declarations that prove it, and Appendix~\ref{app:formal} reproduces the main definitions and statements.

The formalization takes the following classical results from the spin-glass literature as explicit hypotheses:
\begin{enumerate}
\item \emph{The Guerra interpolation.} Theorem~\ref{thm:kstep-hopflax}(b) assumes an interpolating function satisfying the derivative identity provided by Guerra's cascade computation \cite{Guerra2003,BarraDiBiasioGuerra2010} (Remark~\ref{rem:guerra-interp}) and derives the weighted Hamilton--Jacobi sum rule and Hopf--Lax bound.
\item \emph{The Parisi formula.} Theorems~\ref{thm:dirac-hl}(c) and~\ref{thm:klimit} assume that a number $F_{\infty}$ is the infimum of the Parisi functional over atomic measures.
The Parisi formula \cite{Talagrand2006,Panchenko2014} identifies this infimum with $\lim_NF_N$, and this identification is its only use.
\item \emph{The cubic remainder of the two-atom expansion.} Theorem~\ref{thm:twoatom}(c) assumes a cubic bound on the remainder; Remark~\ref{rem:cole-hopf-taylor} outlines how it is obtained from the Cole--Hopf equation.
\end{enumerate}
In Section~\ref{sec:klimit}, general probability measures on a compact interval $[c,d]$ are represented in Lean by their quantile functions.
The \texttt{SpinGlass} project \cite{SpinGlassLean}, a Lean~4 formalization of Talagrand's treatise \cite{TalagrandI}, covers Guerra's interpolation and the replica-symmetric bound for the SK model. The main theorem of \cite{KusuokaNakajima2026} and the note \cite{KusuokaNakajima2026Note} are also formalized in Lean~4. \texttt{GuerraHJ} is independent of these projects. The library, together with the code that produces the figures and the tables, is available at \url{https://github.com/coscienza-research/optimal-transport-guerra-interpolation}.

\section{Sum Rules, Kantorovich Duality and Exponential Families}\label{sec:abstract}

We begin by recording the model-independent calculus used throughout the paper: the sum rule along a line segment, together with the resulting Kantorovich inequality, and the relative-entropy identities along an exponential family.

\subsection{The sum rule along a segment}

\begin{lemma}[Sum rule along a path]\label{lem:sumrule-path}
Let $\varphi,M_1,M_2\colon\R\to\R$ and $t,x,y\in\R$ with $t\neq 0$, and put $v = (x-y)/t$.
\begin{enumerate}
\item[(a)] Assume that $\varphi$ is continuous on $[0,1]$, that $M_1$ and $M_2$ are integrable on $[0,1]$, and that
\[
   \varphi'(s) = -\frac{t}{2}M_2(s)-(x-y)M_1(s)\qquad\text{for }s\in(0,1).
\]
Then
\begin{equation}\label{eq:abstract-sumrule}
   \varphi(1) = \varphi(0)+\frac{(x-y)^2}{2t}-\frac{t}{2}\int_0^1\big(M_2(s)+2vM_1(s)+v^2\big)\,\dd s .
\end{equation}
\item[(b)] If, in addition to the assumptions of (a), $t > 0$ and $M_1(s)^2 \le M_2(s)$ for $s\in[0,1]$, then $\varphi(1)-\varphi(0) \le (x-y)^2/(2t)$.
\item[(c)] Let $S,A,B\colon\R^2\to\R$ and $\gamma(s) = (st,\,y+s(x-y))$.
Assume that for every $s\in(0,1)$ the function $S$ is differentiable at $\gamma(s)$ with $\partial_tS = -\frac12A$ and $\partial_xS = -B$ there, and that $S\circ\gamma$, $A\circ\gamma$ and $B\circ\gamma$ are continuous on $[0,1]$. Then
\[
   S(t,x) = S(0,y)+\frac{(x-y)^2}{2t}-\frac{t}{2}\int_0^1\big(A(\gamma(s))+2vB(\gamma(s))+v^2\big)\,\dd s .
\]
\end{enumerate}
\end{lemma}

\begin{proof}
Since $\frac{(x-y)^2}{2t}-\frac{t}{2}\big(M_2+2vM_1+v^2\big) = -\frac{t}{2}M_2-(x-y)M_1$, the assumption in (a) says that $\varphi'$ is the integrand of \eqref{eq:abstract-sumrule}, and (a) follows from the fundamental theorem of calculus. Under the assumptions of (b), $M_2+2vM_1+v^2 \ge (M_1+v)^2 \ge 0$.
For (c), the chain rule gives $\frac{\dd}{\dd s}S(\gamma(s)) = -\frac{t}{2}A(\gamma(s))-(x-y)B(\gamma(s))$ for $s\in(0,1)$, and (a) applies to $\varphi = S\circ\gamma$, $M_2 = A\circ\gamma$ and $M_1 = B\circ\gamma$.
\end{proof}

If moreover $t > 0$ and $B^2 \le A$ along $\gamma$, part~(b) gives $S(t,x)-S(0,y) \le (x-y)^2/(2t)$. This is the situation of the introduction, with $S = u$, $A = \langle Q^2\rangle$ and $B = \langle Q\rangle$; the integrand is then $\langle(Q+v)^2\rangle$.

An inequality $\psi(x)-\varphi(y) \le c(y,x)$ for all $x,y$ says that $(\varphi,\psi)$ is an admissible pair for the dual Kantorovich problem with cost $c$. Integrated against a coupling it gives weak duality, and equality on the graph of a map certifies that the map is optimal.

\begin{proposition}[Coupling bound and optimality certificate]\label{prop:dual-certificate}
Let $\varphi,\psi\colon\R\to\R$ and $c\colon\R\times\R\to\R$ be Borel functions such that
\begin{equation}\label{eq:admissible}
   \psi(x)-\varphi(y) \le c(y,x)\qquad\text{for all }x,y\in\R .
\end{equation}
\begin{enumerate}
\item[(a)] Let $\pi$ be a Borel measure on $\R^2$ with marginals $\pi_0$ and $\pi_1$. If $\varphi\in L^1(\pi_0)$, $\psi\in L^1(\pi_1)$ and $c\in L^1(\pi)$, then
\[
   \int\psi\,\dd\pi_1-\int\varphi\,\dd\pi_0 \le \int c\,\dd\pi .
\]
\item[(b)] Let $T\colon\R\to\R$ be a Borel function such that $\psi(T(y))-\varphi(y) = c(y,T(y))$ for all $y\in\R$, and let $\rho_0$ be a Borel measure on $\R$ with $\varphi\in L^1(\rho_0)$ and $\psi\in L^1(T_{\#}\rho_0)$. Then every Borel measure $\pi$ on $\R^2$ with marginals $\rho_0$ and $T_{\#}\rho_0$ and with $c\in L^1(\pi)$ satisfies
\[
   \int c(y,T(y))\,\dd\rho_0(y) \le \int c\,\dd\pi .
\]
\end{enumerate}
\end{proposition}

\begin{proof}
For (a), integrate \eqref{eq:admissible} against $\pi$. For (b), integrating the equality on the graph of $T$ against $\rho_0$ gives $\int c(y,T(y))\,\dd\rho_0 = \int\psi\,\dd(T_{\#}\rho_0)-\int\varphi\,\dd\rho_0$, which is at most $\int c\,\dd\pi$ by (a).
\end{proof}

The left-hand side of (b) is the cost of the coupling $(\mathrm{id},T)_{\#}\rho_0$.
Part~(b) therefore says that $T$ is an optimal transport map from $\rho_0$ to $T_{\#}\rho_0$ for the cost $c$, and that the optimal cost is $\int\psi\,\dd(T_{\#}\rho_0)-\int\varphi\,\dd\rho_0$.

\subsection{Relative entropy along an exponential family}
Let $\Omega$ be a nonempty finite set, $p$ a probability vector on $\Omega$ with positive entries, and $\Delta\colon\Omega\to\R$. For $s\in\R$ put
\[
   Z_s = \sum_{\sigma\in\Omega}p(\sigma)e^{s\Delta(\sigma)},\qquad
   \mu_s(\sigma) = \frac{p(\sigma)e^{s\Delta(\sigma)}}{Z_s},\qquad
   \Lambda(s) = \log Z_s,
\]
so that $\mu_0 = p$. We write $\E_s$ and $\Var_s$ for mean and variance under $\mu_s$, and $\KL(\mu\|\nu) = \sum_{\sigma}\mu(\sigma)\log(\mu(\sigma)/\nu(\sigma))$. The curve $s\mapsto\mu_s$ is the exponential geodesic ($e$-geodesic) of information geometry from $\mu_0$ to $\mu_1$ \cite{AmariNagaoka2000}.

\begin{proposition}[Relative entropy along an exponential family]\label{prop:exp-geodesic}
With the notation above,
\begin{align}
   \Lambda(1)& = \E_0[\Delta]+\KL(\mu_0\|\mu_1) = \E_1[\Delta]-\KL(\mu_1\|\mu_0),\label{eq:kl-endpoints}\\
   \KL(\mu_0\|\mu_1)& = \int_0^1(1-s)\Var_s(\Delta)\,\dd s,\qquad
   \KL(\mu_1\|\mu_0) = \int_0^1s\Var_s(\Delta)\,\dd s,\label{eq:kl-integrals}\\
   \KL(\mu_0\|\mu_1)&+\KL(\mu_1\|\mu_0) = \int_0^1\Var_s(\Delta)\,\dd s.\label{eq:jeffreys}
\end{align}
\end{proposition}

\begin{proof}
Since $\log(\mu_0(\sigma)/\mu_1(\sigma)) = \Lambda(1)-\Delta(\sigma)$ and $\mu_0$, $\mu_1$ are probability vectors, $\KL(\mu_0\|\mu_1) = \Lambda(1)-\E_0[\Delta]$ and $\KL(\mu_1\|\mu_0) = \E_1[\Delta]-\Lambda(1)$, which is \eqref{eq:kl-endpoints}. Also, $\Lambda'(s) = \E_s[\Delta]$ and $\frac{\dd}{\dd s}\E_s[\Delta] = \Var_s(\Delta)$, so that
\[
   \frac{\dd}{\dd s}\big(\Lambda(s)+(1-s)\E_s[\Delta]\big) = (1-s)\Var_s(\Delta),\qquad
   \frac{\dd}{\dd s}\big(s\E_s[\Delta]-\Lambda(s)\big) = s\Var_s(\Delta).
\]
Integrating over $[0,1]$ and using $\Lambda(0) = 0$ gives \eqref{eq:kl-integrals}; \eqref{eq:jeffreys} is the sum of the two identities.
\end{proof}

Subtracting the two expressions for $\Lambda(1)$ in \eqref{eq:kl-endpoints} also gives $\KL(\mu_0\|\mu_1)+\KL(\mu_1\|\mu_0) = \E_1[\Delta]-\E_0[\Delta]$.
Proposition~\ref{prop:cw-entropy} applies these identities to Guerra's interpolation for the CW model.

\section{The Curie--Weiss Model}\label{sec:cw}

For $N \ge 1$ let $\Sigma_N = \{-1,1\}^N$ and $m(\sigma) = N^{-1}\sum_{i = 1}^N\sigma_i$. For $t,x\in\R$ define
\begin{gather*}
   Z_N(t,x) = \sum_{\sigma\in\Sigma_N}\exp\Big(N\Big(\frac{t}{2}m(\sigma)^2+x\,m(\sigma)\Big)\Big),\\
   \alpha_N(t,x) = \frac{1}{N}\log Z_N(t,x),\qquad S_N = -\alpha_N,
\end{gather*}
and let $\omega_{t,x}(f) = Z_N(t,x)^{-1}\sum_{\sigma}f(\sigma)\exp\big(N(\frac{t}{2}m(\sigma)^2+x\,m(\sigma))\big)$ be the Gibbs average. Under $\omega_{0,x}$ the spins are independent with mean $\tanh x$. We write $\lch(y) = \log(2\cosh y)$, so that $\lch' = \tanh$ and $\lch'' = \sech^2 = 1-\tanh^2\in(0,1]$.

\subsection{Finite-\texorpdfstring{$N$}{N} equations and the Hopf--Lax limit}

\begin{proposition}[Viscous Hamilton--Jacobi and heat equations]\label{prop:cw-pde}
Let $N \ge 1$ and $t,x\in\R$.
\begin{enumerate}
\item[(a)] $\alpha_N(0,x) = \lch(x)$.
\item[(b)] The derivatives of $S_N$ are
\[
   \partial_tS_N = -\tfrac12\omega_{t,x}(m^2),\qquad
   \partial_xS_N = -\omega_{t,x}(m),\qquad
   \partial_x^2S_N = -N\big(\omega_{t,x}(m^2)-\omega_{t,x}(m)^2\big),
\]
and so $\partial_tS_N+\frac12(\partial_xS_N)^2-\frac{1}{2N}\partial_x^2S_N = 0$.
\item[(c)] $\partial_tZ_N = \frac{1}{2N}\partial_x^2Z_N$.
\end{enumerate}
\end{proposition}

\begin{proof}
Part~(a) is the binomial identity $\sum_{\sigma}e^{xNm(\sigma)} = (e^x+e^{-x})^N$.
Differentiating the finite sum gives $\partial_tZ_N = \frac{N}{2}\sum_{\sigma}m^2e^{N(\frac{t}{2}m^2+xm)}$, $\partial_xZ_N = N\sum_{\sigma}m\,e^{N(\frac{t}{2}m^2+xm)}$ and $\partial_x^2Z_N = N^2\sum_{\sigma}m^2e^{N(\frac{t}{2}m^2+xm)} = 2N\partial_tZ_N$, which is (c). Dividing by $NZ_N$ gives (b).
\end{proof}

Since $\omega_{t,x}(m)^2 \le \omega_{t,x}(m^2)$, the action $S_N$ satisfies \eqref{eq:intro-pde} with $A = \omega_{t,x}(m^2)$ and $B = \omega_{t,x}(m)$, and Lemma~\ref{lem:sumrule-path} applies.

\begin{theorem}[Sum rule, Kantorovich bound and Hopf--Lax limit]\label{thm:cw-hopflax}
Let $N \ge 1$.
\begin{enumerate}
\item[(a)] For $t\neq 0$ and $x,M\in\R$,
\begin{equation}\label{eq:cw-sumrule}
   \alpha_N(t,x) = \lch(x+tM)-\frac{t}{2}M^2+\frac{t}{2}\int_0^1\omega_{st,\,x+(1-s)tM}\big((m-M)^2\big)\,\dd s .
\end{equation}
Let $t > 0$, and let $\pi$ be a Borel measure on $\R^2$ with marginals $\pi_0$ and $\pi_1$ such that $\lch\in L^1(\pi_0)$, $\alpha_N(t,\cdot)\in L^1(\pi_1)$ and $\int(x-y)^2\,\dd\pi(y,x) < \infty$. Then the Kantorovich bound
\[
   \int\lch\,\dd\pi_0-\int\frac{(x-y)^2}{2t}\,\dd\pi(y,x) \le \int\alpha_N(t,\cdot)\,\dd\pi_1
\]
holds; in particular, if $\pi_0,\pi_1$ are probability measures with finite second moments, then $\int\lch\,\dd\pi_0-\mathcal{W}_2^2(\pi_0,\pi_1)/(2t) \le \int\alpha_N(t,\cdot)\,\dd\pi_1$.
\item[(b)] Let $t > 0$, and put $\HL(t,x,y) = \lch(y)-(x-y)^2/(2t)$ and $\HL_{\mathrm{val}}(t,x) = \sup_{y\in\R}\HL(t,x,y)$. For every $x\in\R$,
\begin{equation}\label{eq:cw-rate}
   \HL_{\mathrm{val}}(t,x) \le \alpha_N(t,x) \le \HL_{\mathrm{val}}(t,x)+\frac{\log(N+1)}{N},
\end{equation}
and consequently $\alpha_N(t,x)\to\HL_{\mathrm{val}}(t,x)$ as $N\to\infty$.
\end{enumerate}
\end{theorem}

\begin{proof}
(a) Apply Lemma~\ref{lem:sumrule-path}(c) to $S = S_N$, $A = \omega(m^2)$ and $B = \omega(m)$ (Proposition~\ref{prop:cw-pde}(b)) along the segment from $(0,x+tM)$ to $(t,x)$.
Here $v = -M$ and $A+2vB+v^2 = \omega((m-M)^2)$; as $S_N(0,\cdot) = -\lch$, this gives \eqref{eq:cw-sumrule}. For $t > 0$ the integrand is nonnegative, so $\lch(y)-(x-y)^2/(2t) \le \alpha_N(t,x)$ for all $x,y$.
So $(\varphi,\psi) = (-\lch,-\alpha_N(t,\cdot))$ is an admissible Kantorovich pair satisfying \eqref{eq:admissible} for the cost $(x-y)^2/(2t)$, and Proposition~\ref{prop:dual-certificate}(a) gives the Kantorovich coupling inequality; optimizing over couplings of $(\pi_0,\pi_1)$ gives the $\mathcal{W}_2$ form.

(b) The lower bound is the pointwise inequality just proved.
For the upper bound, group the configurations by the number $k$ of positive spins and put $m_k = (2k-N)/N$, so that $Z_N(t,x) = \sum_{k = 0}^N\binom{N}{k}\exp\big(N(\frac{t}{2}m_k^2+xm_k)\big)$. With $y_k = x+tm_k$ we have $\frac{t}{2}m_k^2+xm_k = m_ky_k-(x-y_k)^2/(2t)$, and $\binom{N}{k}e^{Nm_ky_k} \le (2\cosh y_k)^N$ because the left side is one term of the binomial expansion of the right side.
Each of the $N+1$ terms is therefore at most $e^{N\HL(t,x,y_k)} \le e^{N\HL_{\mathrm{val}}(t,x)}$, and $Z_N(t,x) \le (N+1)e^{N\HL_{\mathrm{val}}(t,x)}$.
\end{proof}

The limit in (b) is the classical variational formula for the Curie--Weiss pressure \cite{EllisNewman1978,Ellis1985}. Laplace's method gives the exact order of the error.

\begin{remark}[Laplace asymptotics]\label{rem:laplace}
For $t > 0$, Theorem~\ref{thm:cw-fleming}(a) ({\em vide infra}) gives $Z_N(t,x) = \sqrt{N/(2\pi t)}\int_{\R}e^{N\HL(t,x,y)}\,\dd y$. The global maximizers $\hat{y}$ of $\HL(t,x,\cdot)$ solve $\hat{y}-t\tanh\hat{y} = x$; there is one of them, except on the half-line $\{x = 0,\ t > 1\}$, where there are two. If $(t,x)\neq(1,0)$, every maximizer is nondegenerate, $1-t\sech^2\hat{y} > 0$, and Laplace's method gives
\[
   N\big(\alpha_N(t,x)-\HL_{\mathrm{val}}(t,x)\big)\longrightarrow\log\sum_{\hat{y}}\big(1-t\sech^2\hat{y}\big)^{-1/2}.
\]
The limit is positive. At $(t,x) = (1,0)$ the maximum is quartic, $\HL(1,0,y) = \log 2-y^4/12+O(y^6)$, and $N(\alpha_N(1,0)-\log 2) = \frac14\log N+O(1)$.
\end{remark}

\subsection{The transport map}\label{sec:cw-brenier}
Define
\[
   T_t(y) = y-t\tanh y,\qquad \Psi_t(y) = \frac12y^2-t\lch(y),
\]
so that $T_t = \Psi_t'$.
Since $\partial_y\HL(t,x,y) = (x-T_t(y))/t$, the critical points of $\HL(t,x,\cdot)$ are the solutions of $T_t(y) = x$. Equivalently, $M = \tanh y$ solves the mean-field equation $M = \tanh(x+tM)$ with $y = x+tM$.

\begin{theorem}[Optimal transport in the Curie--Weiss model]\label{thm:cw-brenier}
\leavevmode
\begin{enumerate}
\item[(a)] Let $0 < t \le 1$. For every $y\in\R$,
\[
   \HL_{\mathrm{val}}(t,T_t(y)) = \HL(t,T_t(y),y) = \lch(y)-\frac{t}{2}\tanh^2y .
\]
Let $\rho_0$ be a finite Borel measure on $\R$ with $\int|y|\,\dd\rho_0(y) < \infty$. Then
\begin{equation}\label{eq:cw-brenier-opt}
   \int\frac{(T_t(y)-y)^2}{2t}\,\dd\rho_0(y) = \frac{t}{2}\int\tanh^2y\,\dd\rho_0(y) = \int\lch\,\dd\rho_0-\int\HL_{\mathrm{val}}(t,\cdot)\,\dd(T_{t\#}\rho_0),
\end{equation}
and every Borel measure $\pi$ on $\R^2$ with marginals $\rho_0$ and $T_{t\#}\rho_0$ and $\int(x-y)^2\,\dd\pi(y,x) < \infty$ satisfies
\[
   \int\frac{(T_t(y)-y)^2}{2t}\,\dd\rho_0(y) \le \int\frac{(x-y)^2}{2t}\,\dd\pi(y,x).
\]
\item[(b)] If $t \le 1$, then $\Psi_t$ is strictly convex, $T_t$ is strictly increasing, and for every $x\in\R$ the equation $M = \tanh(x+tM)$ has at most one solution. If $t \ge 0$, then $T_t$ is surjective. If $t > 0$, every maximizer $y$ of $\HL(t,x,\cdot)$ satisfies $T_t(y) = x$; if $0 < t \le 1$, then $\HL(t,x,\cdot)$ has at most one maximizer.
\item[(c)] Let $t > 1$ and $a = \sqrt{\log t}$.
Then $T_t(a) < 0 = T_t(0)$, so $T_t$ is not monotone and $\Psi_t$ is not convex. There exist a Borel probability measure $\rho_0$ on $\R$ and a Borel measure $\pi$ on $\R^2$ with marginals $\rho_0$ and $T_{t\#}\rho_0$ such that
\[
   \int\frac{(x-y)^2}{2t}\,\dd\pi(y,x) < \int\frac{(T_t(y)-y)^2}{2t}\,\dd\rho_0(y).
\]
Also, $\HL_{\mathrm{val}}(t,0) > \log 2 = \HL(t,0,0)$, and every maximizer $\hat{y}$ of $\HL(t,0,\cdot)$ satisfies $\tanh\hat{y}\neq 0$ and $\tanh\hat{y} = \tanh(t\tanh\hat{y})$.
\end{enumerate}
\end{theorem}

\begin{proof}
(a) Let $x = T_t(y)$, so that $x-y = -t\tanh y$. Expanding $(x-z)^2 = (x-y)^2-2(x-y)(z-y)+(z-y)^2$ and using $\lch(z) \le \lch(y)+\tanh(y)(z-y)+\frac12(z-y)^2$, which holds because $\lch'' \le 1$, we get for every $z\in\R$
\[
   \HL(t,x,z)-\HL(t,x,y) = \lch(z)-\lch(y)-\tanh(y)(z-y)-\frac{(z-y)^2}{2t} \le \Big(\frac12-\frac{1}{2t}\Big)(z-y)^2 \le 0 .
\]
So $y$ maximizes $\HL(t,x,\cdot)$, and $\HL(t,x,y) = \lch(y)-\frac{t}{2}\tanh^2y$.
Put $\varphi = -\lch$ and $\psi = -\HL_{\mathrm{val}}(t,\cdot)$. Then $\psi(x)-\varphi(y) = \lch(y)-\HL_{\mathrm{val}}(t,x) \le (x-y)^2/(2t)$ for all $x,y$, with equality if $x = T_t(y)$. The function $\HL_{\mathrm{val}}(t,x)+x^2/(2t)$ is a supremum of affine functions of $x$, hence convex and continuous, and $\log 2 \le \lch(y) \le \log 2+|y|$; all integrals in \eqref{eq:cw-brenier-opt} are therefore finite, and Proposition~\ref{prop:dual-certificate}(b) gives (a).

(b) $\Psi_t''(y) = 1-t\sech^2y$ is at least $1$ if $t \le 0$ and at least $\tanh^2y$ if $0 < t \le 1$, so it vanishes at most at $y = 0$; hence $\Psi_t$ is strictly convex and $T_t = \Psi_t'$ is strictly increasing. If $M$ solves $M = \tanh(x+tM)$, then $y = x+tM$ satisfies $T_t(y) = x$ and $M = \tanh y$, so two solutions would give two preimages of $x$. For $t \ge 0$, $T_t$ is continuous and $|T_t(y)-y| \le t$, so $T_t$ is surjective.
A maximizer of $\HL(t,x,\cdot)$ is a critical point, hence a solution of $T_t(y) = x$, and for $0 < t \le 1$ there is at most one such solution.

(c) Since $\cosh a \le e^{a^2/2} = \sqrt{t} < t$ and $\sinh a \ge a$, we have $a\cosh a < t\sinh a$, that is, $T_t(a) < 0$. Take $\rho_0 = \frac12(\delta_0+\delta_a)$. The map $T_t$ couples $0$ with $0$ and $a$ with $T_t(a)$; the coupling $\pi = \frac12(\delta_{(0,T_t(a))}+\delta_{(a,0)})$ has the same marginals and a cost smaller by $-aT_t(a)/(2t) > 0$.
For the value at $x = 0$ let $y = \sqrt{t-1}$. Then $\cosh y \ge 1+\frac{y^2}{2} = \frac{t+1}{2}$, and $\log c \ge 1-1/c$ with $c = \frac{t+1}{2}$ gives
\[
   \HL(t,0,y)-\log 2 \ge \log\frac{t+1}{2}-\frac{t-1}{2t} \ge \frac{t-1}{t+1}-\frac{t-1}{2t} > 0 .
\]
A maximizer $\hat{y}$ of $\HL(t,0,\cdot)$ satisfies $\hat{y} = t\tanh\hat{y}$ by (b), and $\hat{y}\neq 0$ because $\HL(t,0,0) = \log 2 < \HL_{\mathrm{val}}(t,0)$. So $M = \tanh\hat{y}\neq 0$ and $M = \tanh(tM)$.
\end{proof}

For $0 < t \le 1$, parts (a) and (b) show that $\HL(t,x,\cdot)$ has exactly one maximizer, $T_t^{-1}(x)$, and that the mean-field equation has exactly one solution, $M = \tanh T_t^{-1}(x)$. The pair $(-\lch,-\HL_{\mathrm{val}}(t,\cdot))$ in the proof of (a) is a pair of Kantorovich potentials: up to sign, they are the pressure at time $0$ and the limiting pressure at time $t$ (Theorem~\ref{thm:cw-hopflax}(b)).
Their contact set $\{(y,x):\lch(y)-\HL_{\mathrm{val}}(t,x) = (x-y)^2/(2t)\}$ consists of the pairs for which $y$ maximizes $\HL(t,x,\cdot)$. For $t \le 1$ it is the graph of $T_t$, that is, the set of pairs $(y,x)$ such that $M = \tanh y$ solves the mean-field equation with $y = x+tM$.

\begin{figure}[ht]
\centering
\includegraphics[width=\textwidth]{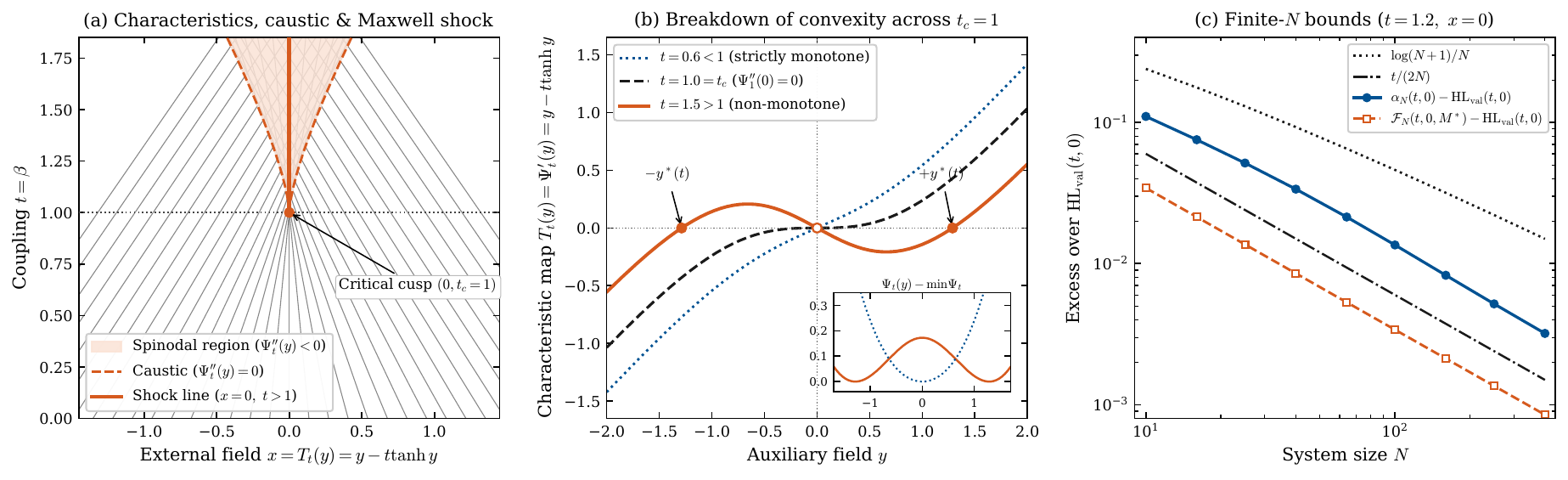}
\caption{The Curie--Weiss model. \textbf{(a)}~Characteristics $x = T_t(y) = y-t\tanh y$ in the $(x,t)$ plane.
The dashed curve is the caustic, the image of $\{\Psi_t''(y) = 0\}$, with its cusp at $(0,1)$; in the shaded region between its branches pass characteristics with $\Psi_t''(y) < 0$. The thick segment is the Maxwell line $\{x = 0,\ t > 1\}$. \textbf{(b)}~$T_t$ for $t = 0.6$, $1$ and $1.5$. For $t = 1.5$ the dots mark $\pm y^{*}$, the two maximizers of $\HL(t,0,\cdot)$, and the open circle marks the critical point $y = 0$.
Inset: $\Psi_t-\min\Psi_t$ for $t = 0.6$ and $t = 1.5$. \textbf{(c)}~At $t = 1.2$, $x = 0$: $\alpha_N(t,0)-\HL_{\mathrm{val}}(t,0)$ and $\mathcal{F}_N(t,0,M^{*})-\HL_{\mathrm{val}}(t,0)$, with $M^{*} = y^{*}/t$, compared with the bounds $\log(N+1)/N$ of Theorem~\ref{thm:cw-hopflax}(b) and $t/(2N)$ of Theorem~\ref{thm:cw-fleming}(c).}
\label{fig:cw-char}
\end{figure}

At $x = 0$ Guerra's bound is $\Gu(t,M) := \lch(tM)-\frac{t}{2}M^2 = \HL(t,0,tM)$.
For $t > 1$ the equation $y = t\tanh y$ has exactly the three solutions $0$ and $\pm y^{*}$, so by (b) and (c) the maximizers of the even function $\HL(t,0,\cdot)$ are $\pm y^{*}$. The two spontaneous magnetizations $\pm M^{*} = \pm y^{*}/t$ satisfy $\Gu(t,\pm M^{*}) = \HL_{\mathrm{val}}(t,0) = \lim_N\alpha_N(t,0)$: they maximize Guerra's bound, which is attained in the limit.
In the $(x,t)$ plane the characteristics $x = T_t(y)$ from $\pm y^{*}$ reach the half-line $\{x = 0,\ t > 1\}$, a Maxwell line on which $\HL(t,0,\cdot)$ has two maximizers (Figure~\ref{fig:cw-char}).

\subsection{Relative entropy and Gaussian shifts}
At $x = 0$ the segment in Theorem~\ref{thm:cw-hopflax}(a) runs from $(0,tM)$ to $(t,0)$.
The Gibbs measures along it, $\mu_s = \omega_{st,(1-s)tM}$, form the exponential family of Section~\ref{sec:abstract} with $p = \omega_{0,tM}$ and $\Delta = N(\frac{t}{2}m^2-tMm)$. Put
\[
   R_N(t,M) = \frac{t}{2}\int_0^1\omega_{st,(1-s)tM}\big((m-M)^2\big)\,\dd s,
\]
so that $\alpha_N(t,0) = \Gu(t,M)+R_N(t,M)$ by \eqref{eq:cw-sumrule}.
For a probability vector $\mu$ on $\Sigma_N$ let $H(\mu) = -\sum_{\sigma}\mu(\sigma)\log\mu(\sigma)$ and let $\mathcal{G}_N(\mu) = \frac{t}{2}\mu(m^2)+\frac{1}{N}H(\mu)$ be the Gibbs functional.

\begin{proposition}[Guerra's remainder, relative entropy and product measures]\label{prop:cw-entropy}
Let $N \ge 1$ and $t\in\R$.
\begin{enumerate}
\item[(a)] For $\psi\in\R$, $\omega_{0,\psi}(m) = \tanh\psi$ and $\omega_{0,\psi}(m^2) = \tanh^2\psi+(1-\tanh^2\psi)/N$.
\item[(b)] Let $t\neq 0$, $M\in\R$, $\mu_0 = \omega_{0,tM}$ and $\mu_1 = \omega_{t,0}$. Then $\frac{1}{N}\KL(\mu_0\|\mu_1) = R_N(t,M)-\frac{t}{2}\,\omega_{0,tM}\big((m-M)^2\big)$, that is,
\begin{equation}\label{eq:cw-rem-kl}
   R_N(t,M)-\frac{1}{N}\KL(\mu_0\|\mu_1) = \frac{t}{2}\big(\tanh(tM)-M\big)^2+\frac{t}{2}\cdot\frac{1-\tanh^2(tM)}{N}.
\end{equation}
In particular, the left side equals $\frac{t}{2}(1-M^2)/N$ if $M = \tanh(tM)$, and is at least $\frac{t}{2}(\tanh(tM)-M)^2$ if $t > 0$.
\item[(c)] For every probability vector $\mu$ on $\Sigma_N$ with positive entries, $\alpha_N(t,0) = \mathcal{G}_N(\mu)+\frac{1}{N}\KL(\mu\|\omega_{t,0})$. For $\psi,M\in\R$,
\begin{align*}
   \mathcal{G}_N(\omega_{0,\psi})& = \lch(\psi)-\psi\tanh\psi+\frac{t}{2}\Big(\tanh^2\psi+\frac{1-\tanh^2\psi}{N}\Big),\\
   \mathcal{G}_N(\omega_{0,tM})& = \Gu(t,M)+\frac{t}{2}\,\omega_{0,tM}\big((m-M)^2\big),\\
   \mathcal{G}_N(\omega_{0,\psi})& \le \Gu(t,\tanh\psi)+\frac{t}{2}\cdot\frac{1-\tanh^2\psi}{N}.
\end{align*}
\end{enumerate}
\end{proposition}

\begin{proof}
(a) Under $\omega_{0,\psi}$ the spins are independent with mean $\tanh\psi$ and variance $1-\tanh^2\psi$.

(b) By \eqref{eq:kl-endpoints}, $\KL(\mu_0\|\mu_1) = \Lambda(1)-\E_0[\Delta]$ with $\Lambda(1) = \log(Z_N(t,0)/Z_N(0,tM)) = N(\alpha_N(t,0)-\lch(tM))$ and $\E_0[\Delta] = N(\frac{t}{2}\mu_0(m^2)-tM\mu_0(m))$. Substituting $\alpha_N(t,0) = \Gu(t,M)+R_N(t,M)$ gives the first identity, and (a) gives $\omega_{0,tM}((m-M)^2) = (\tanh(tM)-M)^2+(1-\tanh^2(tM))/N$.

(c) Since $\log(\mu/\omega_{t,0}) = \log\mu-N\frac{t}{2}m^2+N\alpha_N(t,0)$, we have $\KL(\mu\|\omega_{t,0}) = -H(\mu)-N\frac{t}{2}\mu(m^2)+N\alpha_N(t,0)$.
For the product measure, $H(\omega_{0,\psi}) = N(\lch(\psi)-\psi\tanh\psi)$, and (a) gives the first formula; the second follows from it with $\psi = tM$. For the third put $M = \tanh\psi$ and let $D_{\lch}(z\|\psi) = \lch(z)-\lch(\psi)-\tanh(\psi)(z-\psi) \ge 0$ be the Bregman divergence of $\lch$. The first formula gives
\[
   \mathcal{G}_N(\omega_{0,\psi})+D_{\lch}(tM\|\psi) = \Gu(t,M)+\frac{t}{2}\cdot\frac{1-M^2}{N}.\qedhere
\]
\end{proof}

Part~(b) splits Guerra's remainder into the relative entropy between the endpoints of the interpolation and the mean-square error of the starting product measure. The second term is of order $1/N$ only when $M$ solves the mean-field equation. Part~(c) compares Guerra's bound with the Gibbs functional on product measures. On the product measure with field $tM$, the Gibbs functional exceeds $\Gu(t,M)$ by $\frac{t}{2}\omega_{0,tM}((m-M)^2)$, a term of order one unless $M$ solves the mean-field equation.
On the product measure with magnetization $M$, it is at most $\Gu(t,M)+\frac{t}{2}(1-M^2)/N$. Table~\ref{tab:cw} lists the terms of \eqref{eq:cw-rem-kl} for $N = 30$ and $t = 1.4$.

\begin{remark}[Limits of the Gibbs functional]\label{rem:cw-gibbs-limits}
Letting $N\to\infty$ in Proposition~\ref{prop:cw-entropy}(c), for $t > 0$ and $|M| < 1$,
\begin{align*}
   \lim_N\mathcal{G}_N(\omega_{0,\operatorname{artanh}M})& = \Gu(t,M)-D_{\lch}(tM\|\operatorname{artanh}M) \le \Gu(t,M),\\
   \lim_N\mathcal{G}_N(\omega_{0,tM})& = \Gu(t,M)+\frac{t}{2}\big(\tanh(tM)-M\big)^2 \ge \Gu(t,M).
\end{align*}
Since $\lch$ is strictly convex, each of the two inequalities is an equality if and only if $M = \tanh(tM)$.
\end{remark}

The Hubbard--Stratonovich transform writes $Z_N$ as a Gaussian integral \cite{EllisNewman1978,Ellis1985}.
The Gibbs variational inequality for this integral, restricted to Gaussian shifts, is a static form of Fleming's logarithmic transformation \cite{Fleming1978} and of the Bou\'e--Dupuis formula \cite{BoueDupuis1998}. It gives a lower bound on $\alpha_N$ that lies between Guerra's bound and Guerra's bound plus $t/(2N)$. Let $\gamma_a$ denote the Gaussian measure on $\R$ with mean $a$ and variance $t/N$.

\begin{theorem}[Hubbard--Stratonovich representation and a Gaussian-shift bound]\label{thm:cw-fleming}
Let $N \ge 1$, $t > 0$ and $x\in\R$.
\begin{enumerate}
\item[(a)] $Z_N(t,x) = \int_{\R}(2\cosh y)^N\,\dd\gamma_x(y)$.
\item[(b)] For every $M\in\R$,
\begin{equation}\label{eq:fleming-bound}
   \alpha_N(t,x) \ge \mathcal{F}_N(t,x,M) := \int_{\R}\lch\,\dd\gamma_{x+tM}-\frac{t}{2}M^2,
\end{equation}
and $\frac{1}{N}\KL(\gamma_{x+tM}\|\gamma_x) = \frac{t}{2}M^2$. Equivalently, $\alpha_N(t,x) \ge \int\lch\,\dd\gamma_w-(w-x)^2/(2t)$ for every $w\in\R$.
\item[(c)] For every $M\in\R$,
\begin{equation}\label{eq:fleming-sandwich}
   \lch(x+tM)-\frac{t}{2}M^2 \le \mathcal{F}_N(t,x,M) \le \lch(x+tM)-\frac{t}{2}M^2+\frac{t}{2N}.
\end{equation}
\end{enumerate}
\end{theorem}

\begin{proof}
(a) For each $\sigma$, $e^{tNm(\sigma)^2/2} = \int e^{Nm(\sigma)z}\,\dd\gamma_0(z)$. Summing over $\sigma$ and using $\sum_{\sigma}e^{Nm(\sigma)(x+z)} = (2\cosh(x+z))^N$ gives (a). (b) The density of $\gamma_x$ with respect to $\gamma_{x+tM}$ is $\exp(-NM(y-x-tM)-\frac{Nt}{2}M^2)$, so by (a)
\[
   Z_N(t,x) = \int\exp\Big(N\Big(\lch(y)-M(y-x-tM)-\frac{t}{2}M^2\Big)\Big)\,\dd\gamma_{x+tM}(y),
\]
and Jensen's inequality, with $\int(y-x-tM)\,\dd\gamma_{x+tM} = 0$, gives \eqref{eq:fleming-bound}. The relative entropy of two Gaussian measures with variance $t/N$ and means at distance $tM$ is $(tM)^2N/(2t)$.
The second form of the bound is the first with $w = x+tM$. (c) Integrate $\lch(u)+z\tanh u \le \lch(u+z) \le \lch(u)+z\tanh u+\frac12z^2$, with $u = x+tM$, against the centered Gaussian measure of variance $t/N$.
\end{proof}

\section{The Replica-Symmetric Sherrington--Kirkpatrick Model}\label{sec:sk}

\subsection{Gaussian integration by parts}
Let $Z$ be a standard Gaussian variable, $\iota$ a finite set and $g = (g_k)_{k\in\iota}$ a standard Gaussian vector in $\R^{\iota}$.

\begin{proposition}[Gaussian integration by parts]\label{prop:gaussian-stein}
\leavevmode
\begin{enumerate}
\item[(a)] Let $f\colon\R\to\R$ be differentiable, with $f'$ continuous and $f$, $f'$ bounded. If $Y$ is Gaussian with mean $\mu$ and variance $v \ge 0$, then $\E[(Y-\mu)f(Y)] = v\,\E[f'(Y)]$.
\item[(b)] Let $f\colon\R\to\R$ be twice differentiable, with $f''$ continuous and $f'$, $f''$ bounded. For $h\in\R$ and $w > 0$,
\[
   \frac{\dd}{\dd w}\E f(h+\sqrt{w}Z) = \frac12\E f''(h+\sqrt{w}Z).
\]
\item[(c)] Let $F,DF\colon\R^{\iota}\to\R$ be bounded and continuous, and let $k\in\iota$ be such that $DF$ is the partial derivative of $F$ in the $k$-th coordinate at every point. Then $\E[g_kF(g)] = \E[DF(g)]$.
\item[(d)] Let $\Omega$ be a nonempty finite set, $a\colon\iota\times\Omega\to\R$ and $b\colon\Omega\to\R$. For $g\in\R^{\iota}$ let $\langle f\rangle_g = Z(g)^{-1}\sum_{\sigma\in\Omega}f(\sigma)e^{H_g(\sigma)}$, where $H_g(\sigma) = \sum_ka_k(\sigma)g_k+b(\sigma)$ and $Z(g) = \sum_{\sigma}e^{H_g(\sigma)}$. Then for $k\in\iota$ and $f\colon\Omega\to\R$,
\[
   \E\big[g_k\langle f\rangle_g\big] = \E\big[\langle fa_k\rangle_g-\langle f\rangle_g\langle a_k\rangle_g\big].
\]
\item[(e)] In (d), let $a = a(s)$ depend on $s\in\R$, and assume that for $s$ in a neighbourhood of $s_0$ each $a_k(s,\sigma)$ is differentiable in $s$ with $|\partial_sa_k(s,\sigma)| \le C$. Then $s\mapsto\E\log Z_{a(s)}(g)$ is differentiable at $s_0$, with derivative
\[
   \sum_{k\in\iota}\E\big[\langle a_k\,\partial_sa_k\rangle_g-\langle a_k\rangle_g\langle\partial_sa_k\rangle_g\big]\Big|_{s = s_0} .
\]
\end{enumerate}
\end{proposition}

\begin{proof}
(a) The Gaussian density $\phi_{\mu,v}$ satisfies $(y-\mu)\phi_{\mu,v} = -v\phi_{\mu,v}'$; integrate by parts. (b) Differentiate under the expectation, $\partial_wf(h+\sqrt{w}z) = \frac{z}{2\sqrt{w}}f'(h+\sqrt{w}z)$, and apply (a) to $z\mapsto f'(h+\sqrt{w}z)$. (c) By Fubini's theorem, integrate first in $g_k$ and apply (a). (d) Apply (c) to $F(g) = \langle f\rangle_g$, whose partial derivative in $g_k$ is $\langle fa_k\rangle_g-\langle f\rangle_g\langle a_k\rangle_g$; both are bounded by $\max|f|$ and $2\max|f|\max|a_k|$. (e) The bounds $|\log Z(g)| \le \log|\Omega|+\max|b|+\sum_k\max|a_k|\,|g_k|$ and $|\partial_s\log Z_{a(s)}(g)| \le C\sum_k|g_k|$ allow differentiation under the expectation, and $\partial_s\log Z_{a(s)}(g) = \sum_kg_k\langle\partial_sa_k\rangle_g$; then apply (d) with $f = \partial_sa_k$.
\end{proof}

\subsection{The finite-\texorpdfstring{$N$}{N} subsolution}
Let $N \ge 1$, $h\in\R$ and $t,x \ge 0$, and let $(g_{ij})_{1 \le i,j \le N}$ and $(g_i)_{1 \le i \le N}$ be independent standard Gaussian variables. For $\sigma\in\Sigma_N$ put
\[
   H_N(\sigma) = \sqrt{\frac{t}{2N}}\sum_{i,j = 1}^Ng_{ij}\sigma_i\sigma_j+\sum_{i = 1}^N\big(h+\sqrt{x}\,g_i\big)\sigma_i,
\]
and define
\[
   F_N(h,t,x) = \frac{1}{N}\E\log\sum_{\sigma\in\Sigma_N}e^{H_N(\sigma)},\qquad u_N(h,t,x) = 2F_N(h,t,x)-x-\frac{t}{2}.
\]
The sum runs over ordered pairs, including $i = j$. The diagonal terms do not depend on $\sigma$ and have mean zero, and $(g_{ij}+g_{ji})/\sqrt{2}$ is standard Gaussian, so $F_N$ is the free energy of the SK model with Hamiltonian $\sqrt{t/N}\sum_{i < j}g_{ij}\sigma_i\sigma_j$, inverse temperature $\sqrt{t}$, external field $h$, and an additional Gaussian field of variance $x$.
Let $\langle\cdot\rangle_{t,x}$ be the Gibbs average of two independent replicas $\sigma,\tau$ for given disorder, $R(\sigma,\tau) = N^{-1}\sum_i\sigma_i\tau_i$, and
\[
   B_N(h,t,x) = \E\langle R(\sigma,\tau)\rangle_{t,x},\qquad A_N(h,t,x) = \E\langle R(\sigma,\tau)^2\rangle_{t,x}.
\]
For $w \ge 0$ let $P_0(h,w) = \E\lch(h+\sqrt{w}Z)$ and $u_0(h,w) = 2P_0(h,w)-w$. For $t > 0$ define the replica-symmetric functional and its Hopf--Lax form
\[
   \RS(h,t,x,q) = P_0(h,x+tq)+\frac{t}{4}(1-q)^2,\qquad \hlRS(h,t,x,w) = u_0(h,w)+\frac{(x-w)^2}{2t}.
\]
Expanding the square gives $\hlRS(h,t,x,w) = 2\RS(h,t,x,(w-x)/t)-x-\frac{t}{2}$.

\begin{theorem}[Finite-$N$ subsolution and the replica-symmetric bound]\label{thm:sk-finite}
Let $N \ge 1$ and $h\in\R$.
\begin{enumerate}
\item[(a)] For $t > 0$ and $x \ge 0$, $\partial_tu_N(h,t,x) = -\frac12A_N(h,t,x)$. For $t \ge 0$ and $x > 0$, $\partial_xu_N(h,t,x) = -B_N(h,t,x)$. Always $B_N^2 \le A_N$.
As a result, for $t,x > 0$,
\[
   \partial_tu_N(h,t,x)+\frac12\big(\partial_xu_N(h,t,x)\big)^2 \le 0 .
\]
\item[(b)] For $w \ge 0$, $F_N(h,0,w) = P_0(h,w)$; hence $u_N(h,0,w) = u_0(h,w)$ for every $N$.
\item[(c)] Let $t > 0$ and $x,w \ge 0$ with $x+w > 0$, and put $q = (w-x)/t$. Then
\begin{equation}\label{eq:sk-rs-sumrule}
   F_N(h,t,x) = \RS(h,t,x,q)-\frac{t}{4}\int_0^1\E\big\langle(R(\sigma,\tau)-q)^2\big\rangle_{st,\,w+s(x-w)}\,\dd s \le \RS(h,t,x,q),
\end{equation}
and equivalently
\begin{equation}\label{eq:sk-hopflax-bound}
   u_N(h,t,x) \le u_0(h,w)+\frac{(x-w)^2}{2t} = \hlRS(h,t,x,w).
\end{equation}
\end{enumerate}
\end{theorem}

\begin{proof}
(a) The function $\log\sum_{\sigma}e^{H_N(\sigma)}$ has the form of Proposition~\ref{prop:gaussian-stein}(d), with the $N^2+N$ coordinates $(g_{ij},g_i)$, $a_{ij}(\sigma) = \sqrt{t/(2N)}\,\sigma_i\sigma_j$, $a_i(\sigma) = \sqrt{x}\,\sigma_i$ and $b(\sigma) = h\sum_i\sigma_i$. Proposition~\ref{prop:gaussian-stein}(e) gives
\[
   \partial_tF_N = \frac{1}{4N^2}\sum_{i,j}\E\big[1-\langle\sigma_i\sigma_j\rangle^2\big] = \frac14\big(1-A_N\big),\qquad
   \partial_xF_N = \frac{1}{2N}\sum_i\E\big[1-\langle\sigma_i\rangle^2\big] = \frac12\big(1-B_N\big),
\]
since $N^{-2}\sum_{i,j}\langle\sigma_i\sigma_j\rangle^2 = \langle R^2\rangle$ and $N^{-1}\sum_i\langle\sigma_i\rangle^2 = \langle R\rangle$.
Also, $A_N-B_N^2 = \E\langle(R-B_N)^2\rangle \ge 0$. (b) At $t = 0$ the sum over $\Sigma_N$ factorizes over the sites. (c) The function $s\mapsto u_N(h,st,w+s(x-w))$ is continuous on $[0,1]$, and on $(0,1)$ both $st$ and $w+s(x-w)$ are positive, so (a) gives its derivative there.
Lemma~\ref{lem:sumrule-path}(a), with $M_2 = A_N$, $M_1 = B_N$ and $v = (x-w)/t = -q$, gives \eqref{eq:sk-hopflax-bound} as an identity with remainder $-\frac{t}{2}\int_0^1\E\langle(R-q)^2\rangle\,\dd s$, and the identity $\hlRS(h,t,x,w) = 2\RS(h,t,x,q)-x-\frac{t}{2}$ turns it into \eqref{eq:sk-rs-sumrule}.
\end{proof}

At the corner $x = w = 0$ the field variance vanishes along the whole segment, and (c) does not apply.
There the bound $F_N(h,t,0) \le \RS(h,t,0,0) = \lch(h)+\frac{t}{4}$ is the annealed bound $\E\log\sum_{\sigma}e^{H_N} \le \log\E\sum_{\sigma}e^{H_N}$.

\subsection{The cavity map and the two eigenvalues}
For $h\in\R$, $w \ge 0$ and $Y = h+\sqrt{w}Z$ put
\[
   \bar{q}(h,w) = \E\tanh^2Y,\qquad \bar{q}'(h,w) = \E\big[\sech^2Y\,(1-3\tanh^2Y)\big],
\]
and for $t\in\R$
\begin{align*}
   \Psi_t(h,w)& = \frac12w^2+t\,u_0(h,w), & T_t(h,w)& = w-t\,\bar{q}(h,w),\\
   \LL(h,t,w)& = 1-t\,\bar{q}'(h,w), & \LAT(h,t,w)& = 1-t\,\E\sech^4Y .
\end{align*}
$\LAT$ is the replicon eigenvalue of de~Almeida and Thouless \cite{AT1978}, written in the variable $w = x+tq$.
We call $\LL$ the longitudinal eigenvalue: by Theorem~\ref{thm:lamL-vs-lamAT}(b) below it is $2/t$ times the second derivative of the replica-symmetric functional in $q$, that is, the curvature along the replica-symmetric direction.

\begin{theorem}[Derivatives and the two eigenvalues]\label{thm:lamL-vs-lamAT}
Let $h,t,x\in\R$.
\begin{enumerate}
\item[(a)] For $w > 0$,
\[
   \partial_wP_0(h,w) = \frac12\big(1-\bar{q}(h,w)\big),\qquad \partial_wu_0(h,w) = -\bar{q}(h,w),\qquad \partial_w\bar{q}(h,w) = \bar{q}'(h,w).
\]
\item[(b)] For $q\in\R$ with $x+tq > 0$,
\[
   \partial_q\RS(h,t,x,q) = \frac{t}{2}\big(q-\bar{q}(h,x+tq)\big),\qquad \partial_q^2\RS(h,t,x,q) = \frac{t}{2}\LL(h,t,x+tq).
\]
\item[(c)] For $w \ge 0$,
\begin{equation}\label{eq:lamL-sub-lamAT}
   \LL(h,t,w)-\LAT(h,t,w) = 2t\,\E\big[\tanh^2Y\sech^2Y\big],
\end{equation}
and $\LAT(h,t,w) < \LL(h,t,w)$ if $t > 0$ and $w > 0$.
\end{enumerate}
\end{theorem}

\begin{proof}
(a) Apply Proposition~\ref{prop:gaussian-stein}(b) to $f = \lch$, with $f'' = 1-\tanh^2$, and to $f = \tanh^2$, with $f'' = 2\sech^2(1-3\tanh^2)$. (b) follows from (a) and the chain rule. (c) With $T = \tanh^2u$ we have $\sech^4u-\sech^2u\,(1-3\tanh^2u) = (1-T)^2-(1-T)(1-3T) = 2T(1-T)$.
For $w > 0$ the integrand $\tanh^2Y\sech^2Y$ is positive except at a single value of $Z$.
\end{proof}

By (a), for $w > 0$, $\partial_w\Psi_t(h,w) = T_t(h,w)$ and $\partial_w^2\Psi_t(h,w) = \LL(h,t,w)$: the cavity map $T_t(h,\cdot)$ is the derivative of $\Psi_t(h,\cdot)$, and the longitudinal eigenvalue is its curvature.
By (b), with $w = x+tq$, $\partial_q\RS(h,t,x,q) = \frac12(T_t(h,w)-x)$ and $\partial_w\hlRS(h,t,x,w) = (T_t(h,w)-x)/t$; the critical points are the solutions of the replica-symmetric equation $q = \bar{q}(h,x+tq)$, that is, of $T_t(h,w) = x$. By (c), at every point of the AT line $\LAT = 0$ with $w > 0$ the curvature $\LL$ of the Brenier potential is positive (Table~\ref{tab:sk-at}). At $w = 0$ the gap \eqref{eq:lamL-sub-lamAT} equals $2t\tanh^2h\sech^2h$, which vanishes only for $h = 0$.

\subsection{The Lata{\l}a--Guerra lemma and the absence of caustics}

\begin{theorem}[Lata{\l}a--Guerra lemma and injectivity of the cavity map]\label{thm:latala-guerra}
Let $h\in\R$.
\begin{enumerate}
\item[(a)] For $w > 0$,
\begin{equation}\label{eq:latala-guerra}
   w\,\bar{q}'(h,w) < \bar{q}(h,w).
\end{equation}
So $w\mapsto\bar{q}(h,w)/w$ is strictly decreasing on $(0,\infty)$.
\item[(b)] For $t > 0$ and $w > 0$, $T_t(h,w)/w < \LL(h,t,w)$. In particular, $\LL(h,t,w) > 0$ whenever $T_t(h,w) \ge 0$.
\item[(c)] Let $t > 0$ and $x \ge 0$. The equation $T_t(h,w) = x$ has at most one solution $w > 0$. If $w_0 > 0$ is a solution, then $\hlRS(h,t,x,w_0) < \hlRS(h,t,x,w)$ for every $w > 0$ with $w\neq w_0$.
\end{enumerate}
\end{theorem}

\begin{proof}
(a) Let $a(y) = \tanh^2y-y\tanh y\sech^2y = \sinh y\,(\sinh y\cosh y-y)/\cosh^3y$. Since $\sinh y\cosh y-y$ has the sign of $y$, $a(y) > 0$ for $y\neq 0$. Let $Y = h+\sqrt{w}Z$ and $T = \tanh Y$.
Proposition~\ref{prop:gaussian-stein}(a) applied to $z\mapsto T(1-T^2)$ at $Y = h+\sqrt{w}z$, whose derivative in $Y$ is $(1-T^2)(1-3T^2)$, gives
\[
   w\,\bar{q}'(h,w) = \sqrt{w}\,\E\big[ZT(1-T^2)\big] = \E\big[YT(1-T^2)\big]-h\,\E\big[T(1-T^2)\big],
\]
and so $\bar{q}(h,w)-w\,\bar{q}'(h,w) = \E[a(Y)]+h\,\E[T(1-T^2)]$. The first term is positive. By Proposition~\ref{prop:gaussian-stein}(a) applied to $\tanh^2$, $\E[T(1-T^2)] = \frac{1}{2\sqrt{w}}\E[Z\tanh^2(h+\sqrt{w}Z)]$, and
\[
   \E\big[Z\tanh^2(h+\sqrt{w}Z)\big] = \frac12\E\Big[Z\big(\tanh^2(h+\sqrt{w}Z)-\tanh^2(h-\sqrt{w}Z)\big)\Big]
\]
has the sign of $h$, because $|h+\sqrt{w}z| \ge |h-\sqrt{w}z|$ when $hz \ge 0$. So the second term is nonnegative. The derivative of $\bar{q}(h,w)/w$ is $(w\bar{q}'-\bar{q})/w^2 < 0$.

(b) Multiply \eqref{eq:latala-guerra} by $t/w$: $T_t(h,w)/w = 1-t\bar{q}(h,w)/w < 1-t\bar{q}'(h,w) = \LL(h,t,w)$.

(c) Let $0 < w_1 < w_2$ with $T_t(h,w_1) \ge 0$.
By (a), $1-t\bar{q}(h,w_2)/w_2 > 1-t\bar{q}(h,w_1)/w_1 \ge 0$, hence
\[
   T_t(h,w_2) = w_2\Big(1-t\frac{\bar{q}(h,w_2)}{w_2}\Big) > w_1\Big(1-t\frac{\bar{q}(h,w_1)}{w_1}\Big) = T_t(h,w_1).
\]
So, $T_t(h,\cdot)$ is strictly increasing on $\{w > 0:T_t(h,w) \ge 0\}$, and the value $x \ge 0$ is taken at most once. If $T_t(h,w_0) = x$, the same argument gives $T_t(h,w) > x$ for $w > w_0$ and $T_t(h,w) < x$ for $0 < w < w_0$.
Since $\partial_w\hlRS(h,t,x,w) = (T_t(h,w)-x)/t$, the function $\hlRS(h,t,x,\cdot)$ is strictly decreasing on $(0,w_0]$ and strictly increasing on $[w_0,\infty)$.
\end{proof}

In terms of $q = (w-x)/t$, part~(c) says that a solution $q > -x/t$ of the replica-symmetric equation $q = \bar{q}(h,x+tq)$ is unique and is the strict minimizer of $\RS(h,t,x,\cdot)$ on $\{q:x+tq > 0\}$, so that it gives the best bound in \eqref{eq:sk-rs-sumrule}.

\begin{remark}[The physical half-line and the zero-field case]\label{rem:sk-boundary}
Let $t > 0$. By Theorem~\ref{thm:latala-guerra}(a) the ratio $T_t(h,w)/w = 1-t\bar{q}(h,w)/w$ is strictly increasing in $w > 0$, so $I = \{w > 0:T_t(h,w) \ge 0\}$ is an interval, unbounded above. On $I$ the map $T_t(h,\cdot)$ is strictly increasing with values in $[0,\infty)$, and it is the derivative of $\Psi_t(h,\cdot)$, which is strictly convex there by Theorem~\ref{thm:latala-guerra}(b).
Since $0 \le \bar{q} \le 1$, $T_t(h,\cdot)$ maps a probability measure $\rho$ on $I$ with finite second moment to one with finite second moment, and, being monotone, it is an optimal transport map from $\rho$ to $T_t(h,\cdot)_{\#}\rho$ for the quadratic cost \cite[Chapter~2]{Santambrogio2015}. The cavity map thus transports physical cavity variances to physical field variances optimally, even beyond the AT line.

The endpoint $w = 0$ is excluded in Theorem~\ref{thm:latala-guerra}(c).
Since $T_t(h,0) = -t\tanh^2h$, it solves $T_t(h,w) = x \ge 0$ only if $h = x = 0$. If $h\neq 0$, or $h = 0$ and $x > 0$, then $T_t(h,0) < x$ and $T_t(h,w)\to\infty$, so a solution $w_0 > 0$ of $T_t(h,w) = x$ exists and, by continuity of $\hlRS$ at $w = 0$, minimizes $\hlRS(h,t,x,\cdot)$ strictly on $[0,\infty)$. If $h = x = 0$, then $w = 0$ solves $T_t(0,w) = 0$. Since $\bar{q}(0,w) < w$ for $w > 0$, there is no other solution for $t \le 1$, and $w = 0$ is the minimizer.
For $t > 1$, $\hlRS(0,t,0,w) = 2\log 2+\frac{1-t}{2t}w^2+O(w^3)$, so $w = 0$ is a local maximum on $[0,\infty)$, and the minimizer is the unique positive solution $w^{*}$. For $t = 1.5$: $w^{*}\approx 0.28689$, $q^{*} = w^{*}/t\approx 0.19126$, $\hlRS(0,t,0,w^{*})\approx 1.38282 < 2\log 2\approx 1.38629$, $\LL\approx 0.3045$ and $\LAT\approx-0.0406$ at $w^{*}$.

For $h = 0$ the characteristic $x = T_t(0,w) = w-t\bar{q}(0,w)$ issued from $w > 0$ is a straight line in the $(x,t)$ plane.
It reaches $x = 0$ at time $t = w/\bar{q}(0,w) > 1$ and leaves the physical domain there. By Theorem~\ref{thm:latala-guerra}(c), characteristics issued from different points $w > 0$ do not meet in $\{x \ge 0\}$. The characteristic issued from $w = 0$ is the half-line $\{x = 0\}$, and each of the others meets it at the time it leaves the physical domain.
The half-line $\{x = 0,\ t > 1\}$ is nevertheless not a Maxwell line: the competitor of $w^{*}$ there is the endpoint $w = 0$, a local maximum of $\hlRS$, and not a second minimizer.
\end{remark}

\section{Splitting a Dirac Mass: the Replicon as a Wasserstein Second Variation}\label{sec:twoatom}

By Theorem~\ref{thm:latala-guerra}, the transport of Dirac masses does not see the AT line.
In this section we leave the curve of Dirac masses: we split $\delta_{\bar{m}}$ into two atoms with the same barycenter and compute the second variation of a surrogate of the two-atom Parisi functional along the splitting.

\subsection{The Cole--Hopf layer and a Riccati identity}
For $m\neq 0$, $\tau \ge 0$ and $y\in\R$ let
\[
   \mathcal{T}_{m,\tau}(y) = \frac{1}{m}\log\E\exp\big(m\lch(y+\sqrt{\tau}Z)\big),
\]
the Cole--Hopf transform \cite{Hopf1950,Cole1951} of $\lch$.
Factoring $e^{m\lch(y)}$ out of the expectation gives $\mathcal{T}_{m,\tau}(y) = \lch(y)+\frac{1}{m}\log\E\exp\big(m(\lch(y+\sqrt{\tau}Z)-\lch(y))\big)$, which is finite because $|\lch(y+\sqrt{\tau}z)-\lch(y)| \le \sqrt{\tau}|z|$. For $m = 1$, since $e^{\lch(u)} = e^u+e^{-u}$ and $\E e^{\pm(y+\sqrt{\tau}Z)} = e^{\pm y}e^{\tau/2}$, the Cole--Hopf transform reduces to the linear shift
\begin{equation}\label{eq:cole-hopf-one}
   \mathcal{T}_{1,\tau}(y) = \lch(y)+\frac{\tau}{2}.
\end{equation}

Let $L_1 = \tanh$, $L_2 = \sech^2$, $L_3 = -2\tanh\sech^2$ and $L_4 = -2\sech^2(1-3\tanh^2)$ be the derivatives of orders $1$ to $4$ of $\lch$, and put
\[
   u_1(m,y) = \frac12\big(L_2+mL_1^2\big)(y),\qquad
   u_2(m,y) = \frac18\big(L_4+4mL_1L_3+2mL_2^2+4m^2L_1^2L_2\big)(y).
\]
These are the coefficients of $\tau$ and $\tau^2$ in the expansion of $\mathcal{T}_{m,\tau}(y)$ (Remark~\ref{rem:cole-hopf-taylor}).

\begin{lemma}[Riccati identities]\label{lem:riccati-collapse}
For $m,y\in\R$:
\begin{enumerate}
\item[(a)] $L_2 = 1-L_1^2$, $L_3 = -2L_1L_2$, $L_4 = -2L_2^2-2L_1L_3$, and
\begin{equation}\label{eq:riccati-quartic}
   L_4+4L_1L_3+4L_1^2L_2 = -2L_2^2.
\end{equation}
\item[(b)] $u_1(m,y) = \frac12\big(1-(1-m)\tanh^2y\big) = \frac{m}{2}+\frac{1-m}{2}L_2(y)$, and $\frac14\big(\partial_y^2u_1+2mL_1\partial_yu_1\big)(m,y) = u_2(m,y)$.
\item[(c)] $u_2(m,y) = \dfrac{(1-m)^2}{8}L_4(y)-\dfrac{m(1-m)}{4}\sech^4y$.
\end{enumerate}
\end{lemma}

\begin{proof}
All identities are polynomial identities in $\tanh y$, obtained from the Riccati equation $\tanh' = 1-\tanh^2$.
For (b), $\partial_yu_1 = \frac{1-m}{2}L_3$ and $\partial_y^2u_1 = \frac{1-m}{2}L_4$; the difference $u_2-\frac{1-m}{8}(L_4+2mL_1L_3)$ equals $\frac{m}{8}(L_4+2L_1L_3+2L_2^2)+\frac{m^2}{4}L_1(L_3+2L_1L_2) = 0$ by (a). Similarly, (c) follows from (a) by comparing the coefficients of $1$, $m$ and $m^2$.
\end{proof}

By (c), $u_2$ contains $L_4$ only with the factor $(1-m)^2$, and this part cancels in the second variation computed below.
What remains is $-\frac{m(1-m)}{4}\sech^4$, and $\E\sech^4(h+\sqrt{w}Z)$ is the expectation that appears in $\LAT$.

\subsection{Splitting a Dirac mass}
Let $0 < m < 1$ and $\bar{m}\in\R$. For $\eps \ge 0$ let
\[
   a_{\eps} = \bar{m}-(1-m)\eps,\qquad b_{\eps} = \bar{m}+m\eps,\qquad \nu_{\eps} = m\delta_{a_{\eps}}+(1-m)\delta_{b_{\eps}} .
\]
The barycenter of $\nu_{\eps}$ is $\bar{m}$, and $\nu_0 = \delta_{\bar{m}}$.
The quantile function of $\nu_{\eps}$ equals $a_{\eps}$ on $(0,m]$ and $b_{\eps}$ on $(m,1)$, so it is affine in $\eps$. Since on the line $\W$ is the $L^2$ distance between quantile functions, $\eps\mapsto\nu_{\eps}$ is a geodesic of constant speed, with $\W^2(\delta_{\bar{m}},\nu_{\eps}) = m(1-m)\eps^2$. More generally, the only coupling of $\delta_q$ with $\nu = m\delta_a+(1-m)\delta_b$ is the product coupling (Theorem~\ref{thm:dirac-hl}(a)), and if $\nu$ has barycenter $\bar{m} = ma+(1-m)b$, then
\begin{equation}\label{eq:pythagoras}
   \W^2(\delta_q,\nu) = m(a-q)^2+(1-m)(b-q)^2 = (\bar{m}-q)^2+m(1-m)(b-a)^2.
\end{equation}
Translations of $\delta_q$ and splittings at fixed barycenter are therefore orthogonal directions at $\delta_q$.

For $m\in\R$, $a \le b$ and $x+ta \ge 0$ the two-atom Parisi functional is
\[
   \mathcal{P}_2(a,b) = \mathcal{P}_2(h,t,x,m,a,b) = \E\,\mathcal{T}_{m,t(b-a)}\big(h+\sqrt{x+ta}\,Z\big)+\frac{t}{4}\Big((1-b)^2-m\big(b^2-a^2\big)\Big),
\]
the case $K = 1$ of the functional of Section~\ref{sec:kstep} (Proposition~\ref{prop:pk-cases}). Its last term is the sum of the contribution $\frac{t}{2}(1-b)$ of the top layer, where the Parisi parameter is $1$ by~\eqref{eq:cole-hopf-one}, and of Guerra's overlap term $-\frac{t}{4}(1-ma^2-(1-m)b^2)$. Along the splitting, expanding the squares gives
\[
   \frac{t}{4}\Big((1-b_{\eps})^2-m\big(b_{\eps}^2-a_{\eps}^2\big)\Big) = \frac{t}{4}(1-\bar{m})^2-\frac{mt}{2}\eps+\frac{t}{4}m(1-m)\eps^2.
\]
Put $\bar{w} = x+t\bar{m}$, $\tau = t\eps$ and $w_{\eps} = x+ta_{\eps} = \bar{w}-(1-m)t\eps$, and define the second-order surrogate
\begin{multline*}
   \mathcal{J}_2(\eps) = P_0(h,w_{\eps})+\tau\,\E u_1\big(m,h+\sqrt{w_{\eps}}Z\big)+\tau^2\,\E u_2\big(m,h+\sqrt{\bar{w}}Z\big)\\
   +\frac{t}{4}(1-\bar{m})^2-\frac{mt}{2}\eps+\frac{t}{4}m(1-m)\eps^2.
\end{multline*}
The formula defines $\mathcal{J}_2(\eps)$ whenever $w_{\eps} \ge 0$ and $\bar{w} \ge 0$; it is obtained from $\mathcal{P}_2(a_{\eps},b_{\eps})$ by replacing the Cole--Hopf layer $\E\,\mathcal{T}_{m,\tau}(h+\sqrt{w_{\eps}}Z)$ with its Taylor polynomial of degree two in $\tau$, with the coefficient of $\tau^2$ evaluated at $\bar{w}$ instead of $w_{\eps}$.
We write $\mathcal{J}_2(\eps) = \mathcal{J}_2(h,t,x,m,\bar{m},\eps)$ when the other parameters vary.

\begin{theorem}[Second variation along the splitting]\label{thm:twoatom}
Let $h,t,x,m,\bar{m}\in\R$ and $\bar{w} = x+t\bar{m}$.
\begin{enumerate}
\item[(a)] If $m\neq 0$, then $\mathcal{P}_2(q,q) = \RS(h,t,x,q)$ whenever $x+tq \ge 0$. Also, $\mathcal{J}_2(0) = \RS(h,t,x,\bar{m})$, and if $\bar{w} > 0$, then
\begin{equation}\label{eq:jet-derivs}
   \mathcal{J}_2'(0) = 0,\qquad \mathcal{J}_2''(0) = \frac{t}{2}\LAT(h,t,\bar{w})\,m(1-m).
\end{equation}
\item[(b)] Let $q,a,b\in\R$ with $w_0 = x+tq > 0$, let $\bar{m} = ma+(1-m)b$, and put
\[
   \mathcal{D}(s) = \RS\big(h,t,x,q+s(\bar{m}-q)\big)+\frac{t}{4}\LAT(h,t,w_0)\,m(1-m)\,s^2(b-a)^2.
\]
Then $\mathcal{D}'(0) = \partial_q\RS(h,t,x,q)\,(\bar{m}-q)$ and
\[
   \mathcal{D}''(0) = \frac{t}{2}\LL(h,t,w_0)\,(\bar{m}-q)^2+\frac{t}{2}\LAT(h,t,w_0)\,m(1-m)(b-a)^2.
\]
At $h = 0$ and $w = 0$, $\LL(0,t,0) = \LAT(0,t,0) = 1-t$; both are negative for $t > 1$.
\item[(c)] Suppose that $t > 0$, $0 < m < 1$ and $\LAT(h,t,\bar{w}) < 0$. If $g\colon\R\to\R$ and $C > 0$ satisfy
\[
   \Big|g(\eps)-\RS(h,t,x,\bar{m})-\frac{t}{4}\LAT(h,t,\bar{w})\,m(1-m)\eps^2\Big| \le C\eps^3
\]
for $0 < \eps < \frac{t}{4}|\LAT(h,t,\bar{w})|\,m(1-m)/C$, then $g(\eps) < \RS(h,t,x,\bar{m})$.
Similarly, if $q,\eta\in\R$ with $\eta\neq 0$ and $\Lambda := \LAT(h,t,x+tq) < 0$, and if $f\colon\R\to\R$ and $C > 0$ satisfy $|\eta| < -\frac{t}{10}\Lambda/C$, $f(0) = \RS(h,t,x,q)$, and $\big|f(s)-\RS(h,t,x,q)-\frac{t}{4}\Lambda s^2\eta^2\big| \le Cs^3|\eta|^3$ for $s\in\{\frac12,1\}$, then $f(\frac12) > \frac12\big(f(0)+f(1)\big)$.
\end{enumerate}
\end{theorem}

\begin{proof}
(a) At $a = b = q$ we have $\tau = 0$ and $\mathcal{T}_{m,0} = \lch$, so $\mathcal{P}_2(q,q) = P_0(h,x+tq)+\frac{t}{4}(1-q)^2$, and at $\eps = 0$ the definition of $\mathcal{J}_2$ gives $\mathcal{J}_2(0) = \RS(h,t,x,\bar{m})$. Let $w_{\eps} > 0$. By Lemma~\ref{lem:riccati-collapse}(b), $\E u_1(m,h+\sqrt{w_{\eps}}Z) = \frac12(1-(1-m)\bar{q}(h,w_{\eps}))$, and Theorem~\ref{thm:lamL-vs-lamAT}(a) gives
\[
   \mathcal{J}_2'(\eps) = t^2\eps\Big(\frac{(1-m)^2}{2}\bar{q}'(h,w_{\eps})+2\,\E u_2\big(m,h+\sqrt{\bar{w}}Z\big)\Big)+\frac{t}{2}m(1-m)\eps,
\]
because the terms of order zero in $\eps$, namely $-\frac{(1-m)t}{2}(1-\bar{q}(h,w_{\eps}))$ from $P_0$, $\frac{t}{2}(1-(1-m)\bar{q}(h,w_{\eps}))$ from the $\tau$ term and $-\frac{mt}{2}$, add up to zero. So $\mathcal{J}_2'(0) = 0$ and, $\bar{q}'(h,\cdot)$ being continuous,
\[
   \mathcal{J}_2''(0) = t^2\Big(\frac{(1-m)^2}{2}\bar{q}'(h,\bar{w})+2\,\E u_2\big(m,h+\sqrt{\bar{w}}Z\big)\Big)+\frac{t}{2}m(1-m).
\]
Since $\E L_4(h+\sqrt{\bar{w}}Z) = -2\bar{q}'(h,\bar{w})$, Lemma~\ref{lem:riccati-collapse}(c) gives $2\E u_2 = -\frac{(1-m)^2}{2}\bar{q}'(h,\bar{w})-\frac{m(1-m)}{2}\E\sech^4(h+\sqrt{\bar{w}}Z)$, and $\mathcal{J}_2''(0) = \frac{t}{2}m(1-m)\big(1-t\,\E\sech^4(h+\sqrt{\bar{w}}Z)\big)$.

(b) Where $x+t(q+s(\bar{m}-q)) > 0$, Theorem~\ref{thm:lamL-vs-lamAT}(b) gives $\mathcal{D}'(s) = \partial_q\RS(h,t,x,q+s(\bar{m}-q))(\bar{m}-q)+\frac{t}{2}\LAT(h,t,w_0)m(1-m)(b-a)^2s$ and $\partial_q^2\RS = \frac{t}{2}\LL$. At $h = 0$ and $w = 0$ the Gaussian variable disappears, and $\bar{q}'(0,0) = 1 = \sech^40$.

(c) Under the first set of assumptions, $g(\eps)-\RS(h,t,x,\bar{m}) \le \eps^2\big(C\eps-\frac{t}{4}|\LAT|m(1-m)\big) < 0$.
Under the second, $f(\frac12)-\frac12(f(0)+f(1)) \ge \eta^2\big(-\frac{t}{16}\Lambda-\frac58C|\eta|\big) > 0$.
\end{proof}

\begin{figure}[ht]
\centering
\includegraphics[width=\textwidth]{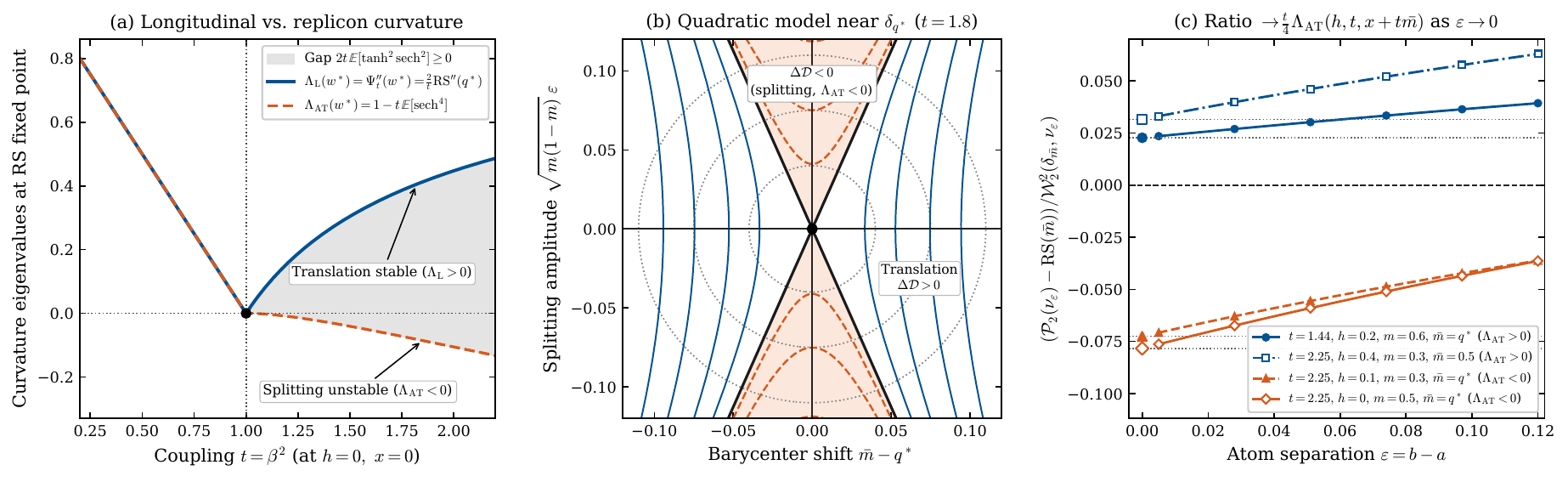}
\caption{Translation and splitting of a Dirac mass. \textbf{(a)}~Along the zero-field branch ($h = x = 0$) of minimizers $w^{*}(t)$ of $\hlRS$ ($w^{*} = 0$ for $t \le 1$, Remark~\ref{rem:sk-boundary}): $\LL(w^{*})$ (solid) and $\LAT(w^{*})$ (dashed), with the gap $2t\,\E[\tanh^2\sech^2]$ shaded.
For $t > 1$, $\LL > 0 > \LAT$. \textbf{(b)}~The quadratic form $\frac{t}{4}(\LL u^2+\LAT v^2)$ at $t = 1.8$, $h = x = 0$ and $w = w^{*}$, in the coordinates $u = \bar{m}-q^{*}$ and $v = \sqrt{m(1-m)}\,\eps$ (Theorem~\ref{thm:twoatom}(b)).
The dotted circles are $\W$-spheres around $\delta_{q^{*}}$; the form is negative in the shaded region. \textbf{(c)}~$(\mathcal{P}_2(\nu_{\eps})-\RS(\bar{m}))/\W^2(\delta_{\bar{m}},\nu_{\eps})$ for the exact two-atom functional, as a function of $\eps$, for the four parameter sets in the legend; the markers at $\eps = 0$ are $\frac{t}{4}\LAT(h,t,x+t\bar{m})$.}
\label{fig:long-vs-at}
\end{figure}

By (a), $\partial_{\eps}\mathcal{J}_2$ vanishes at $\eps = 0$ for every $\bar{m}$ with $\bar{w} > 0$, so the mixed derivative $\partial_{\bar{m}}\partial_{\eps}\mathcal{J}_2$ vanishes at $\eps = 0$, while $\mathcal{J}_2(h,t,x,m,\bar{m},0) = \RS(h,t,x,\bar{m})$ has second derivative $\frac{t}{2}\LL(h,t,\bar{w})$ in $\bar{m}$ (Theorem~\ref{thm:lamL-vs-lamAT}(b)).
By \eqref{eq:pythagoras}, the coordinates $(\bar{m},\sqrt{m(1-m)}\,\eps)$ are orthonormal for $\W$ at $\delta_{\bar{m}}$, and in these coordinates the Hessian of $\mathcal{J}_2$ at $\eps = 0$ is diagonal with entries $\frac{t}{2}\LL$ and $\frac{t}{2}\LAT$. Part~(b) states the same for the quadratic model along an arbitrary displacement interpolation from $\delta_q$.
The longitudinal eigenvalue is the curvature along translations of the Dirac mass, and the replicon eigenvalue is the curvature along splittings (Figure~\ref{fig:long-vs-at}).

At a solution $q^{*}$ of the replica-symmetric equation, $\partial_q\RS(h,t,x,q^{*}) = 0$, so $\delta_{q^{*}}$ is a critical point in both directions. If $t > 0$ and $\LAT < 0$, the splitting is a direction of negative curvature.
Applying part~(c) to $g(\eps) = \mathcal{P}_2(a_{\eps},b_{\eps})$ gives Toninelli's instability \cite{Toninelli2002}, and applying it to the values of $\mathcal{P}_2$ along the $\W$-geodesic $\mu_s = \frac12(\delta_{q-s\eta}+\delta_{q+s\eta})$ shows that $\mathcal{P}_2$ is not displacement convex \cite{McCann1997} along it. Numerically, at $h = x = 0$ and $t = 1.1,1.2,\dots,2.4$, Figure~\ref{fig:rsb-hierarchy}(c) exhibits two-atom measures at which $\mathcal{P}_2$ is smaller than $\min_q\RS(0,t,0,q)$.

\begin{remark}[The cubic remainder]\label{rem:cole-hopf-taylor}
For $m\neq 0$ the function $X(\tau,y) = \mathcal{T}_{m,\tau}(y)$ solves $\partial_{\tau}X = \frac12\big(\partial_y^2X+m(\partial_yX)^2\big)$ with $X(0,\cdot) = \lch$, because $e^{mX}$ solves the heat equation \cite{Hopf1950,Cole1951}. Differentiating the equation at $\tau = 0$ gives $\partial_{\tau}X(0,\cdot) = u_1(m,\cdot)$ and $\frac12\partial_{\tau}^2X(0,\cdot) = u_2(m,\cdot)$;
Lemma~\ref{lem:riccati-collapse}(b) is the second identity written in terms of $u_1$. As the derivatives of $\lch$ of orders $1$ to $6$ are bounded and $w\mapsto\E u_2(m,h+\sqrt{w}Z)$ is Lipschitz on $[0,\infty)$, the third $\tau$-derivative of $X$ is uniformly bounded, which yields the cubic remainder estimate $|\mathcal{P}_2(a_{\eps},b_{\eps})-\mathcal{J}_2(\eps)| \le C\eps^3$ for small $\eps \ge 0$, locally uniformly in the parameters \cite{Toninelli2002}.
Together with~\eqref{eq:jet-derivs}, this verifies the hypotheses of Theorem~\ref{thm:twoatom}(c) for $\mathcal{P}_2$ at small $\eps$ and $\eta$. Table~\ref{tab:twoatom} compares the finite-difference ratio $(\mathcal{P}_2(\nu_{\eps})-\RS(h,t,x,\bar{m}))/\W^2(\delta_{\bar{m}},\nu_{\eps})$ with $\frac{t}{4}\LAT(h,t,\bar{w})$.
\end{remark}

\section{\texorpdfstring{$K$}{K}-Step Replica Symmetry Breaking and the Hopf--Lax Formula at Dirac Data}\label{sec:kstep}

In this section the order parameter is a finitely supported probability measure on $[0,1]$, as in Guerra's bound with $K$ steps of replica symmetry breaking \cite{Guerra2003} and in its Hamilton--Jacobi form \cite{BarraDiBiasioGuerra2010}.
The calculus of Section~\ref{sec:abstract} applies coordinatewise, with one weight for each atom. Started from a Dirac mass, Guerra's bound becomes a Hopf--Lax inequality on the Wasserstein space of the line, and the Parisi formula becomes a Hopf--Lax formula over atomic measures (Theorem~\ref{thm:dirac-hl}). Section~\ref{sec:klimit} lets $K\to\infty$.

\subsection{Weighted subsolutions}
Let $K \ge 0$ and $w\in[0,\infty)^{K+1}$. For $t > 0$ and $x,y\in\R^{K+1}$ put
\[
   c_{t,w}(y,x) = \sum_{a = 0}^Kw_a\frac{(x_a-y_a)^2}{2t}.
\]
When all weights are positive, $c_{t,w}(y,x) = t\,L\big((x-y)/t\big)$ is the action of the straight path from $y$ to $x$ in time $t$ for the Lagrangian $L(v) = \frac12\sum_{a = 0}^Kw_av_a^2$, whose Legendre transform is the Hamiltonian $H(p) = \frac12\sum_{a = 0}^Kp_a^2/w_a$.
If a function $u$ on $(0,\infty)\times\R^{K+1}$ has derivatives $\partial_tu = -\frac12\sum_{a = 0}^Kw_aM_{2,a}$ and $\partial_{x_a}u = -w_aM_{1,a}$ with $M_{1,a}^2 \le M_{2,a}$ for each $a$, then $\partial_tu+H(\nabla_xu) = -\frac12\sum_{a = 0}^Kw_a(M_{2,a}-M_{1,a}^2) \le 0$, so $u$ is a subsolution of $\partial_tu+H(\nabla_xu) = 0$.
Integrating along the segment from $(0,y)$ to $(t,x)$ yields the following weighted version of Lemma~\ref{lem:sumrule-path}, together with the elementary variance identity used below.

\begin{theorem}[Weighted sum rule and variance decomposition]\label{thm:weighted-hj}
Let $K \ge 0$ and $w\in\R^{K+1}$.
\begin{enumerate}
\item[(a)] Let $t\neq 0$, $x,y\in\R^{K+1}$ and $v_a = (x_a-y_a)/t$. Let $\varphi\colon\R\to\R$ be continuous on $[0,1]$, let $M_{1,a},M_{2,a}\colon\R\to\R$ be integrable on $[0,1]$, and assume that
\[
   \varphi'(s) = -\frac{t}{2}\sum_{a = 0}^Kw_aM_{2,a}(s)-\sum_{a = 0}^Kw_a(x_a-y_a)M_{1,a}(s)\qquad\text{for }s\in(0,1).
\]
Then
\begin{equation}\label{eq:weighted-sumrule}
   \varphi(1) = \varphi(0)+\sum_{a = 0}^Kw_a\frac{(x_a-y_a)^2}{2t}-\frac{t}{2}\int_0^1\sum_{a = 0}^Kw_a\big(M_{2,a}(s)+2v_aM_{1,a}(s)+v_a^2\big)\,\dd s .
\end{equation}
If moreover $t > 0$, $w_a \ge 0$ and $M_{1,a}(s)^2 \le M_{2,a}(s)$ for all $a$ and all $s\in[0,1]$, then $\varphi(1)-\varphi(0) \le c_{t,w}(y,x)$.
\item[(b)] If $\sum_{a = 0}^Kw_a = 1$, $q\in\R^{K+1}$ and $\bar{q} = \sum_{a = 0}^Kw_aq_a$, then for every $r\in\R$
\[
   \sum_{a = 0}^Kw_a(q_a-r)^2 = (\bar{q}-r)^2+\sum_{a = 0}^Kw_a(q_a-\bar{q})^2.
\]
\end{enumerate}
\end{theorem}

\begin{proof}
(a) Since
\[
   \sum_{a = 0}^Kw_a\frac{(x_a-y_a)^2}{2t}-\frac{t}{2}\sum_{a = 0}^Kw_a\big(M_{2,a}+2v_aM_{1,a}+v_a^2\big) = -\frac{t}{2}\sum_{a = 0}^Kw_aM_{2,a}-\sum_{a = 0}^Kw_a(x_a-y_a)M_{1,a},
\]
\eqref{eq:weighted-sumrule} follows from the fundamental theorem of calculus, as in Lemma~\ref{lem:sumrule-path}(a). Under the additional assumptions, $M_{2,a}+2v_aM_{1,a}+v_a^2 \ge (M_{1,a}+v_a)^2 \ge 0$. (b) Expand $(q_a-r)^2 = \big((q_a-\bar{q})+(\bar{q}-r)\big)^2$ and use $\sum_{a = 0}^Kw_a(q_a-\bar{q}) = 0$.
\end{proof}

\subsection{The \texorpdfstring{$K$}{K}-step Parisi functional and Guerra's bound}
In the rest of this section $t > 0$, $x \ge 0$ and $h\in\R$. For $g\colon\R\to\R$, $m\neq 0$ and $\tau \ge 0$ let
\[
   (\mathcal{T}_{m,\tau}g)(y) = \frac{1}{m}\log\E\exp\big(mg(y+\sqrt{\tau}Z)\big),
\]
so that $\mathcal{T}_{m,\tau}\lch$ is the Cole--Hopf layer $\mathcal{T}_{m,\tau}$ of Section~\ref{sec:twoatom}. The expectation is finite when $g$ is Lipschitz.

\begin{figure}[ht]
\centering
\includegraphics[width=\textwidth]{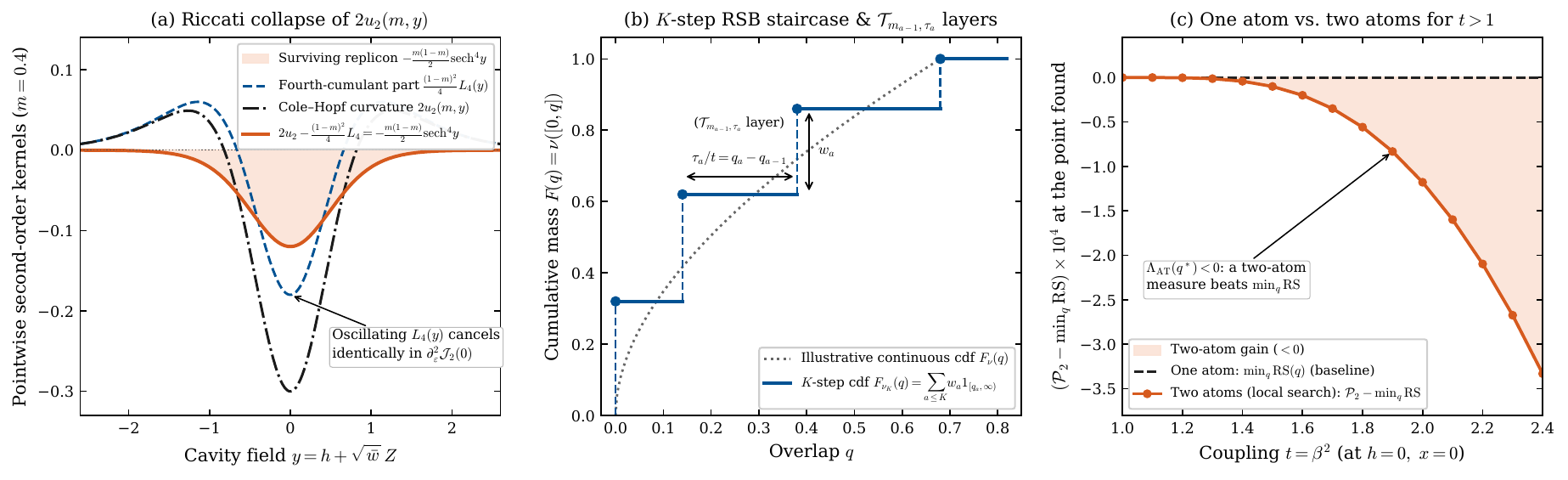}
\caption{Second variation and $K$-step cascades. \textbf{(a)}~The decomposition of Lemma~\ref{lem:riccati-collapse}(c) at $m = 0.4$, $2u_2(m,y) = \frac{(1-m)^2}{4}L_4(y)-\frac{m(1-m)}{2}\sech^4y$, as a function of $y$.
In $\mathcal{J}_2''(0)$ the expectation of the first term cancels against the contribution of the shift $w_{\eps} = \bar{w}-(1-m)t\eps$ of the cavity variance (proof of Theorem~\ref{thm:twoatom}(a)); the second term (shaded) produces $\LAT$. \textbf{(b)}~The distribution function $q\mapsto\nu([0,q])$ of a $K$-step order parameter $\nu = \sum_aw_a\delta_{q_a}$ (staircase), next to an illustrative continuous distribution function (dotted).
The jump at $q_a$ is $w_a$; between two consecutive atoms the cascade applies a Cole--Hopf layer whose parameter is the height of the step and whose variance is $t$ times its length. \textbf{(c)}~Numerical evaluation of $10^4\big(\inf_{m,a,b}\mathcal{P}_2(a,b)-\RS(0,t,0,q^{*})\big)$ at $h = x = 0$ and $t = 1.1,1.2,\dots,2.4$, where $q^{*}$ is the positive solution of the replica-symmetric equation (the marker at $t = 1$ is $0$).}
\label{fig:rsb-hierarchy}
\end{figure}

A \emph{$K$-step order parameter} ($K \ge 0$) is a probability measure $\nu = \sum_{a = 0}^Kw_a\delta_{q_a}$ with weights $w_a > 0$ and atoms $0 \le q_0 \le \dots \le q_K \le 1$; every finitely supported probability measure on $[0,1]$ can be written in this way. Its Parisi parameters are $m_a = w_0+\dots+w_a$ for $0 \le a < K$; they lie in $(0,1)$. Let $y_a = x+tq_a$ be the cavity variances.
Define the cascade $X_K = \lch$ and $X_a = \mathcal{T}_{m_a,\,y_{a+1}-y_a}X_{a+1}$ for $a = K-1,\dots,0$ (Figure~\ref{fig:rsb-hierarchy}(b)); each $X_a$ is $1$-Lipschitz (Lemma~\ref{lem:layer}(a)). Set
\[
   \Phi_{t,x}(\nu) = \E X_0\big(h+\sqrt{y_0}\,Z\big),\qquad
   \mathcal{P}(\nu) = \Phi_{t,x}(\nu)+\frac{t}{4}\Big(1-2q_K+\sum_{a = 0}^Kw_aq_a^2\Big).
\]
An atom repeated in the list contributes a layer with $\tau = 0$, and $\mathcal{T}_{m,0}$ is the identity; thus $\Phi_{t,x}(\nu)$ and $\mathcal{P}(\nu)$ depend only on $\nu$ and not on the way it is written.

Above $q_K$ the Parisi parameter is $1$, and by~\eqref{eq:cole-hopf-one} the corresponding layer $\mathcal{T}_{1,t(1-q_K)}$ adds $\frac{t}{2}(1-q_K)$ to the cascade, while Guerra's overlap term $-\frac{t}{4}\big(1-\sum_aw_aq_a^2\big)$ equals $-\frac{t}{4}\sum_{a = 0}^Km_a(q_{a+1}^2-q_a^2)$ with $m_K = 1$ and $q_{K+1} = 1$ (Proposition~\ref{prop:pk-cases}(c)), which is the form used in \cite{Guerra2003}; their sum is $\frac{t}{4}\big(1-2q_K+\sum_aw_aq_a^2\big)$.

\begin{proposition}[One and two atoms]\label{prop:pk-cases}
\leavevmode
\begin{enumerate}
\item[(a)] For $0 \le q \le 1$, $\mathcal{P}(\delta_q) = \RS(h,t,x,q)$.
\item[(b)] For $0 < m < 1$ and $0 \le a \le b \le 1$, $\mathcal{P}\big(m\delta_a+(1-m)\delta_b\big) = \mathcal{P}_2(a,b)$, the two-atom functional of Section~\ref{sec:twoatom}.
\item[(c)] Let $m_{-1},m_0,\dots,m_K$ and $q_0,\dots,q_{K+1}$ be real numbers with $m_{-1} = 0$, $m_K = 1$ and $q_{K+1} = 1$, and put $w_a = m_a-m_{a-1}$. Then
\[
   \sum_{a = 0}^Km_a\big(q_{a+1}^2-q_a^2\big) = 1-\sum_{a = 0}^Kw_aq_a^2.
\]
\end{enumerate}
\end{proposition}

\begin{proof}
(a) For $K = 0$ the cascade is $X_0 = \lch$, so $\Phi_{t,x}(\delta_q) = P_0(h,x+tq)$, and $\frac{t}{4}(1-2q+q^2) = \frac{t}{4}(1-q)^2$. (b) For $K = 1$ the cascade is the single layer $\mathcal{T}_{m,t(b-a)}\lch$, with outer variance $x+ta$, and $\frac{t}{4}\big(1-2b+ma^2+(1-m)b^2\big) = \frac{t}{4}\big((1-b)^2-m(b^2-a^2)\big)$. (c) Shifting the index in the first half of the sum gives $\sum_{a = 0}^Km_aq_{a+1}^2-\sum_{a = 0}^Km_aq_a^2 = m_Kq_{K+1}^2-\sum_{a = 0}^K(m_a-m_{a-1})q_a^2$.
\end{proof}

For $\Phi\in\R$ and $w,q,y\in\R^{K+1}$ put
\[
   \mathcal{P}_K(\Phi,w,q) = \Phi+\frac{t}{4}\Big(1-2q_K+\sum_{a = 0}^Kw_aq_a^2\Big),\qquad
   \hlK(\Phi,w,y) = 2\Phi-y_K+\sum_{a = 0}^Kw_a\frac{(y_a-x)^2}{2t},
\]
so that $\mathcal{P}(\nu) = \mathcal{P}_K(\Phi_{t,x}(\nu),w,q)$. For $K = 0$, $w_0 = 1$ and $\Phi = P_0(h,y_0)$, $\hlK$ is the functional $\hlRS(h,t,x,y_0)$ of Section~\ref{sec:sk}.

\begin{theorem}[$K$-step Hopf--Lax form of Guerra's bound]\label{thm:kstep-hopflax}
Let $K \ge 0$, $\Phi\in\R$ and $w,q\in\R^{K+1}$, and put $y_a = x+tq_a$.
\begin{enumerate}
\item[(a)] $2\mathcal{P}_K(\Phi,w,q)-x-\frac{t}{2} = \hlK(\Phi,w,y)$. If $\sum_aw_a = 1$, then with $\bar{q} = \sum_aw_aq_a$ and $V = \sum_aw_a(q_a-\bar{q})^2$,
\begin{equation}\label{eq:pk-var}
   \mathcal{P}_K(\Phi,w,q) = \Phi+\frac{t}{4}\big((1-\bar{q})^2+V\big)-\frac{t}{2}\,(q_K-\bar{q}).
\end{equation}
\item[(b)] Let $N \ge 1$, let $w_a \ge 0$ with $\sum_aw_a = 1$, and let $\varphi,M_{1,a},M_{2,a}\colon\R\to\R$, $0 \le a \le K$, satisfy the following conditions:
\begin{itemize}
\item $\varphi$ is continuous on $[0,1]$, $\varphi(0) = 2\Phi-y_K$ and $\varphi(1) = u_N(h,t,x)$;
\item $M_{1,a}$ and $M_{2,a}$ are integrable on $[0,1]$, and $M_{1,a}(s)^2 \le M_{2,a}(s)$ for $s\in[0,1]$;
\item $\varphi'(s) = -\frac{t}{2}\sum_{a = 0}^Kw_aM_{2,a}(s)+t\sum_{a = 0}^Kw_aq_aM_{1,a}(s)$ for $s\in(0,1)$.
\end{itemize}
Let $I = \int_0^1\sum_{a = 0}^Kw_a\big(M_{2,a}(s)-2q_aM_{1,a}(s)+q_a^2\big)\,\dd s$. Then
\begin{equation}\label{eq:kstep-sumrule}
   u_N(h,t,x) = \hlK(\Phi,w,y)-\frac{t}{2}\,I,\qquad
   F_N(h,t,x) = \mathcal{P}_K(\Phi,w,q)-\frac{t}{4}\,I .
\end{equation}
For every $s\in[0,1]$ the integrand of $I$ is at least $\sum_aw_a\big(M_{1,a}(s)-q_a\big)^2 \ge 0$; consequently
\begin{equation}\label{eq:kstep-ub}
   F_N(h,t,x) \le \mathcal{P}_K(\Phi,w,q),\qquad u_N(h,t,x) \le \hlK(\Phi,w,y).
\end{equation}
In particular, if $\nu = \sum_aw_a\delta_{q_a}$ is a $K$-step order parameter and $\Phi = \Phi_{t,x}(\nu)$, then $F_N(h,t,x) \le \mathcal{P}(\nu)$.
\end{enumerate}
\end{theorem}

\begin{proof}
(a) Since $\sum_aw_a(y_a-x)^2/(2t) = \frac{t}{2}\sum_aw_aq_a^2$, both sides of the first identity equal $2\Phi-x-tq_K+\frac{t}{2}\sum_aw_aq_a^2$.
Formula~\eqref{eq:pk-var} follows from $\sum_aw_aq_a^2 = \bar{q}^2+V$, the case $r = 0$ of Theorem~\ref{thm:weighted-hj}(b).

(b) Apply Theorem~\ref{thm:weighted-hj}(a) from the point $y$ to the point with all coordinates equal to $x$. Then $x-y_a = -tq_a$ and $v_a = -q_a$, the assumption on $\varphi'$ is that of Theorem~\ref{thm:weighted-hj}(a), and since $\varphi(0)+\sum_aw_a(y_a-x)^2/(2t) = \hlK(\Phi,w,y)$, \eqref{eq:weighted-sumrule} is the first identity in~\eqref{eq:kstep-sumrule}.
The second follows from the first, from~(a) and from $u_N = 2F_N-x-\frac{t}{2}$. Finally, $M_{2,a}-2q_aM_{1,a}+q_a^2 = (M_{2,a}-M_{1,a}^2)+(M_{1,a}-q_a)^2$.
\end{proof}

\begin{remark}[Guerra's interpolation]\label{rem:guerra-interp}
Let $\nu = \sum_aw_a\delta_{q_a}$ be a $K$-step order parameter. Guerra's interpolation \cite{Guerra2003}, in the Hamilton--Jacobi form of \cite{BarraDiBiasioGuerra2010}, provides functions with the properties assumed in Theorem~\ref{thm:kstep-hopflax}(b) for $\Phi = \Phi_{t,x}(\nu)$.
Let $\tilde{\alpha}(s)$, $0 \le s \le 1$, be Guerra's interpolating pressure for $\nu$, with the layer of parameter $1$ above $q_K$ included in the cascade, so that $\tilde{\alpha}(1) = F_N(h,t,x)$ and $\tilde{\alpha}(0) = \Phi_{t,x}(\nu)+\frac{t}{2}(1-q_K)$.
Then $\varphi(s) = 2\tilde{\alpha}(s)-x-t+\frac{st}{2}$ has the prescribed values at $0$ and $1$, $M_{1,a}(s)$ and $M_{2,a}(s)$ are the first two moments of the overlap of two replicas at level $a$ of the cascade, at time $s$ and averaged over the disorder, and Guerra's formula for $\tilde{\alpha}'(s)$ becomes the identity for $\varphi'$.
\end{remark}

\subsection{Dirac data}\label{sec:dirac}
For $E\subset\R$ we denote by $\Ps(E)$ the set of Borel probability measures on $\R$ supported in $E$.
For a finitely supported measure $\mu = \sum_{a = 0}^Kw_a\delta_{y_a}\in\Ps([0,\infty))$, with $w_a > 0$ and $0 \le y_0 \le \dots \le y_K$, let $X_K = \lch$ and $X_a = \mathcal{T}_{m_a,\,y_{a+1}-y_a}X_{a+1}$ with $m_a = w_0+\dots+w_a$, as above, and put
\[
   \Phi_0(\mu) = \E X_0\big(h+\sqrt{y_0}\,Z\big),\qquad \psi(\mu) = 2\Phi_0(\mu)-y_K.
\]
If $\nu$ is a $K$-step order parameter and $\nu_t = \sum_aw_a\delta_{x+tq_a}$ is its image under $q\mapsto x+tq$, then $\Phi_0(\nu_t) = \Phi_{t,x}(\nu)$ and $\psi(\nu_t) = 2\Phi_{t,x}(\nu)-y_K$. The function $\psi$ depends neither on $t$ nor on $x$, and $\psi(\delta_w) = 2P_0(h,w)-w = u_0(h,w)$: it extends the initial datum $u_0(h,\cdot)$ of Section~\ref{sec:sk} from Dirac masses to atomic measures.
For $\mu\in\Ps(\R)$ let $\Pi(\delta_x,\mu)$ be the set of Borel measures on $\R^2$ with marginals $\delta_x$ and $\mu$, and
\[
   \W^2(\delta_x,\mu) = \inf\Big\{\int(y-x')^2\,\dd\rho(x',y):\rho\in\Pi(\delta_x,\mu)\Big\}.
\]

\begin{theorem}[Transport from a Dirac mass and the Parisi formula]\label{thm:dirac-hl}
\leavevmode
\begin{enumerate}
\item[(a)] \emph{Trivial transport.} Let $\mu$ be a Borel measure on $\R$.
Every Borel measure $\rho$ on $\R^2$ with marginals $\delta_x$ and $\mu$ is the image of $\mu$ under $y\mapsto(x,y)$. If $\mu = \sum_{a = 0}^Kw_a\delta_{y_a}$ with $w_a \ge 0$ and $\sum_aw_a = 1$, then $\delta_x\otimes\mu\in\Pi(\delta_x,\mu)$ and every $\rho\in\Pi(\delta_x,\mu)$ has cost $\int(y-x')^2\,\dd\rho(x',y) = \sum_aw_a(y_a-x)^2$. So $\W^2(\delta_x,\mu) = \sum_aw_a(y_a-x)^2$; for $\mu = \nu_t$ this is $t^2\sum_aw_aq_a^2$.
\item[(b)] \emph{Hopf--Lax form.} For every $K$-step order parameter $\nu$,
\[
   2\mathcal{P}(\nu)-x-\frac{t}{2} = \psi(\nu_t)+\frac{\W^2(\delta_x,\nu_t)}{2t}.
\]
\item[(c)] \emph{Infimum over atomic measures.} If $F_{\infty}\in\R$ is the infimum of $\mathcal{P}(\nu)$ over all $K$-step order parameters $\nu$, $K \ge 0$, then $u_{\infty} := 2F_{\infty}-x-\frac{t}{2}$ is the infimum of $\psi(\nu_t)+\W^2(\delta_x,\nu_t)/(2t)$ over the same $\nu$.
\item[(d)] \emph{One and two atoms.} For $w \ge 0$, $\psi(\delta_w)+(w-x)^2/(2t) = \hlRS(h,t,x,w)$.
Let $0 < m < 1$ and $\bar{m}\in\R$ with $\bar{w} = x+t\bar{m} > 0$. The function $\eps\mapsto 2\mathcal{J}_2(h,t,x,m,\bar{m},\eps)-x-\frac{t}{2}$ of Section~\ref{sec:twoatom} equals $\hlRS(h,t,x,\bar{w})$ at $\eps = 0$, and its first and second derivatives at $\eps = 0$ are $0$ and $t\LAT(h,t,\bar{w})\,m(1-m)$.
\end{enumerate}
\end{theorem}

\begin{proof}
(a) The first marginal of $\rho$ is $\delta_x$, so $x' = x$ for $\rho$-almost every $(x',y)$, and $\rho$ is the image of its second marginal $\mu$ under $y\mapsto(x,y)$.
For $\mu = \sum_aw_a\delta_{y_a}$ the cost of $\rho$ is therefore $\int(y-x)^2\,\dd\mu = \sum_aw_a(y_a-x)^2$, and $\delta_x\otimes\mu$ has the marginals $\delta_x$ and $\mu$ because $\sum_aw_a = 1$. (b) Combine Theorem~\ref{thm:kstep-hopflax}(a), with $\Phi = \Phi_{t,x}(\nu)$, and~(a). (c) By~(b) the two functionals differ by the increasing affine map $z\mapsto 2z-x-\frac{t}{2}$, which maps greatest lower bounds to greatest lower bounds. (d) The first identity is the definition of $\hlRS$, since $\psi(\delta_w) = u_0(h,w)$.
The second follows from Theorem~\ref{thm:twoatom}(a) and $\hlRS(h,t,x,\bar{w}) = 2\RS(h,t,x,\bar{m})-x-\frac{t}{2}$.
\end{proof}

In~(d) the two terms of the Hopf--Lax functional pull in opposite directions. Let $\nu_{\eps,t}$ be the image of the split measure $\nu_{\eps}$ of Section~\ref{sec:twoatom} under $q\mapsto x+tq$. Its transport cost is $\W^2(\delta_x,\nu_{\eps,t})/(2t) = \frac{t}{2}\big(\bar{m}^2+m(1-m)\eps^2\big)$, which contributes $t\,m(1-m)$ to the second derivative in~(d).
The rest, $2\mathcal{J}_2-x-\frac{t}{2}-\W^2(\delta_x,\nu_{\eps,t})/(2t)$, is the second-order surrogate of the initial datum $\psi(\nu_{\eps,t})$, and it contributes $-t^2\,\E\sech^4(h+\sqrt{\bar{w}}Z)\,m(1-m)$. Splitting a Dirac mass thus increases the transport cost and lowers the initial datum, and $\LAT(h,t,\bar{w}) < 0$ means that the second effect prevails.

Part~(c) applies to the SK model through the Parisi formula.
The Gaussian part of the Hamiltonian $H_N$ of Section~\ref{sec:sk} has covariance $N\xi(R(\sigma,\tau))$ with $\xi(r) = xr+\frac{t}{2}r^2$, so $F_N(h,t,x)$ is the free energy of a mixed $p$-spin model with $p\in\{1,2\}$ and external field $h$. For this model $\mathcal{P}$ is the Parisi functional written in the variables of Section~\ref{sec:sk}: the cavity variances are $y_a = \xi'(q_a)$, and $q\xi'(q)-\xi(q) = \frac{t}{2}q^2$ produces Guerra's overlap term.
The Parisi formula, due to Talagrand for $x = 0$ \cite{Talagrand2006} and to Panchenko in general \cite{Panchenko2014}, states that $\lim_NF_N(h,t,x)$ is the infimum of $\mathcal{P}$ over all $K$-step order parameters. So~(c) holds with $F_{\infty} = \lim_NF_N(h,t,x)$, and then $u_{\infty} = \lim_Nu_N(h,t,x)$.

\begin{remark}[Relation to the Hamilton--Jacobi equation of Mourrat]\label{rem:mourrat}
In Talagrand's statement of the Parisi formula \cite{Talagrand2006} the infimum runs over finitely many steps of replica symmetry breaking, that is, over the atomic measures of Theorem~\ref{thm:dirac-hl}(c). Mourrat showed that the Parisi formula can be recast as the solution of a Hamilton--Jacobi equation on the Wasserstein space of probability measures on $\R_{+}$ \cite{Mourrat2022}; see \cite{DominguezMourrat2024}.
Theorem~\ref{thm:dirac-hl} is the case of a Dirac initial datum, where the transport problem is trivial, with the infimum restricted to atomic measures. Section~\ref{sec:klimit} removes this restriction, and Remark~\ref{rem:mourrat-dict} compares the result with \cite{Mourrat2022,MourratPanchenko2020}. The initial datum $\psi(\nu_t) = 2\Phi_{t,x}(\nu)-y_K$ involves the largest atom $y_K$ of $\nu_t$, not its barycenter.
By~\eqref{eq:pk-var}, replacing $y_K$ by the barycenter would change $\mathcal{P}$ by $\frac{t}{2}(q_K-\bar{q})$; along the splitting of Section~\ref{sec:twoatom} this is $\frac{t}{2}m\eps$, and the first variation in Theorem~\ref{thm:twoatom}(a) would be $\frac{t}{2}m$ instead of $0$.
\end{remark}

\subsection{The limit \texorpdfstring{$K\to\infty$}{K to infinity} at Dirac data}\label{sec:klimit}
To pass from finitely supported measures (Theorem~\ref{thm:dirac-hl}(c)) to all probability measures in $\Ps([x,x+t])$, we combine the $\Wone$-Lipschitz continuity of $\psi$ (Proposition~\ref{prop:psi-lip}) with equal-mass quantile discretization.

Let $0 \le c \le d$.
A finitely supported measure $\nu\in\Ps([c,d])$ is written $\nu = \sum_{a = 0}^Kw_a\delta_{y_a}$ with weights $w_a > 0$ and atoms $c \le y_0 \le \dots \le y_K \le d$, and $\Phi_0(\nu)$ and $\psi(\nu) = 2\Phi_0(\nu)-y_K$ are defined as in Section~\ref{sec:dirac}. For any $\mu\in\Ps([c,d])$ let $Q_{\mu}(u) = \inf\{y \ge c:\mu([c,y]) \ge u\}$ for $0 < u \le 1$, and $Q_{\mu}(u) = d$ for $u > 1$.
The quantile function $Q_{\mu}$ is nondecreasing with values in $[c,d]$, the image of the Lebesgue measure on $(0,1)$ under $Q_{\mu}$ is $\mu$, and the monotone coupling $(Q_{\mu}(U),Q_{\mu'}(U))$, with $U$ uniform on $(0,1)$, is optimal for every convex cost on the line \cite[Theorem~6.0.2]{AGS2008}.
So \begin{equation}\label{eq:quantile-dist}
   \Wone(\mu,\mu') = \int_0^1\big|Q_{\mu}(u)-Q_{\mu'}(u)\big|\,\dd u,\qquad
   \W^2(\delta_x,\mu) = \int_0^1\big(Q_{\mu}(u)-x\big)^2\,\dd u;
\end{equation}
the second identity also follows from Theorem~\ref{thm:dirac-hl}(a). The \emph{equal-mass discretization} of $\mu$ at level $n \ge 0$ is
\[
   \mu^{(n)} = \frac{1}{n+1}\sum_{i = 0}^n\delta_{Q_{\mu}((i+1)/(n+1))}.
\]

\begin{lemma}[Quantiles and equal-mass discretization]\label{lem:quantile}
Let $\mu,\mu'\in\Ps([c,d])$ and let $n \ge 0$.
\begin{enumerate}
\item[(a)] If $\nu = \sum_{a = 0}^Kw_a\delta_{y_a}$ is finitely supported on $[c,d]$ and $s_a = w_0+\dots+w_{a-1}$, then $Q_{\nu} = y_a$ on $(s_a,s_{a+1}]$ for $a = 0,\dots,K$, and $\int_0^1f(Q_{\nu}(u))\,\dd u = \sum_aw_af(y_a)$ for every $f\colon\R\to\R$.
\item[(b)] $Q_{\mu}(u) \le Q_{\mu^{(n)}}(u) \le Q_{\mu}\big(u+\frac{1}{n+1}\big)$ for $0 < u \le 1$, and $\Wone(\mu^{(n)},\mu) \le (d-c)/(n+1)$.
\item[(c)] If $c = x$ and $t > 0$, then $\big|\W^2(\delta_x,\mu)-\W^2(\delta_x,\mu')\big|/(2t) \le \frac{d-x}{t}\,\Wone(\mu,\mu')$.
\end{enumerate}
\end{lemma}

\begin{proof}
(a) For $u\in(s_a,s_{a+1}]$ we have $\nu([c,y_a]) \ge s_{a+1} \ge u$, while $\nu([c,y]) \le s_a < u$ for $y < y_a$; thus $Q_{\nu}(u) = y_a$, and the integral splits over the intervals $(s_a,s_{a+1}]$, of lengths $w_a$. (b) By~(a), $Q_{\mu^{(n)}}(u) = Q_{\mu}((j+1)/(n+1))$ for $u\in(j/(n+1),(j+1)/(n+1)]$, and both inequalities follow from the monotonicity of $Q_{\mu}$. With $\delta = 1/(n+1)$, therefore,
\[
   \Wone(\mu^{(n)},\mu) \le \int_0^1\big(Q_{\mu}(u+\delta)-Q_{\mu}(u)\big)\,\dd u = \int_1^{1+\delta}Q_{\mu}(u)\,\dd u-\int_0^{\delta}Q_{\mu}(u)\,\dd u \le \delta\,(d-c).
\]
(c) For $p,q\in[x,d]$, $|(p-x)^2-(q-x)^2| = |p-q|\,(p+q-2x) \le 2(d-x)\,|p-q|$; integrate along the monotone coupling and use~\eqref{eq:quantile-dist}.
\end{proof}

To establish the $\Wone$-Lipschitz continuity of $\psi$, we apply the layers $\mathcal{T}_{m,\tau}$ of Section~\ref{sec:kstep} to $1$-Lipschitz functions $g\colon\R\to\R$, with $0 \le m \le 1$. For $m = 0$ we put $(\mathcal{T}_{0,\tau}g)(y) = \E[g(y+\sqrt{\tau}Z)]$, the heat semigroup. All these expectations are finite, since $|g(y+\sqrt{\tau}z)-g(y)| \le \sqrt{\tau}|z|$.

\begin{lemma}[Cole--Hopf layers of a Lipschitz function]\label{lem:layer}
Let $g,g_1,g_2\colon\R\to\R$ be $1$-Lipschitz, $0 \le m \le m' \le 1$, and $\tau,\tau_1,\tau_2 \ge 0$.
\begin{enumerate}
\item[(a)] $\mathcal{T}_{m,\tau}g$ is $1$-Lipschitz, $\mathcal{T}_{m,0}g = g$, and $\mathcal{T}_{m,\tau}(g+k) = \mathcal{T}_{m,\tau}g+k$ for every constant $k$. If $g_1 \le g_2+k$ on $\R$, then $\mathcal{T}_{m,\tau}g_1 \le \mathcal{T}_{m,\tau}g_2+k$ on $\R$; in particular $|g_1-g_2| \le k$ implies $|\mathcal{T}_{m,\tau}g_1-\mathcal{T}_{m,\tau}g_2| \le k$.
\item[(b)] $\mathcal{T}_{m,\tau_1}\mathcal{T}_{m,\tau_2}g = \mathcal{T}_{m,\tau_1+\tau_2}g$.
\item[(c)] $0 \le \mathcal{T}_{m',\tau}g-\mathcal{T}_{m,\tau}g \le \frac{\tau}{2}(m'-m)$.
\end{enumerate}
\end{lemma}

The constant in~(c) is attained by $g(y) = y$, for which $\mathcal{T}_{m,\tau}g = g+\frac{m\tau}{2}$.

\begin{proof}
(a) If $g_1(y_1+\sqrt{\tau}z) \le g_2(y_2+\sqrt{\tau}z)+k$ for all $z$, then $\mathcal{T}_{m,\tau}g_1(y_1) \le \mathcal{T}_{m,\tau}g_2(y_2)+k$: for $m > 0$ exponentiate, integrate, take logarithms and divide by $m$; for $m = 0$ integrate. With $g_1 = g_2 = g$ and $k = |y_1-y_2|$ this is the Lipschitz bound, and with $y_1 = y_2$ it is the comparison; the other two claims are immediate.

(b) Let $Z'$ be a standard Gaussian variable independent of $Z$.
For $m > 0$, $\exp(m\,\mathcal{T}_{m,\tau_2}g(y)) = \E[\exp(mg(y+\sqrt{\tau_2}Z'))]$, so by Fubini's theorem
\begin{align*}
   \E\big[\exp\big(m\,\mathcal{T}_{m,\tau_2}g(y+\sqrt{\tau_1}Z)\big)\big]& = \E\big[\exp\big(mg(y+\sqrt{\tau_1}Z+\sqrt{\tau_2}Z')\big)\big]\\
   & = \E\big[\exp\big(mg(y+\sqrt{\tau_1+\tau_2}\,Z)\big)\big],
\end{align*}
because $\sqrt{\tau_1}Z+\sqrt{\tau_2}Z'$ has the law of $\sqrt{\tau_1+\tau_2}\,Z$. For $m = 0$ the same computation runs without exponentials.

(c) Fix $y$ and $\delta \ge 0$, and let $X = g(y+\sqrt{\delta}Z)-g(y)$, so that $|X| \le \sqrt{\delta}|Z|$. The cumulant generating function $\chi(r) = \log\E[e^{rX}]$ is finite and smooth, $\chi(0) = 0$, and $\mathcal{T}_{m,\delta}g(y) = g(y)+\chi(m)/m$ for $m > 0$, $\mathcal{T}_{0,\delta}g(y) = g(y)+\chi'(0)$. For $0 \le r \le 1$, $\chi''(r)$ is the variance of $X$ under the law with density $e^{rX}/\E[e^{rX}]$, hence
\[
   0 \le \chi''(r) \le \frac{\E[X^2e^{rX}]}{\E[e^{rX}]} \le \delta\,\kappa(\delta),\qquad
   \kappa(\delta) = \frac{\E\big[Z^2e^{\sqrt{\delta}|Z|}\big]}{\E\big[e^{-\sqrt{\delta}|Z|}\big]}.
\]
The functions $\chi$ and $\frac12\delta\kappa(\delta)r^2-\chi$ are convex on $[0,1]$ and vanish at $0$, so their chord slopes from $0$ are nondecreasing: $r\mapsto\chi(r)/r$ and $r\mapsto\frac12\delta\kappa(\delta)r-\chi(r)/r$ are nondecreasing on $(0,1]$, with limits $\chi'(0)$ and $-\chi'(0)$ at $r = 0$.
As a result, \begin{equation}\label{eq:layer-one-step}
   0 \le \mathcal{T}_{m',\delta}g-\mathcal{T}_{m,\delta}g \le \frac{\delta\,\kappa(\delta)}{2}\,(m'-m).
\end{equation}
Now let $n \ge 1$ and $\delta = \tau/n$, and write $A_k = \mathcal{T}_{m,k\delta}g$ and $A'_k = \mathcal{T}_{m',k\delta}g$. By~(b), $A_{k+1} = \mathcal{T}_{m,\delta}A_k$ and $A'_{k+1} = \mathcal{T}_{m',\delta}A'_k$, so
\[
   A'_{k+1}-A_{k+1} = \big(\mathcal{T}_{m',\delta}A'_k-\mathcal{T}_{m,\delta}A'_k\big)+\big(\mathcal{T}_{m,\delta}A'_k-\mathcal{T}_{m,\delta}A_k\big).
\]
The function $A'_k$ is $1$-Lipschitz by~(a), so the first bracket lies in $[0,\frac12\delta\kappa(\delta)(m'-m)]$ by~\eqref{eq:layer-one-step}; once $A'_k \ge A_k$, the second lies in $[0,\sup(A'_k-A_k)]$ by~(a). By induction, $0 \le A'_n-A_n \le \frac12\tau\kappa(\tau/n)(m'-m)$. Finally $\kappa(\tau/n)\to\E[Z^2] = 1$ as $n\to\infty$, by dominated convergence.
\end{proof}

A single step gives~\eqref{eq:layer-one-step} with $\delta = \tau$, and $\kappa(\tau) > 1$ for $\tau > 0$; the subdivision removes this factor.

\begin{proposition}[Guerra's continuity estimate]\label{prop:psi-lip}
Let $0 \le c \le d$. Then
\begin{equation}\label{eq:psi-lip}
   |\psi(\nu)-\psi(\nu')| \le \Wone(\nu,\nu')\qquad\text{for all finitely supported $\nu,\nu'\in\Ps([c,d])$.}
\end{equation}
\end{proposition}

This is a form of Guerra's continuity estimate for the Parisi functional \cite{Guerra2003}; in the form used here it corresponds to \cite[Proposition~2.1]{Mourrat2022} for a single spin with law $P_1(\sigma)\propto e^{h\sigma}$ on $\{-1,1\}$ at time $0$ (see Remark~\ref{rem:mourrat-dict}).

\begin{proof}
For a list $\ell = ((v_0,z_0),\dots,(v_n,z_n))$ of weights $v_k \ge 0$ and atoms $z_0 \le \dots \le z_n$ in $[s,d]$, and a mass $m \ge 0$ with $m+v_0+\dots+v_n \le 1$, let $C_{m,s}(\ell)$ be the cascade that starts at time $s$ with mass $m$ and adds the weight $v_k$ at time $z_k$:
\[
   C_{m,s}(\ell) = \mathcal{T}_{m,\,z_0-s}\,\mathcal{T}_{m+v_0,\,z_1-z_0}\cdots\mathcal{T}_{m+v_0+\dots+v_{n-1},\,z_n-z_{n-1}}\,\mathcal{T}_{m+v_0+\dots+v_n,\,d-z_n}\lch.
\]
It is $1$-Lipschitz by Lemma~\ref{lem:layer}(a).
If $\nu = \sum_{a = 0}^Kw_a\delta_{y_a}$ and $\ell_{\nu} = ((w_a,y_a))_a$, then the masses of $C_{0,0}(\ell_{\nu})$ are $0$ on $[0,y_0]$, $m_a$ on $[y_a,y_{a+1}]$, and $1$ on $[y_K,d]$; here we use $c \ge 0$. Since $\mathcal{T}_{1,d-y_K}\lch = \lch+\frac12(d-y_K)$ by~\eqref{eq:cole-hopf-one}, Lemma~\ref{lem:layer}(a) gives
\[
   \psi(\nu) = 2\,C_{0,0}(\ell_{\nu})(h)-d.
\]

\emph{Step 1: splitting atoms.} Replacing $(v_k,z_k)$ by $(v',z_k),(v'',z_k)$ with $v'+v'' = v_k$ inserts the layer $\mathcal{T}_{m+v_0+\dots+v_{k-1}+v',\,0}$, which is the identity. So, $C_{m,s}$ does not change.

\emph{Step 2: equal weights.} If $\ell = ((v_k,z_k))_k$ and $\ell' = ((v_k,z'_k))_k$ carry the same weights, then
\begin{equation}\label{eq:casc-move}
   \big|C_{m,s}(\ell)-C_{m,s}(\ell')\big| \le \frac12\sum_{k = 0}^nv_k\,|z_k-z'_k|.
\end{equation}
We argue by induction on $n$; the case of empty lists is trivial. By symmetry let $z_0 \le z'_0$, let $R,R'$ be the lists with the first pair removed, and let $m_1 = m+v_0$.
Then $C_{m,s}(\ell) = \mathcal{T}_{m,z_0-s}C_{m_1,z_0}(R)$, and by Lemma~\ref{lem:layer}(b), $C_{m,s}(\ell') = \mathcal{T}_{m,z_0-s}\mathcal{T}_{m,z'_0-z_0}C_{m_1,z'_0}(R')$ and $C_{m_1,z_0}(R') = \mathcal{T}_{m_1,z'_0-z_0}C_{m_1,z'_0}(R')$. So, by the induction hypothesis and Lemma~\ref{lem:layer}(c) with the masses $m \le m_1 \le 1$,
\[
   \big|C_{m_1,z_0}(R)-\mathcal{T}_{m,z'_0-z_0}C_{m_1,z'_0}(R')\big|
 \le \big|C_{m_1,z_0}(R)-C_{m_1,z_0}(R')\big|+\frac{z'_0-z_0}{2}\,v_0
 \le \frac12\sum_{k = 0}^nv_k|z_k-z'_k|,
\]
and~\eqref{eq:casc-move} follows from Lemma~\ref{lem:layer}(a) applied to $\mathcal{T}_{m,z_0-s}$.

\emph{Step 3: a common refinement.} Let $\nu,\nu'$ be finitely supported on $[c,d]$, and let $0 = r_0 < r_1 < \dots < r_J = 1$ list the partial sums of the weights of $\nu$ and of $\nu'$.
By Lemma~\ref{lem:quantile}(a), $Q_{\nu}$ and $Q_{\nu'}$ are constant on each interval $(r_k,r_{k+1}]$, say equal to $z_k$ and $z'_k$. Splitting the atoms of $\nu$ and $\nu'$ along these intervals turns $\ell_{\nu}$ and $\ell_{\nu'}$ into $\ell = ((r_{k+1}-r_k,z_k))_k$ and $\ell' = ((r_{k+1}-r_k,z'_k))_k$. By Steps~1 and~2 and by~\eqref{eq:quantile-dist},
\begin{align*}
   |\psi(\nu)-\psi(\nu')|& = 2\,\big|C_{0,0}(\ell)(h)-C_{0,0}(\ell')(h)\big| \le \sum_k(r_{k+1}-r_k)\,|z_k-z'_k|\\
   & = \int_0^1\big|Q_{\nu}-Q_{\nu'}\big|\,\dd u = \Wone(\nu,\nu').\qedhere
\end{align*}
\end{proof}

Step~3 is the north-west corner rule, which builds the monotone coupling of $\nu$ and $\nu'$.
Between consecutive atoms the mass of the cascade $C_{0,0}(\ell_{\nu})$ is the distribution function $s\mapsto\nu([c,s])$, and $\int_0^1|Q_{\nu}-Q_{\nu'}|\,\dd u = \int_c^d|\nu([c,s])-\nu'([c,s])|\,\dd s$. In this form Proposition~\ref{prop:psi-lip} is an $L^1$ continuity estimate for the Parisi functional in its functional order parameter.

\begin{proposition}[Extension of $\psi$]\label{prop:psi-ext}
Let $0 \le c \le d$.
For every $\mu\in\Ps([c,d])$ the limit $\bar\psi(\mu) = \lim_{n\to\infty}\psi(\mu^{(n)})$ exists, and:
\begin{enumerate}
\item[(a)] $|\bar\psi(\mu)-\bar\psi(\mu')| \le \Wone(\mu,\mu')$ for all $\mu,\mu'\in\Ps([c,d])$;
\item[(b)] $\bar\psi(\nu) = \psi(\nu)$ for every finitely supported $\nu\in\Ps([c,d])$;
\item[(c)] if $\Psi\colon\Ps([c,d])\to\R$ is Lipschitz for $\Wone$ and agrees with $\psi$ on finitely supported measures, then $\Psi = \bar\psi$.
\end{enumerate}
\end{proposition}

\begin{proof}
By Proposition~\ref{prop:psi-lip}, the triangle inequality for $\Wone$ and Lemma~\ref{lem:quantile}(b), $|\psi(\mu^{(n)})-\psi(\mu^{(k)})| \le (d-c)\big(\frac{1}{n+1}+\frac{1}{k+1}\big)$, so $(\psi(\mu^{(n)}))_n$ is a Cauchy sequence. In the same way $|\psi(\mu^{(n)})-\psi(\mu'^{(n)})| \le \Wone(\mu,\mu')+2(d-c)/(n+1)$, which gives~(a) as $n\to\infty$. For~(b), $|\psi(\nu^{(n)})-\psi(\nu)| \le (d-c)/(n+1)$.
For~(c), if $\Psi$ is $L'$-Lipschitz, then $|\Psi(\mu)-\psi(\mu^{(n)})| = |\Psi(\mu)-\Psi(\mu^{(n)})| \le L'(d-c)/(n+1)$.
\end{proof}

\begin{theorem}[The limit $K\to\infty$ at a Dirac datum]\label{thm:klimit}
Let $t > 0$ and $x \ge 0$. Suppose, as in Theorem~\ref{thm:dirac-hl}(c), that $F_{\infty}$ is the infimum of $\mathcal{P}$ over all $K$-step order parameters, $K \ge 0$, and put $u_{\infty} = 2F_{\infty}-x-\frac{t}{2}$.
\begin{enumerate}
\item[(a)] \emph{Finitely supported measures.} $u_{\infty}$ is the infimum of $\psi(\nu)+\W^2(\delta_x,\nu)/(2t)$ over the finitely supported probability measures $\nu$ on $[x,x+t]$.
\item[(b)] \emph{Equal masses.} For $n \ge 0$ let $V_n$ be the infimum of $\psi(\nu)+\W^2(\delta_x,\nu)/(2t)$ over the measures $\nu = \frac{1}{n+1}\sum_{i = 0}^n\delta_{y_i}$ with $x \le y_0 \le \dots \le y_n \le x+t$, that is, the infimum of $\mathrm{HL}_n(\Phi_0(\nu),w,y)$ (Theorem~\ref{thm:kstep-hopflax}) over these measures, with the equal weights $w_i = \frac{1}{n+1}$. Then
\begin{equation}\label{eq:eqmass-rate}
   u_{\infty} \le V_n \le u_{\infty}+\frac{2t}{n+1},
\end{equation}
and in particular $V_n\to u_{\infty}$ as $n\to\infty$.
\item[(c)] \emph{All measures.}
\[
   u_{\infty} = \inf\Big\{\bar\psi(\mu)+\frac{\W^2(\delta_x,\mu)}{2t}:\mu\in\Ps([x,x+t])\Big\}.
\]
\item[(d)] \emph{Larger supports.} For every $B \ge x+t$, $u_{\infty}$ is also the infimum of $\psi(\nu)+\W^2(\delta_x,\nu)/(2t)$ over the finitely supported probability measures on $[x,B]$, and of $\bar\psi(\mu)+\W^2(\delta_x,\mu)/(2t)$ over $\mu\in\Ps([x,B])$.
As a result, $u_{\infty}$ is the infimum of $\bar\psi(\mu)+\W^2(\delta_x,\mu)/(2t)$ over all compactly supported probability measures $\mu$ on $[x,\infty)$.
\end{enumerate}
\end{theorem}

\begin{proof}
(a) The map $\nu\mapsto\nu_t$ of Section~\ref{sec:dirac} is a bijection from the finitely supported probability measures on $[0,1]$ onto those on $[x,x+t]$, with inverse $y\mapsto(y-x)/t$ on the atoms, and $2\mathcal{P}(\nu)-x-\frac{t}{2} = \psi(\nu_t)+\W^2(\delta_x,\nu_t)/(2t)$ by Theorem~\ref{thm:dirac-hl}(b).
So~(a) follows from Theorem~\ref{thm:dirac-hl}(c).

(b) Equal-mass measures are finitely supported, so $V_n \ge u_{\infty}$ by~(a). Conversely, let $\nu$ be finitely supported on $[x,x+t]$. Its discretization $\nu^{(n)}$ is one of the measures that define $V_n$, and $\Wone(\nu^{(n)},\nu) \le t/(n+1)$ by Lemma~\ref{lem:quantile}(b). By Proposition~\ref{prop:psi-lip} and Lemma~\ref{lem:quantile}(c) with $d = x+t$,
\[
   V_n \le \psi(\nu^{(n)})+\frac{\W^2(\delta_x,\nu^{(n)})}{2t} \le \psi(\nu)+\frac{\W^2(\delta_x,\nu)}{2t}+\frac{2t}{n+1},
\]
and taking the infimum over $\nu$ gives~\eqref{eq:eqmass-rate} by~(a).

(c) Let $\mu\in\Ps([x,x+t])$. By~(a), $\psi(\mu^{(n)})+\W^2(\delta_x,\mu^{(n)})/(2t) \ge u_{\infty}$ for every $n$, and by Proposition~\ref{prop:psi-ext} and Lemma~\ref{lem:quantile}(b),(c) the left-hand side tends to $\bar\psi(\mu)+\W^2(\delta_x,\mu)/(2t)$. So, $u_{\infty}$ is a lower bound.
It is the greatest lower bound by~(a), since the finitely supported measures belong to $\Ps([x,x+t])$ and $\bar\psi = \psi$ on them (Proposition~\ref{prop:psi-ext}(b)).

(d) Let $\nu = \sum_aw_a\delta_{y_a}$ be finitely supported on $[x,B]$, and let $\tilde\nu = \sum_aw_a\delta_{\min(y_a,x+t)}$ be its truncation at $x+t$. By~\eqref{eq:quantile-dist} and Lemma~\ref{lem:quantile}(a), $\Wone(\nu,\tilde\nu) = D := \sum_aw_a(y_a-x-t)_{+}$, so $\psi(\tilde\nu) \le \psi(\nu)+D$ by Proposition~\ref{prop:psi-lip}.
For $y \ge x+t$,
\[
   \frac{(y-x)^2}{2t} = \frac{t}{2}+(y-x-t)+\frac{(y-x-t)^2}{2t},
\]
hence $\W^2(\delta_x,\tilde\nu)/(2t)+D \le \W^2(\delta_x,\nu)/(2t)$. Adding the two inequalities shows that truncation does not increase $\psi+\W^2(\delta_x,\cdot)/(2t)$. Since $\tilde\nu$ lives on $[x,x+t]$ and the finitely supported measures on $[x,x+t]$ are among those on $[x,B]$, the first claim follows from~(a). The second follows from the first as in~(c), now with Lemma~\ref{lem:quantile}(c) for $d = B$.
For the last claim, a compactly supported probability measure $\mu$ on $[x,\infty)$ lives on some $[x,B]$ with $B \ge x+t$, and $\bar\psi(\mu)$ does not depend on $B$, because $\mu^{(n)}$ does not.
\end{proof}

Part~(d) is Step~4 of the proof of \cite[Proposition~4.1]{Mourrat2022}, written in our variables. The rate~\eqref{eq:eqmass-rate} compares two variational problems and does not involve $N$; the Parisi formula enters only through the identification $F_{\infty} = \lim_NF_N(h,t,x)$ of Section~\ref{sec:dirac}.

\begin{remark}[Mourrat's normalization]\label{rem:mourrat-dict}
We compare our formulation with that of \cite{Mourrat2022}, denoting the variables of \cite{Mourrat2022} with a subscript $\mathrm{M}$.
Mourrat considers a centered Gaussian field $H_N$ on $\R^N$ with $\E[H_N(\sigma)H_N(\tau)] = N\xi(\sigma\cdot\tau/N)$, which for the SK model is $\xi(r) = r^2$, a reference measure $P_N$, the enriched free energy $\bar{F}_N(t_{\mathrm{M}},\mu)$ of \cite[(2.7)]{Mourrat2022}, and its initial condition $\psi_{\mathrm{M}}(\mu) = \lim_N\bar{F}_N(0,\mu)$. For $P_N$ uniform on $\{-1,1\}^N$, his Theorem~1.1, in the Hopf--Lax form of his Proposition~4.1, states that
\[
   \lim_{N\to\infty}-\frac{1}{N}\E\log\int\exp\big(\sqrt{2t_{\mathrm{M}}}\,H_N(\sigma)-Nt_{\mathrm{M}}\big)\,\dd P_N(\sigma)
 = \sup_{\mu}\Big\{\psi_{\mathrm{M}}(\mu)-t_{\mathrm{M}}\int\xi^{*}\Big(\frac{s}{t_{\mathrm{M}}}\Big)\,\dd\mu(s)\Big\},
\]
the supremum running over the probability measures $\mu$ on $\R_{+}$ with $\int\xi^{*}\,\dd\mu < \infty$, where $\xi^{*}(s) = \sup_{r \ge 0}(rs-\xi(r)) = \frac14s_{+}^2$.
Take for $P_N$ the product of copies of $P_1(\sigma) = e^{h\sigma}/(2\cosh h)$ on $\{-1,1\}$, which satisfies his standing assumption~(2.1). The two normalizations are related as follows:
\begin{itemize}
\item \emph{Time.} $\sqrt{2t_{\mathrm{M}}}\,H_N$ has covariance $2t_{\mathrm{M}}NR^2$, while the two-body term of $H_N$ in Section~\ref{sec:sk}, with coupling $\sqrt{t/(2N)}$, has covariance $\frac{t}{2}NR^2$; thus $t = 4t_{\mathrm{M}}$.
\item \emph{Initial datum.} The cascade field of $\bar{F}_N$ at level $\ell$ has variance $2(q_{\ell}-q_{\ell-1})$, and the parameter attached to the layer $(q_{\ell-1},q_{\ell}]$ is $\mu([0,q_{\ell-1}])$, matching $m_a$ on $(y_a,y_{a+1}]$. Evaluating $\psi_{\mathrm{M}} = \bar{F}_1(0,\cdot)$ \cite[Proposition~3.1(1)]{Mourrat2022} by \cite[Lemma~2.2]{Mourrat2022} gives
\[
   \psi(\nu) = 2\lch(h)-2\psi_{\mathrm{M}}(\mu),\qquad\text{$\nu$ the image of $\mu$ under $q\mapsto 2q$;}
\]
the term $-q_k|\sigma|^2$ of \cite[(2.7)]{Mourrat2022} produces $-y_K$, and the normalization of $P_1$ produces $\lch(h)$. By Proposition~\ref{prop:psi-ext}(c), the identity passes to $\bar\psi$ and to the continuous extension of $\psi_{\mathrm{M}}$.
\item \emph{Free energy.} The field $\sqrt{x}\,g_i$, with its compensator $-\frac{x}{2}$, is Mourrat's enrichment at $\mu = \delta_{x/2}$; hence $u_N(h,t,x) = 2\lch(h)-2\bar{F}_N(t/4,\delta_{x/2})$.
\item \emph{Distances and costs.} $\Wone(\nu,\nu') = 2\Wone(\mu,\mu')$, so the estimate $|\psi_{\mathrm{M}}(\mu)-\psi_{\mathrm{M}}(\mu')| \le \Wone(\mu,\mu')$ of \cite[(2.8)]{Mourrat2022} corresponds to Proposition~\ref{prop:psi-lip} on every $[c,d]\subset[0,\infty)$.
Also, $t_{\mathrm{M}}\int\xi^{*}((s-\frac{x}{2})/t_{\mathrm{M}})\,\dd\mu(s) = \frac{1}{4t}\int(y-x)_{+}^2\,\dd\nu(y)$, so that
\[
   2\lch(h)-2\Big\{\psi_{\mathrm{M}}(\mu)-t_{\mathrm{M}}\int\xi^{*}\Big(\frac{s-x/2}{t_{\mathrm{M}}}\Big)\,\dd\mu(s)\Big\} = \psi(\nu)+\frac{1}{2t}\int(y-x)_{+}^2\,\dd\nu(y),
\]
which is $\psi(\nu)+\W^2(\delta_x,\nu)/(2t)$ when $\nu$ lives on $[x,\infty)$.
\end{itemize}
As a result, at $x = h = 0$, Theorem~\ref{thm:klimit}(d) is Mourrat's Theorem~1.1 for the SK model, with the supremum restricted to compactly supported measures. The two suprema agree: for a measure of unbounded support in his class, his truncation inequality \cite[(4.5)]{Mourrat2022}, the counterpart of Theorem~\ref{thm:klimit}(d) once $\psi_{\mathrm{M}}$ is extended by continuity, produces a compactly supported measure with no smaller value.
For $x > 0$ or $h\neq 0$, Theorem~\ref{thm:klimit} expresses $\lim_N\bar{F}_N(t_{\mathrm{M}},\delta_{x/2})$ as a Hopf--Lax formula started at $\delta_{x/2}$. This is the case $\mu = \delta_{x/2}$ of \cite[Conjecture~2.5]{Mourrat2022}, which Mourrat and Panchenko proved for product reference measures and initial measures of bounded support \cite[Theorem~1.1]{MourratPanchenko2020}.
For a Dirac initial measure, $\bar{F}_N(t_{\mathrm{M}},\delta_{x/2}) = \lch(h)-\frac12u_N(h,4t_{\mathrm{M}},x)$ is an affine function of the free energy of the mixed model of Section~\ref{sec:dirac}, and this case also follows from the Parisi formula for that model \cite{Panchenko2014}, which is the route taken here.
In the Hopf--Lax formula \cite[(1.7)]{Mourrat2022} started at $\delta_{x/2}$ the supremum runs over the measures $\mu$ on $\R_{+}$ with $\int\xi^{*}\,\dd\mu < \infty$ (in \cite{MourratPanchenko2020} it is written, with the opposite sign, as an infimum over measures of bounded support). Mass below $x/2$ costs nothing, since $\xi^{*}$ vanishes on $(-\infty,0]$; lifting that mass to $x/2$ leaves the cost unchanged and does not decrease $\psi_{\mathrm{M}}$ \cite[Lemma~2.4]{Mourrat2022}, so the two formulas agree.
Finally, at $x = 0$ the truncation in~(d), which preserves equal masses, shows that $V_n = 2\lch(h)-2f^{(n+1)}(t_{\mathrm{M}},\delta_0)$ for the finite-dimensional approximations $f^{(k)}$ of \cite[Section~5]{Mourrat2022}, and~\eqref{eq:eqmass-rate} reads $0 \le f(t_{\mathrm{M}},\delta_0)-f^{(k)}(t_{\mathrm{M}},\delta_0) \le 4t_{\mathrm{M}}/k$, where $f(t_{\mathrm{M}},\delta_0) = \lim_N\bar{F}_N(t_{\mathrm{M}},\delta_0)$.
In this section the Hopf--Lax structure is exact for finitely many atoms, by the identity of Theorem~\ref{thm:kstep-hopflax}(a) and the triviality of transport from a Dirac mass (Theorem~\ref{thm:dirac-hl}(a)), and the limit $K\to\infty$ is a density argument based on Proposition~\ref{prop:psi-lip}.
\end{remark}

\section{Conclusions and Outlook}\label{sec:conclusions}

Throughout this work, we have established that Guerra's interpolation translates directly into the language of quadratic optimal transport. 
By expressing the finite-$N$ free energy as a mechanical action, one obtains a classical subsolution of the inviscid Hamilton--Jacobi equation. Along a straight segment, Guerra's sum rule admits a geometric reading: the quadratic term subtracted by Guerra corresponds to the transport cost, the remainder represents a duality gap, and Guerra's bound enforces Kantorovich admissibility uniformly in the volume. 
The variational optimization over order parameters thus yields a Hopf--Lax formula, with the self-consistency condition defining a Brenier transport map. On the underlying configuration space, this path traces an exponential family whose remainder equals a relative entropy.

The two archetypal mean-field models exhibit sharply contrasting behaviors. 
In the Curie--Weiss model the transport map $T_t(y) = y-t\tanh y$ remains optimal precisely up to the critical point. Beyond it the characteristics cross on a Maxwell line, whose two maximizers represent the two signs of spontaneous magnetization: the ferromagnetic transition corresponds directly to the breakdown of optimal transport.

In the replica-symmetric Sherrington--Kirkpatrick model, instead, the Lata\l{}a--Guerra
inequality keeps the cavity map injective and its Brenier potential strictly convex on
the physical half-line, even beyond the de~Almeida--Thouless instability line. Its curvature, the
longitudinal eigenvalue, strictly exceeds the replicon one, so transporting Dirac masses
does not detect the instability, which lives in an orthogonal
direction of Wasserstein space: the geodesic that splits a Dirac mass into two atoms at
fixed barycenter. Along it the replicon eigenvalue appears as a second variation, and
its sign governs Toninelli's instability and the failure of displacement convexity of
the Parisi functional. In this language, replica symmetry breaking is reflected in the
fact that, beyond the de~Almeida--Thouless line, splitting an atom of the overlap
distribution lowers the objective. 

For $K$ steps of replica symmetry breaking, Guerra's
bounds take Hopf--Lax form over atomic measures and, in the limit $K\to\infty$, they recover, at Dirac
data, the Wasserstein Hopf--Lax representation of the Parisi formula due to Mourrat
and to Mourrat and Panchenko \cite{Mourrat2022,MourratPanchenko2020}, with an explicit
rate for equal-mass approximations.

Several directions remain open.
\begin{itemize}
\item \emph{Beyond Dirac data.} From a Dirac mass every coupling is the product one,
so the transport in Section~\ref{sec:kstep} is elementary. In the multi-dimensional
Hamilton--Jacobi formulation of Barra, Di~Biasio and Guerra \cite{BarraDiBiasioGuerra2010}, each step of
replica symmetry breaking carries its own spatial coordinate. Atomic measures then appear as genuine
initial data, and the associated mass matrix should give a non-trivial Kantorovich
duality between overlap distributions.
\item \emph{Splitting beyond replica symmetry.} Splitting one atom of a $K$-step Parisi measure,
rather than a single Dirac mass, should reproduce the stability conditions of Parisi
theory, in particular the replicon-type condition of Jagannath and Tobasco
\cite{JagannathTobasco2017}, and would characterize the Parisi measure as a point where no
splitting direction has negative curvature.
\item \emph{Finite $N$ and entropic transport.} For the Curie--Weiss model the
finite-$N$ action solves a viscous Hamilton--Jacobi equation with viscosity $1/N$
(Proposition~\ref{prop:cw-pde}), which suggests reading the finite-$N$ problem as an entropic
(Schr\"odinger) transport problem whose small-noise limit is the thermodynamic limit.
\item \emph{Multi-species models and neural networks.} In bipartite and multi-species
systems, such as the Hopfield model of associative memory, the kinetic term of the
Hamilton--Jacobi equation is no longer convex \cite{Mourrat2021Nonconvex}.
Transport then becomes a saddle problem
with several splitting directions\footnote{For completeness, we note that in neural network models such as Hopfield architectures: (i) because of the saddle-point nature of the variational principle, standard interpolation does not yield a remainder with a definite sign, an obstacle only recently overcome through generalized interpolation schemes \cite{Diego1}; (ii) while the microscopic Hamiltonian requires interacting neurons, the interactions need not be strictly pairwise: many-body extensions admit analogous interpolating treatments, allowing modern dense neural networks \cite{Peter,Krotov1,Krotov2,Lowe} to be incorporated into this framework \cite{Eli_NN2020,Diego2,Linda_JSP2022}.}; extending the dictionary to these models, where the
correct form of replica symmetry breaking is not always known, is a natural perspective
for applications to artificial intelligence.
\end{itemize}

\newpage

\appendix

\section{Statistical-Mechanics Background and Notation}\label{app:nutshell}

For readers coming from partial differential equations or optimal transport, we briefly recall the standard statistical-mechanics notation for the Curie--Weiss and Sherrington--Kirkpatrick models and how it relates to the parameters $(t,x)$ used in the main text; see \cite{Ellis1985,MPV1987,TalagrandI,TalagrandII,PanchenkoBook} for comprehensive treatments.

Given a configuration space $\Sigma_N = \{-1,1\}^N$, a Hamiltonian $H_N\colon\Sigma_N\to\R$, and an inverse temperature $\beta \ge 0$, the Boltzmann--Gibbs measure and the pressure per spin are
\[
   \omega(\sigma) = \frac{e^{-\beta H_N(\sigma)}}{Z_N(\beta)},\qquad
   Z_N(\beta) = \sum_{\sigma\in\Sigma_N}e^{-\beta H_N(\sigma)},\qquad
   \alpha_N(\beta) = \frac{1}{N}\log Z_N(\beta),
\]
which satisfies the Gibbs variational principle $\alpha_N(\beta) = \frac{1}{N}\sup_{\mu}\{-\beta\mu(H_N)+\Ent(\mu)\}$ over probability measures $\mu$ on $\Sigma_N$.

\subsection{The Curie--Weiss model}\label{app:cw}
The Curie--Weiss (CW) Hamiltonian in an external field $h\in\R$ is
\begin{equation}\label{app:cwH}
   H_N(\sigma) = -\frac{1}{2N}\sum_{i,j = 1}^N\sigma_i\sigma_j-h\sum_{i = 1}^N\sigma_i
 = -N\Big(\frac{m(\sigma)^2}{2}+h\,m(\sigma)\Big),\qquad
   m(\sigma) = \frac{1}{N}\sum_{i = 1}^N\sigma_i\in[-1,1].
\end{equation}
The limiting pressure is given by the variational formula \cite{EllisNewman1978,Ellis1985}
\begin{equation}\label{app:cwsol}
   \alpha(\beta,h) = \sup_{M\in[-1,1]}\Big\{\log\big(2\cosh(\beta(M+h))\big)-\frac{\beta}{2}M^2\Big\},
\end{equation}
whose maximizers solve the self-consistency equation $M = \tanh(\beta(M+h))$. At $h = 0$, $M = 0$ is the unique solution for $\beta \le 1$ (paramagnetic phase), whereas for $\beta > \beta_c = 1$ two nonzero maximizers $\pm M(\beta)$ appear (ferromagnetic phase).
Setting $t = \beta$ and $x = \beta h$ gives the partition function $Z_N(t,x)$ of Section~\ref{sec:cw}, and with $y = x+tM$ the variational formula~\eqref{app:cwsol} becomes the Hopf--Lax formula of Theorem~\ref{thm:cw-hopflax}(b), with stationarity condition $x = T_t(y) = y-t\tanh y$.

\subsection{The Sherrington--Kirkpatrick model}\label{app:sk}
The Sherrington--Kirkpatrick (SK) model \cite{SK1975} has the random Hamiltonian
\begin{equation}\label{app:skH}
   H_N(\sigma;J) = -\frac{1}{\sqrt{N}}\sum_{1 \le i < j \le N}J_{ij}\,\sigma_i\sigma_j-h\sum_{i = 1}^N\sigma_i ,
\end{equation}
where the couplings $J_{ij}$ are independent standard Gaussian variables (quenched disorder).
The quenched pressure is $\alpha_N(\beta,h) = \frac{1}{N}\E\log Z_N(\beta,h;J)$, whose thermodynamic limit was established by Guerra and Toninelli \cite{GuerraToninelli}, and the order parameter is the law of the overlap $R(\sigma,\tau) = \frac{1}{N}\sum_{i = 1}^N\sigma_i\tau_i\in[-1,1]$ of two independent replicas drawn from the Gibbs measure $\omega_J$ under the quenched product measure $\langle\cdot\rangle = \E(\omega_J\otimes\omega_J)(\cdot)$.

In the replica-symmetric (RS) regime the overlap concentrates at a deterministic value $q\in[0,1]$, and the limiting pressure is given by the RS functional
\begin{equation}\label{app:rs}
   \alpha_{\RS}(\beta,h;q) = \E\,\log\big(2\cosh(\beta h+\beta\sqrt{q}\,Z)\big)+\frac{\beta^2}{4}(1-q)^2,\qquad
   q = \E\,\tanh^2\big(\beta h+\beta\sqrt{q}\,Z\big),
\end{equation}
with $Z\sim\mathcal{N}(0,1)$. At $h = 0$ and $\beta < 1$ the solution is $q = 0$, and the limiting pressure is the annealed value $\log 2+\beta^2/4$ \cite{ALR1987}. De~Almeida and Thouless \cite{AT1978} identified the instability region of the RS solution as
\begin{equation}\label{app:at}
   \beta^2\,\E\,\sech^4\big(\beta h+\beta\sqrt{q}\,Z\big) > 1,
\end{equation}
beyond which the RS value exceeds the limiting pressure \cite{Toninelli2002}.
In Section~\ref{sec:sk} we set $t = \beta^2$, write $h$ in place of $\beta h$, and introduce an auxiliary Gaussian field $\sqrt{x}\,g_i$ of variance $x \ge 0$ (with $x = 0$ at the physical point); the interaction $\sqrt{t/(2N)}\sum_{i,j}g_{ij}\sigma_i\sigma_j$ differs from~\eqref{app:skH} only by the mean-zero diagonal terms, while $(g_{ij}+g_{ji})/\sqrt{2}$ are standard Gaussian.
In these variables $F_N(\beta h,\beta^2,0) = \alpha_N(\beta,h)$, $\RS(h,t,0,q) = \alpha_{\RS}(\beta,h;q)$, the RS equation is $q = \bar{q}(h,tq)$, and~\eqref{app:at} is $\LAT(h,t,tq) < 0$.

\subsection{Replica symmetry breaking and the Parisi formula}\label{app:rsb}
In Parisi's ansatz \cite{Parisi1979,Parisi1980,MPV1987}, at the $K$-th step of replica symmetry breaking (RSB) the overlap distribution is approximated by a finitely supported probability measure $\nu = \sum_{a = 0}^Kw_a\delta_{q_a}$ on $[0,1]$ with cumulative weights $m_a = w_0+\dots+w_a$, and the associated Parisi functional $\mathcal{P}(\nu)$ is defined by the backward Cole--Hopf cascade of Section~\ref{sec:kstep}.
Guerra proved the upper bound $\alpha_N(\beta,h) \le \mathcal{P}(\nu)$ for all $N$ and $\nu$ \cite{Guerra2003}, and Talagrand \cite{Talagrand2006} and Panchenko \cite{Panchenko2014} proved the Parisi formula
\begin{equation}\label{app:parisi}
   \lim_{N\to\infty}\alpha_N(\beta,h) = \inf_{\nu}\mathcal{P}(\nu),
\end{equation}
whose unique minimizer over $\Ps([0,1])$ is the Parisi measure \cite{AuffingerChen2015,ACZ2020}.
Sections~\ref{sec:twoatom} and~\ref{sec:kstep} analyze the second variation of the two-atom functional along Wasserstein splittings and recast Guerra's $K$-step bound and the limit $K\to\infty$ at Dirac data in Hopf--Lax form \cite{Mourrat2022,MourratPanchenko2020}.

\section{Numerical Tables}\label{app:num}

The tables below are computed using 240-node Gauss--Hermite quadrature and exact binomial sums, and illustrate the results of Sections~\ref{sec:cw}, \ref{sec:sk} and~\ref{sec:twoatom}.

\begin{table}[ht]
\centering\small
\caption{The terms of~\eqref{eq:cw-rem-kl} and the bound~\eqref{eq:fleming-bound} for $N = 30$, $t = 1.4$ and $x = 0$.
The third row is the positive solution $M^{*}\approx 0.814529$ of $M = \tanh(tM)$, at which $\frac{t}{2}(\tanh(tM)-M)^2$ vanishes.}\label{tab:cw}
\begin{tabular}{cccccc}
\toprule
$M$ & $\Gu(t,M)$ & $\mathcal{F}_N(t,0,M)$ & $R_N(t,M)$ & $\KL(\mu_0\|\mu_1)/N$ & $\frac{t}{2}(\tanh(tM)-M)^2$ \\
\midrule
$0.300000$ & $0.7158699$ & $0.7352863$ & $0.0938437$ & $0.0676098$ & $0.0065769$ \\
$0.700000$ & $0.7687810$ & $0.7790350$ & $0.0409327$ & $0.0288607$ & $0.0019712$ \\
$0.814529\ (M^{*})$ & $0.7732417$ & $0.7812681$ & $0.0364719$ & $0.0286192$ & $0.0000000$ \\
\bottomrule
\end{tabular}
\end{table}

\begin{table}[ht]
\centering\small
\caption{The replica-symmetric SK model at $h = 0$, with an additional Gaussian field of variance $x > 0$, along the AT line.
For each $x$, the pair $(t_{\mathrm{AT}},w^{*})$ solves $T_t(0,w^{*}) = x$ and $\LAT(0,t,w^{*}) = 0$; at the computed points $|\LAT(0,t_{\mathrm{AT}},w^{*})| < 10^{-13}$. The last column is the curvature $\LL = \partial_w^2\Psi_t(0,w^{*})$ (Theorem~\ref{thm:lamL-vs-lamAT}).}\label{tab:sk-at}
\begin{tabular}{ccccc}
\toprule
$x$ & $t_{\mathrm{AT}}(x)$ & $w^{*}$ & $q^{*} = (w^{*}-x)/t_{\mathrm{AT}}$ & $\LL(0,t_{\mathrm{AT}},w^{*})$ \\
\midrule
$0.02$ & $1.54419$ & $0.37124$ & $0.22746$ & $0.385905$ \\
$0.05$ & $1.76095$ & $0.57005$ & $0.29532$ & $0.481814$ \\
$0.10$ & $1.98607$ & $0.80580$ & $0.35537$ & $0.560547$ \\
$0.20$ & $2.28515$ & $1.16450$ & $0.42207$ & $0.641312$ \\
$0.40$ & $2.68619$ & $1.72615$ & $0.49369$ & $0.720078$ \\
$0.80$ & $3.22928$ & $2.63308$ & $0.56764$ & $0.792400$ \\
\bottomrule
\end{tabular}
\end{table}

\begin{table}[ht]
\centering\footnotesize
\setlength{\tabcolsep}{3.5pt}
\caption{The two-atom functional at $x = 0$ (Theorem~\ref{thm:twoatom} and Remark~\ref{rem:cole-hopf-taylor}): the ratio $(\mathcal{P}_2(\nu_{\eps})-\RS(h,t,0,\bar{m}))/\W^2(\delta_{\bar{m}},\nu_{\eps})$ at $\eps = 0.04$ and $\eps = 0.01$, its Richardson extrapolation to $\eps = 0$ from $\eps\in\{0.01,0.005,0.0025\}$, and the coefficient $\frac{t}{4}\LAT(h,t,t\bar{m})$. The mark $q^{*}$ indicates that $\bar{m}$ solves the replica-symmetric equation.}\label{tab:twoatom}
\begin{tabular}{ccccccccc}
\toprule
$t$ & $h$ & $\bar{m}$ & $m$ & $\eps = 0.04$ & $\eps = 0.01$ & Extrapolated & $\frac{t}{4}\LAT(h,t,t\bar{m})$ & Rel.\ error \\
\midrule
$2.25$ & $0.40$ & $0.50000$ & $0.30$ & $+0.043181$ & $+0.034632$ & $+0.031616$ & $+0.031616$ & $3.62\times 10^{-6}$ \\
$2.25$ & $0.00$ & $0.35260\ (q^{*})$ & $0.50$ & $-0.062960$ & $-0.074360$ & $-0.078369$ & $-0.078369$ & $1.59\times 10^{-6}$ \\
$1.44$ & $0.20$ & $0.22490\ (q^{*})$ & $0.60$ & $+0.028689$ & $+0.024279$ & $+0.022751$ & $+0.022751$ & $9.57\times 10^{-7}$ \\
$2.25$ & $0.10$ & $0.35705\ (q^{*})$ & $0.30$ & $-0.059042$ & $-0.069029$ & $-0.072569$ & $-0.072569$ & $1.96\times 10^{-6}$ \\
$0.64$ & $0.30$ & $0.60000$ & $0.80$ & $+0.099119$ & $+0.098711$ & $+0.098573$ & $+0.098573$ & $1.06\times 10^{-9}$ \\
\bottomrule
\end{tabular}
\end{table}
\clearpage
\section{Lean Declarations}\label{app:lean}

The library \texttt{GuerraHJ} is in the directory \texttt{lean/} of the repository \url{https://github.com/coscienza-research/optimal-transport-guerra-interpolation}; the toolchain is \texttt{leanprover/lean4:v4.28.0}, with Mathlib \texttt{v4.28.0}.
It has 18 modules, the axiom check \lean{GuerraHJ/Axioms.lean} and a root file, 7{,}891 lines in total, with 519 theorems and lemmas and no \texttt{sorry}. The axiom check runs \texttt{\#print axioms} on 125 principal theorems; each depends only on \texttt{propext}, \texttt{Classical.choice} and \texttt{Quot.sound}. Table~\ref{tab:lean-modules} describes the modules, and Tables~\ref{tab:lean-dict} and~\ref{tab:lean-dict-2} list, for each numbered statement, the declarations that prove it.
Declarations are in the namespaces \texttt{GuerraHJ}, \texttt{GuerraHJ.CW} and \texttt{GuerraHJ.SK}.

The external inputs of Section~\ref{sec:formal} appear in Lean as follows. Guerra's $K$-step interpolation is the structure \lean{GuerraKStepInterpolation}, an argument of \lean{guerra_kstep_bound} and of the other statements of Theorem~\ref{thm:kstep-hopflax}(b).
The Parisi formula is the hypothesis \texttt{IsGLB (Set.range \dots parisiP \dots) Finf}, with $F_{\infty}$ a real number, in the statements of Theorems~\ref{thm:dirac-hl}(c) and~\ref{thm:klimit}. The cubic remainder bound is the hypothesis \texttt{hrem} of \lean{toninelli_instability} and the hypotheses \texttt{h12}, \texttt{h1} of \lean{not_displacement_convex_of_expansion} in Theorem~\ref{thm:twoatom}(c).
In \lean{Parisi/KLimit.lean} a probability measure on $[c,d]$ is represented by its quantile function via the structure \lean{MonoQuantile}, and the right-hand sides of~\eqref{eq:quantile-dist} serve as the definitions of $\Wone$ and $\W^2(\delta_x,\cdot)$. Appendix~\ref{app:formal} reproduces the main definitions and statements.

\begin{table}[ht]
\centering\footnotesize
\renewcommand{\arraystretch}{1.05}
\caption{Modules of the Lean~4 library \texttt{GuerraHJ}.}\label{tab:lean-modules}
\begin{tabular}{L{3.8cm}L{9.7cm}}
\toprule
\textbf{Module} & \textbf{Contents} \\
\midrule
\lean{Basic/Hyperbolic.lean} & Smoothness, convexity and Taylor bounds for $\lch(y) = \log(2\cosh y)$. \\
\lean{Basic/TanhPoly.lean} & Polynomials in $\tanh y$ and their derivatives;
Riccati identities for $L_1,\dots,L_4$. \\
\lean{Abstract/SumRule.lean} & Sum rules along paths and segments, Kantorovich inequalities, optimality certificates. \\
\lean{Abstract/ExpGeodesic.lean} & Exponential families: relative entropy identities and the symmetrized divergence. \\
\lean{CurieWeiss/Finite.lean} & Finite-$N$ CW partition function, viscous Hamilton--Jacobi and heat equations, Guerra's sum rule. \\
\lean{CurieWeiss/HopfLax.lean} & Binomial reduction and the bound $\log(N+1)/N$ for the Hopf--Lax limit. \\
\lean{CurieWeiss/Brenier.lean} & Optimality of $T_t$ for $0 < t \le 1$; monotonicity and uniqueness; non-optimality and spontaneous magnetization for $t > 1$. \\
\lean{CurieWeiss/Entropy.lean} & Relative-entropy decomposition of Guerra's remainder;
Gibbs functional of product measures. \\
\lean{CurieWeiss/Fleming.lean} & Hubbard--Stratonovich identity and the Gaussian-shift bound. \\
\lean{Gaussian/Stein.lean} & Gaussian integration by parts in one and several variables; the variance derivative. \\
\lean{Gaussian/HeatExp.lean} & Gaussian integration by parts and the variance derivative for integrands of exponential growth. \\
\lean{Gaussian/Interpolation.lean} & Finite Gibbs measures with Gaussian disorder; derivatives of $\E\log Z$. \\
\lean{SK/Finite.lean} & Finite-$N$ SK action, its derivatives, the subsolution property, the RS sum rule and the Hopf--Lax bound. \\
\lean{SK/RS.lean} & Hopf--Lax form of the RS functional, the gap $\LL-\LAT$, the Lata{\l}a--Guerra lemma, monotonicity of the cavity map. \\
\lean{Parisi/TwoAtom.lean} & Cole--Hopf layer; coefficients $u_1,u_2$; two-atom functional, its second-order surrogate and second variation; $K$-step cascade and Parisi functional;
Guerra's $K$-step sum rule; couplings with a Dirac mass;
Hopf--Lax form at Dirac data. \\
\lean{Parisi/KLimit.lean} & Quantile functions, equal-mass discretization and continuous extension of $\psi$. \\
\lean{Parisi/ColeHopfLayer.lean} & Cole--Hopf layers of $1$-Lipschitz functions: comparison, semigroup law, cumulant generating function, the mass bound. \\
\lean{Parisi/GuerraLipschitz.lean} & Cascades along lists of atoms, splitting of atoms, the north-west corner rule, the $\Wone$-Lipschitz bound for $\psi$ and the limit $K\to\infty$ at Dirac data. \\
\lean{Axioms.lean} & \texttt{\#print axioms} for 125 principal theorems. \\
\bottomrule
\end{tabular}
\end{table}

\begin{table}[ht]
\centering\scriptsize
\renewcommand{\arraystretch}{1.04}
\caption{Numbered statements of Sections~\ref{sec:abstract}--\ref{sec:sk} and their Lean~4 declarations.}\label{tab:lean-dict}
\begin{tabular}{L{2.5cm}L{3.3cm}L{7.5cm}}
\toprule
\textbf{Statement} & \textbf{Module} & \textbf{Lean~4 declarations} \\
\midrule
Lemma~\ref{lem:sumrule-path} & \lean{Abstract/SumRule} & \lean{sumrule_path}, \lean{kantorovich_path}, \lean{sumrule_segment} \\
Proposition~\ref{prop:dual-certificate} & \lean{Abstract/SumRule} & \lean{kantorovich_coupling}, \lean{map_optimal_of_certificate} \\
Proposition~\ref{prop:exp-geodesic} & \lean{Abstract/ExpGeodesic} & \lean{Phi_one_eq}, \lean{KL_01_integral}, \lean{KL_10_integral}, \lean{jeffreys} \\
Proposition~\ref{prop:cw-pde} & \lean{CurieWeiss/Finite} & \lean{alpha_zero}, \lean{viscous_HJ}, \lean{heat_equation} \\
Theorem~\ref{thm:cw-hopflax} & \lean{CurieWeiss/Finite}, \lean{HopfLax} & \lean{guerra_sumrule}, \lean{kantorovich_cw_coupling}, \lean{HL_le_alpha}, \lean{alpha_le}, \lean{tendsto_alpha} \\
Theorem~\ref{thm:cw-brenier} & \lean{CurieWeiss/Brenier} & \lean{HL_on_graph}, \lean{brenier_cw}, \lean{brenier_cw_value}, \lean{strictConvexOn_Psi}, \lean{strictMono_Tmap}, \lean{cw_unique}, \lean{surjective_Tmap}, \lean{maximizer_Tmap}, \lean{HL_max_unique}, \lean{Tmap_neg}, \lean{not_monotone_Tmap}, \lean{not_convexOn_Psi}, \lean{not_optimal_Tmap}, \lean{log_two_lt_HL_zero}, \lean{spontaneous_magnetization} \\
Proposition~\ref{prop:cw-entropy} & \lean{CurieWeiss/Entropy} & \lean{gibbs_zero_mag}, \lean{gibbs_zero_mag_sq}, \lean{KL_cw_guerra}, \lean{guerra_remainder_vs_KL}, \lean{guerra_remainder_vs_KL_fixed}, \lean{guerra_remainder_gap}, \lean{gibbs_variational}, \lean{G_prod}, \lean{G_prod_eq_guerra}, \lean{G_prod_mag_le_guerra} \\
Theorem~\ref{thm:cw-fleming} & \lean{CurieWeiss/Fleming} & \lean{hubbard_stratonovich}, \lean{fleming_bound}, \lean{fleming_bound_hopfLax}, \lean{fleming_entropy_cost}, \lean{guerra_le_fleming}, \lean{fleming_le_guerra} \\
Proposition~\ref{prop:gaussian-stein} & \lean{Gaussian/Stein}, \lean{Interpolation} & \lean{stein}, \lean{hasDerivAt_integral_gaussian_sqrt}, \lean{stein_coord}, \lean{integral_coord_mul_linGibbs}, \lean{hasDerivAt_integral_log_linZ} \\
Theorem~\ref{thm:sk-finite} & \lean{SK/Finite} & \lean{hasDerivAt_u_t}, \lean{hasDerivAt_u_x}, \lean{B_sq_le_A}, \lean{subsolution}, \lean{F_zero}, \lean{guerra_sum_rule}, \lean{guerra_rs_bound}, \lean{u_le_hopfLax} \\
Theorem~\ref{thm:lamL-vs-lamAT} & \lean{SK/RS} & \lean{hasDerivAt_P0}, \lean{hasDerivAt_u0}, \lean{hasDerivAt_qbar}, \lean{hasDerivAt_RS}, \lean{hasDerivAt_dRS}, \lean{LamL_sub_LamAT}, \lean{LamAT_lt_LamL} \\
Theorem~\ref{thm:latala-guerra} & \lean{SK/RS} & \lean{lgA_pos}, \lean{latala_guerra}, \lean{strictAntiOn_qbar_div}, \lean{Tmap_div_lt_LamL}, \lean{LamL_pos_of_Tmap_nonneg}, \lean{Tmap_injOn}, \lean{hlRS_lt} \\
\bottomrule
\end{tabular}
\end{table}

\begin{table}[ht]
\centering\scriptsize
\renewcommand{\arraystretch}{1.04}
\caption{Numbered statements of Sections~\ref{sec:twoatom} and~\ref{sec:kstep} and their Lean~4 declarations.}\label{tab:lean-dict-2}
\begin{tabular}{L{2.5cm}L{3.3cm}L{7.5cm}}
\toprule
\textbf{Statement} & \textbf{Module} & \textbf{Lean~4 declarations} \\
\midrule
Lemma~\ref{lem:riccati-collapse} & \lean{Basic/TanhPoly}, \lean{TwoAtom} & \lean{coleHopf_eq_lch_add}, \lean{coleHopf_one}, \lean{riccati_quartic}, \lean{u1_eq_one_sub}, \lean{u2_eq_coleHopf_pde}, \lean{u2_eq_riccati_collapse} \\
Theorem~\ref{thm:twoatom} & \lean{Parisi/TwoAtom} & \lean{P2_diag}, \lean{twoAtom_energy_eq}, \lean{P2_split_sub_jet}, \lean{P2Jet_zero}, \lean{hasDerivAt_P2Jet}, \lean{hasDerivAt_P2Jet_zero}, \lean{hasDerivAt_dP2Jet_zero}, \lean{twoAtom_expansion_variance}, \lean{toninelli_instability}, \lean{hasDerivAt_dilation}, \lean{hasDerivAt_dilation_zero}, \lean{hasDerivAt_dDilation_zero}, \lean{w2_sq_twoAtom_pythagoras}, \lean{not_displacement_convex_splitting}, \lean{not_displacement_convex_of_expansion}, \lean{LamL_zero_eq}, \lean{LamAT_zero_eq}, \lean{LamL_LamAT_zero_neg} \\
Theorem~\ref{thm:weighted-hj} & \lean{Parisi/TwoAtom} & \lean{weighted_hj_subsolution}, \lean{weighted_sumrule_path}, \lean{weighted_kantorovich_path}, \lean{weighted_pythagoras} \\
Proposition~\ref{prop:pk-cases} & \lean{Parisi/TwoAtom} & \lean{parisiP_one_atom}, \lean{parisiP_two_atom}, \lean{second_moment_ladder} \\
Theorem~\ref{thm:kstep-hopflax} & \lean{Parisi/TwoAtom} & \lean{parisiK_eq_hopfLax}, \lean{parisiPK_variance_decomp}, \lean{guerra_kstep_action_sum_rule}, \lean{guerra_kstep_sum_rule}, \lean{guerra_kstep_bound}, \lean{guerra_kstep_bound_parisiP}, \lean{guerra_kstep_hopfLax}, \lean{guerra_kstep_bias_variance} \\
Theorem~\ref{thm:dirac-hl} & \lean{Parisi/TwoAtom} & \lean{coupling_dirac_eq}, \lean{cost_coupling_dirac}, \lean{transportCosts_dirac_atomic}, \lean{parisiP_eq_hopfLax}, \lean{hopfLax_dirac_of_parisi}, \lean{hopfLax_one_atom}, \lean{hopfLax_twoAtom_jet} \\
Lemma~\ref{lem:quantile} & \lean{Parisi/KLimit} & \lean{Ladder.quantile_eq}, \lean{Ladder.integral_comp_quantile}, \lean{MonoQuantile.ladder_quantile_bounds}, \lean{MonoQuantile.integral_abs_ladder_sub_le}, \lean{MonoQuantile.abs_costQ_sub_le} \\
Lemma~\ref{lem:layer} & \lean{Parisi/ColeHopfLayer} & \lean{layer_le_of_le}, \lean{layer_lip1}, \lean{layer_zero_var}, \lean{layer_add_const}, \lean{abs_layer_sub_le}, \lean{layer_add}, \lean{layer_eq_cgf}, \lean{iteratedDeriv_two_cgf_incr_le}, \lean{chordSlope_bounds}, \lean{layer_mass_step}, \lean{tendsto_kap}, \lean{layer_mass_bound} \\
Proposition~\ref{prop:psi-lip} & \lean{Parisi/GuerraLipschitz} & \lean{Ladder.psi_eq_casc}, \lean{Refines.casc_eq}, \lean{abs_casc_sub_le}, \lean{exists_common_refinement}, \lean{integral_abs_lq_sub}, \lean{Ladder.quantile_eq_lq}, \lean{psiW1Lipschitz_of_nonneg} \\
Proposition~\ref{prop:psi-ext} & \lean{Parisi/KLimit} & \lean{MonoQuantile.cauchySeq_psiSeq}, \lean{MonoQuantile.tendsto_psiSeq}, \lean{MonoQuantile.abs_psiExt_sub_le}, \lean{psiExt_toMono}, \lean{psiExt_unique} \\
Theorem~\ref{thm:klimit} & \lean{Parisi/KLimit}, \lean{GuerraLipschitz} & \lean{isGLB_ladder_of_parisi}, \lean{kLimit_eqMass_of_nonneg}, \lean{kLimit_hopfLax_of_nonneg}, \lean{kLimit_hopfLax_wide_of_nonneg}, \lean{kLimit_hopfLax_compact_of_nonneg} \\
\bottomrule
\end{tabular}
\end{table}

\clearpage
\section{Formal Statements}\label{app:formal}

This appendix reproduces the main definitions of \texttt{GuerraHJ} and the statements of some principal theorems; proofs are omitted, and each statement ends with \texttt{:=}. In Lean, \texttt{√} is \texttt{Real.sqrt}, which vanishes on negative numbers, and $0^{-1} = 0$.
So \texttt{coleHopf} and \texttt{coleHopfOp} vanish for $m = 0$, a case that never occurs in the statements, where Parisi parameters lie in $(0,1)$; the layers of Section~\ref{sec:klimit} are defined separately and are the heat semigroup for $m = 0$. \texttt{P2} is defined for all real $a,b$ and agrees with $\mathcal{P}_2$ for $a \le b$ and $x+ta \ge 0$. \texttt{HL} is a supremum (\texttt{⨆}) of a family that is bounded above.
Configurations of the CW model are subsets of $\{1,\dots,N\}$, the set of positive spins, and those of the SK model are functions \texttt{Fin N → Bool}. The disorder of the SK model is indexed by \texttt{(Fin N × Fin N) ⊕ Fin N}: ordered pairs, diagonal included, and single sites.

\subsection*{Curie--Weiss model (Section~\ref{sec:cw})}

\begin{Verbatim}[fontsize=\scriptsize]
def lch (y : ℝ) : ℝ := Real.log (2 * Real.cosh y)

def mk (N k : ℕ) : ℝ := (2 * (k : ℝ) - N) / N

def eK (N : ℕ) (t x : ℝ) (k : ℕ) : ℝ := (N : ℝ) * (t * mk N k ^ 2 / 2 + x * mk N k)

def mag (N : ℕ) (s : Finset (Fin N)) : ℝ := mk N s.card

def expo (N : ℕ) (t x : ℝ) (s : Finset (Fin N)) : ℝ := eK N t x s.card

def Z (N : ℕ) (t x : ℝ) : ℝ := ∑ s : Finset (Fin N), Real.exp (expo N t x s)

def alpha (N : ℕ) (t x : ℝ) : ℝ := Real.log (Z N t x) / N

def HL (t x : ℝ) : ℝ := ⨆ y : ℝ, (lch y - (x - y) ^ 2 / (2 * t))

def Tmap (t y : ℝ) : ℝ := y - t * Real.tanh y

theorem HL_le_alpha (N : ℕ) (hN : N ≠ 0) {t : ℝ} (ht : 0 < t) (x : ℝ) : HL t x ≤ alpha N t x :=

theorem alpha_le (N : ℕ) (hN : N ≠ 0) {t : ℝ} (ht : 0 < t) (x : ℝ) :
    alpha N t x ≤ HL t x + Real.log ((N : ℝ) + 1) / N :=

theorem brenier_cw {t : ℝ} (ht0 : 0 < t) (ht1 : t ≤ 1) (ρ0 : Measure ℝ) [IsFiniteMeasure ρ0]
    (hρ : Integrable (fun y => y) ρ0) (P : Measure (ℝ × ℝ))
    (hP0 : P.map Prod.fst = ρ0) (hP1 : P.map Prod.snd = ρ0.map (Tmap t))
    (hc : Integrable (fun p : ℝ × ℝ => (p.2 - p.1) ^ 2 / (2 * t)) P) :
    ∫ y, (Tmap t y - y) ^ 2 / (2 * t) ∂ρ0 ≤ ∫ p, (p.2 - p.1) ^ 2 / (2 * t) ∂P :=

theorem not_optimal_Tmap {t : ℝ} (ht : 1 < t) :
    ∃ (ρ0 : Measure ℝ) (P : Measure (ℝ × ℝ)), IsProbabilityMeasure ρ0 ∧
      P.map Prod.fst = ρ0 ∧ P.map Prod.snd = ρ0.map (Tmap t) ∧
      ∫ p, (p.2 - p.1) ^ 2 / (2 * t) ∂P < ∫ y, (Tmap t y - y) ^ 2 / (2 * t) ∂ρ0 :=
\end{Verbatim}

\subsection*{Replica-symmetric SK model (Section~\ref{sec:sk})}

\begin{Verbatim}[fontsize=\scriptsize]
abbrev gaussPi (ι : Type*) [Fintype ι] : Measure (ι → ℝ) :=
  Measure.pi fun _ : ι => gaussianReal 0 1

abbrev Conf (N : ℕ) := Fin N → Bool

def sgn : Bool → ℝ
  | true => 1
  | false => -1

def spin (σ : Conf N) (i : Fin N) : ℝ := sgn (σ i)

abbrev Idx (N : ℕ) := (Fin N × Fin N) ⊕ Fin N

def coef (α β : ℝ) : Idx N → Conf N → ℝ :=
  Sum.elim (fun p σ => α * (spin σ p.1 * spin σ p.2)) (fun i σ => β * spin σ i)

def field (h : ℝ) (σ : Conf N) : ℝ := h * ∑ i, spin σ i

def linE (a : ι → Ω → ℝ) (b : Ω → ℝ) (g : ι → ℝ) (σ : Ω) : ℝ := ∑ k, a k σ * g k + b σ

def linZ (a : ι → Ω → ℝ) (b : Ω → ℝ) (g : ι → ℝ) : ℝ := ∑ σ, Real.exp (linE a b g σ)

def logZ (h α β : ℝ) : ℝ :=
  ∫ g, Real.log (linZ (coef N α β) (field N h) g) ∂(gaussPi (Idx N))

def F (h t x : ℝ) : ℝ := logZ N h (√(t / (2 * N))) (√x) / N

def u (h t x : ℝ) : ℝ := 2 * F N h t x - x - t / 2

def P0 (h w : ℝ) : ℝ := ∫ z, lch (h + √w * z) ∂(gaussianReal 0 1)

def u0 (h w : ℝ) : ℝ := 2 * P0 h w - w

def RS (h t x q : ℝ) : ℝ := P0 h (x + t * q) + t / 4 * (1 - q) ^ 2

def hlRS (h t x w : ℝ) : ℝ := u0 h w + (x - w) ^ 2 / (2 * t)

def qbar (h w : ℝ) : ℝ := ∫ z, Real.tanh (h + √w * z) ^ 2 ∂(gaussianReal 0 1)

def dqbar (h w : ℝ) : ℝ :=
  ∫ z, (1 - Real.tanh (h + √w * z) ^ 2) * (1 - 3 * Real.tanh (h + √w * z) ^ 2)
    ∂(gaussianReal 0 1)

def Tmap (h t w : ℝ) : ℝ := w - t * qbar h w

def LamL (h t w : ℝ) : ℝ := 1 - t * dqbar h w

def LamAT (h t w : ℝ) : ℝ :=
  1 - t * ∫ z, (1 - Real.tanh (h + √w * z) ^ 2) ^ 2 ∂(gaussianReal 0 1)

theorem subsolution (h : ℝ) {t x : ℝ} (ht : 0 < t) (hx : 0 < x) :
    ∃ ut ux, HasDerivAt (fun t => u N h t x) ut t ∧ HasDerivAt (fun x => u N h t x) ux x
      ∧ ut + ux ^ 2 / 2 ≤ 0 :=

theorem u_le_hopfLax (h : ℝ) {t x w : ℝ} (ht : 0 < t) (hx : 0 ≤ x) (hw : 0 ≤ w)
    (hxw : 0 < x + w) : u N h t x ≤ u N h 0 w + (x - w) ^ 2 / (2 * t) :=

theorem LamL_sub_LamAT (h t w : ℝ) :
    LamL h t w - LamAT h t w
      = 2 * t * ∫ z, Real.tanh (h + √w * z) ^ 2 * (1 - Real.tanh (h + √w * z) ^ 2)
          ∂(gaussianReal 0 1) :=

theorem latala_guerra (h : ℝ) {w : ℝ} (hw : 0 < w) : w * dqbar h w < qbar h w :=

theorem Tmap_injOn (h : ℝ) {t x w₁ w₂ : ℝ} (ht : 0 < t) (hx : 0 ≤ x) (hw₁ : 0 < w₁)
    (hw₂ : 0 < w₂) (h₁ : Tmap h t w₁ = x) (h₂ : Tmap h t w₂ = x) : w₁ = w₂ :=

theorem hlRS_lt (h : ℝ) {t x w₀ w : ℝ} (ht : 0 < t) (hx : 0 ≤ x) (hw₀ : 0 < w₀)
    (hT : Tmap h t w₀ = x) (hw : 0 < w) (hne : w ≠ w₀) : hlRS h t x w₀ < hlRS h t x w :=
\end{Verbatim}

\subsection*{Two atoms (Section~\ref{sec:twoatom})}

\begin{Verbatim}[fontsize=\scriptsize]
def L1 (y : ℝ) : ℝ := Real.tanh y

def L2 (y : ℝ) : ℝ := 1 - Real.tanh y ^ 2

def L3 (y : ℝ) : ℝ := -2 * Real.tanh y * (1 - Real.tanh y ^ 2)

def L4 (y : ℝ) : ℝ := -2 * (1 - Real.tanh y ^ 2) * (1 - 3 * Real.tanh y ^ 2)

def coleHopf (m τ y : ℝ) : ℝ :=
  m⁻¹ * Real.log (∫ z, Real.exp (m * lch (y + √τ * z)) ∂(gaussianReal 0 1))

def P2 (h t x m a b : ℝ) : ℝ :=
  (∫ z, coleHopf m (t * (b - a)) (h + √(x + t * a) * z) ∂(gaussianReal 0 1))
    + t / 4 * ((1 - b) ^ 2 - m * (b ^ 2 - a ^ 2))

def u1 (m y : ℝ) : ℝ := (L2 y + m * L1 y ^ 2) / 2

def u2 (m y : ℝ) : ℝ :=
  (L4 y + 4 * m * L1 y * L3 y + 2 * m * L2 y ^ 2 + 4 * m ^ 2 * L1 y ^ 2 * L2 y) / 8

def Eu1 (h m w : ℝ) : ℝ := ∫ z, u1 m (h + √w * z) ∂(gaussianReal 0 1)

def Eu2 (h m w : ℝ) : ℝ := ∫ z, u2 m (h + √w * z) ∂(gaussianReal 0 1)

def P2Jet (h t x mbar m ε : ℝ) : ℝ :=
  let w := x + t * mbar
  let wa := w - (1 - m) * t * ε
  P0 h wa + (t * ε) * Eu1 h m wa + (t * ε) ^ 2 * Eu2 h m w
    + (t / 4 * (1 - mbar) ^ 2 - m * t / 2 * ε + t / 4 * m * (1 - m) * ε ^ 2)

theorem u2_eq_riccati_collapse (m y : ℝ) :
    u2 m y = (1 - m) ^ 2 / 8 * L4 y - m * (1 - m) / 4 * L2 y ^ 2 :=

theorem Eu2_eq (h m w : ℝ) :
    Eu2 h m w = -(1 - m) ^ 2 / 4 * dqbar h w
      - m * (1 - m) / 4 * ∫ z, (1 - Real.tanh (h + √w * z) ^ 2) ^ 2 ∂(gaussianReal 0 1) :=

theorem P2_diag (h t x : ℝ) {m : ℝ} (hm : m ≠ 0) (q : ℝ) : P2 h t x m q q = RS h t x q :=

theorem P2_split_sub_jet (h t x mbar m ε : ℝ) :
    P2 h t x m (mbar - (1 - m) * ε) (mbar + m * ε) - P2Jet h t x mbar m ε
      = (∫ z, coleHopf m (t * ε) (h + √(x + t * mbar - (1 - m) * t * ε) * z)
            ∂(gaussianReal 0 1))
        - (P0 h (x + t * mbar - (1 - m) * t * ε)
          + (t * ε) * Eu1 h m (x + t * mbar - (1 - m) * t * ε)
          + (t * ε) ^ 2 * Eu2 h m (x + t * mbar)) :=

theorem P2Jet_zero (h t x mbar m : ℝ) : P2Jet h t x mbar m 0 = RS h t x mbar :=

theorem hasDerivAt_P2Jet_zero (h t x mbar m : ℝ) (hw : 0 < x + t * mbar) :
    HasDerivAt (P2Jet h t x mbar m) 0 0 :=
\end{Verbatim}

\subsection*{\texorpdfstring{$K$}{K}-step functional and Dirac data (Sections~\ref{sec:kstep}--\ref{sec:dirac})}

\begin{Verbatim}[fontsize=\scriptsize]
def coleHopfOp (m τ : ℝ) (g : ℝ → ℝ) (y : ℝ) : ℝ :=
  m⁻¹ * Real.log (∫ z, Real.exp (m * g (y + √τ * z)) ∂(gaussianReal 0 1))

def cumWeights {K : ℕ} (w : Fin (K + 1) → ℝ) : Fin K → ℝ :=
  fun a => ∑ c ∈ Finset.Iic a.castSucc, w c

def parisiCascade : List (ℝ × ℝ) → ℝ → ℝ
  | [], y => lch y
  | (m, τ) :: rest, y => coleHopfOp m τ (parisiCascade rest) y

def cascadeSlabs {K : ℕ} (m : Fin K → ℝ) (y : Fin (K + 1) → ℝ) : List (ℝ × ℝ) :=
  List.ofFn fun a : Fin K => (m a, y a.succ - y a.castSucc)

def parisiPhi0 (h w0 : ℝ) (slabs : List (ℝ × ℝ)) : ℝ :=
  ∫ z, parisiCascade slabs (h + √w0 * z) ∂(gaussianReal 0 1)

def cascadePhi {K : ℕ} (h : ℝ) (m : Fin K → ℝ) (y : Fin (K + 1) → ℝ) : ℝ :=
  parisiPhi0 h (y 0) (cascadeSlabs m y)

def parisiPK {K : ℕ} (t Φ : ℝ) (w q : Fin (K + 1) → ℝ) : ℝ :=
  Φ + t / 4 * (1 - 2 * q (Fin.last K) + ∑ a, w a * q a ^ 2)

def parisiP {K : ℕ} (h t x : ℝ) (w q : Fin (K + 1) → ℝ) : ℝ :=
  parisiPK t (cascadePhi h (cumWeights w) (fun a => x + t * q a)) w q

def u0K {K : ℕ} (Φ : ℝ) (y : Fin (K + 1) → ℝ) : ℝ :=
  2 * Φ - y (Fin.last K)

def hlK {K : ℕ} (t x Φ : ℝ) (w y : Fin (K + 1) → ℝ) : ℝ :=
  u0K Φ y + ∑ a, w a * (y a - x) ^ 2 / (2 * t)

structure GuerraKStepInterpolation (N K : ℕ) (h t x Φ : ℝ) (w q : Fin (K + 1) → ℝ) where
  φ : ℝ → ℝ
  M1 : Fin (K + 1) → ℝ → ℝ
  M2 : Fin (K + 1) → ℝ → ℝ
  hw_nonneg : ∀ a, 0 ≤ w a
  hw_sum : ∑ a, w a = 1
  hφ_zero : φ 0 = u0K Φ (fun a => x + t * q a)
  hφ_one : φ 1 = u N h t x
  hφ_cont : ContinuousOn φ (Icc 0 1)
  hM1_int : ∀ a, IntervalIntegrable (M1 a) volume 0 1
  hM2_int : ∀ a, IntervalIntegrable (M2 a) volume 0 1
  hJensen : ∀ a, ∀ s ∈ Icc (0 : ℝ) 1, M1 a s ^ 2 ≤ M2 a s
  hφ_deriv : ∀ s ∈ Ioo (0 : ℝ) 1, HasDerivAt φ
    (-(t / 2) * (∑ a, w a * M2 a s) - ∑ a, w a * (x - (x + t * q a)) * M1 a s) s

theorem guerra_kstep_bound {N K : ℕ} {h t x Φ : ℝ} {w q : Fin (K + 1) → ℝ}
    (G : GuerraKStepInterpolation N K h t x Φ w q) (ht : 0 < t) :
    F N h t x ≤ parisiPK t Φ w q :=

structure AtomicLadder where
  K : ℕ
  w : Fin (K + 1) → ℝ
  q : Fin (K + 1) → ℝ
  w_pos : ∀ a, 0 < w a
  w_sum : ∑ a, w a = 1
  q_mono : Monotone q
  q_nonneg : ∀ a, 0 ≤ q a
  q_le_one : ∀ a, q a ≤ 1

def psiAtomic {K : ℕ} (h : ℝ) (w y : Fin (K + 1) → ℝ) : ℝ :=
  2 * cascadePhi h (cumWeights w) y - y (Fin.last K)

theorem hopfLax_dirac_of_parisi (h : ℝ) {t : ℝ} (ht : t ≠ 0) (x Finf : ℝ)
    (hP : IsGLB (Set.range fun ν : AtomicLadder => parisiP h t x ν.w ν.q) Finf) :
    IsGLB (Set.range fun ν : AtomicLadder =>
        psiAtomic h ν.w (fun a => x + t * ν.q a)
          + ∑ a, ν.w a * ((x + t * ν.q a) - x) ^ 2 / (2 * t))
      (2 * Finf - x - t / 2) :=
\end{Verbatim}

\subsection*{The limit \texorpdfstring{$K\to\infty$}{K to infinity} (Section~\ref{sec:klimit})}

\begin{Verbatim}[fontsize=\scriptsize]
structure Ladder (a b : ℝ) where
  K : ℕ
  w : ℕ → ℝ
  y : ℕ → ℝ
  w_pos : ∀ i, 0 < w i
  w_sum : ∑ i ∈ Finset.range (K + 1), w i = 1
  y_mono : Monotone y
  y_mem : ∀ i, y i ∈ Icc a b

def mass (ν : Ladder a b) (j : ℕ) : ℝ := ∑ i ∈ Finset.range j, ν.w i

def quantile (ν : Ladder a b) (u : ℝ) : ℝ :=
  ν.y 0 + ∑ i ∈ Finset.range ν.K, (ν.y (i + 1) - ν.y i) * if ν.mass (i + 1) < u then 1 else 0

def psi (h : ℝ) (ν : Ladder a b) : ℝ :=
  psiAtomic h (fun i : Fin (ν.K + 1) => ν.w i) (fun i => ν.y i)

def cost (x t : ℝ) (ν : Ladder a b) : ℝ :=
  ∑ i ∈ Finset.range (ν.K + 1), ν.w i * ((ν.y i - x) ^ 2 / (2 * t))

def PsiW1Lipschitz (h L a b : ℝ) : Prop :=
  ∀ ν ν' : Ladder a b,
    |ν.psi h - ν'.psi h| ≤ L * ∫ u in (0 : ℝ)..1, |ν.quantile u - ν'.quantile u|

structure MonoQuantile (a b : ℝ) where
  Q : ℝ → ℝ
  mono : Monotone Q
  mem : ∀ u, Q u ∈ Icc a b

def ladder (Q : MonoQuantile a b) (n : ℕ) : Ladder a b where
  K := n
  w := fun _ => 1 / ((n : ℝ) + 1)
  y := fun i => Q.Q (((i : ℝ) + 1) / ((n : ℝ) + 1))
  w_pos := fun _ => by positivity
  w_sum := by
    show ∑ _i ∈ Finset.range (n + 1), 1 / ((n : ℝ) + 1) = 1
    rw [Finset.sum_const, Finset.card_range, nsmul_eq_mul, Nat.cast_add, Nat.cast_one,
      mul_one_div_cancel (by positivity)]
  y_mono := fun i j hij => Q.mono (div_le_div_of_nonneg_right
    (by exact_mod_cast Nat.add_le_add_right hij 1) (by positivity))
  y_mem := fun _ => Q.mem _

def psiSeq (h : ℝ) (Q : MonoQuantile a b) (n : ℕ) : ℝ := (Q.ladder n).psi h

def psiExt (h : ℝ) (Q : MonoQuantile a b) : ℝ := limUnder atTop (Q.psiSeq h)

def costQ {a b : ℝ} (x t : ℝ) (Q : MonoQuantile a b) : ℝ :=
  ∫ u in (0 : ℝ)..1, (Q.Q u - x) ^ 2 / (2 * t)

def eqMassValues (h t x : ℝ) (n : ℕ) : Set ℝ :=
  Set.range fun ν : {ν : Ladder x (x + t) // ν.K = n ∧ ∀ i, ν.w i = 1 / ((n : ℝ) + 1)} =>
    ν.1.psi h + ν.1.cost x t

theorem psiW1Lipschitz_of_nonneg (h : ℝ) {a b : ℝ} (ha : 0 ≤ a) : PsiW1Lipschitz h 1 a b :=

theorem kLimit_eqMass_of_nonneg (h : ℝ) {t : ℝ} (ht : 0 < t) {x : ℝ} (hx : 0 ≤ x) (Finf : ℝ)
    (hP : IsGLB (Set.range fun ν : AtomicLadder => parisiP h t x ν.w ν.q) Finf) :
    (∀ n : ℕ, 2 * Finf - x - t / 2 ≤ sInf (eqMassValues h t x n) ∧
      sInf (eqMassValues h t x n) ≤ 2 * Finf - x - t / 2 + 2 * t / ((n : ℝ) + 1)) ∧
    Tendsto (fun n => sInf (eqMassValues h t x n)) atTop (N (2 * Finf - x - t / 2)) :=

theorem kLimit_hopfLax_compact_of_nonneg (h : ℝ) {t : ℝ} (ht : 0 < t) {x : ℝ} (hx : 0 ≤ x)
    (Finf : ℝ) (hP : IsGLB (Set.range fun ν : AtomicLadder => parisiP h t x ν.w ν.q) Finf) :
    IsGLB (Set.range fun p : (Σ B : {B : ℝ // x + t ≤ B}, MonoQuantile x B.1) =>
      p.2.psiExt h + p.2.costQ x t) (2 * Finf - x - t / 2) :=
\end{Verbatim}

\end{document}